\documentclass{article}
\usepackage{authblk}
\usepackage{blindtext}
\usepackage{todonotes}

\usepackage{newtxtext,newtxmath, amsfonts, bm,mathtools, float}

\usepackage[subrefformat=parens,labelformat=parens]{subfig}

\mathtoolsset{showonlyrefs=true}
\usepackage[utf8]{inputenc}
\usepackage[boxed,lined,linesnumbered]{algorithm2e}
\usepackage{amsmath, textcomp, paralist, bm, natbib}
\usepackage{array}
\usepackage{wrapfig}
\usepackage{multirow}
\usepackage{tabularx}
\usepackage{booktabs}
\usepackage{float}
\usepackage{enumitem}
\DeclareMathAlphabet{\mathcal}{OMS}{cmsy}{m}{n}
\graphicspath{{Graphics/}}
\usepackage{arydshln}
\usepackage{xcolor}
\usepackage[hyphens,spaces,obeyspaces]{url}
\usepackage{indentfirst}
\usepackage{pdflscape}
\usepackage[flushleft]{threeparttable}

\usepackage{pifont}

\usepackage[margin=3cm]{geometry}
\usepackage[margin=2cm]{caption}
\usepackage{booktabs}

\usepackage{tikz}
\usetikzlibrary{positioning}

\SetKwInOut{Parameter}{parameter}
				
\newcommand{\var}{\text{var}}
\newcommand{\cov}{\text{cov}}
\newcommand{\binomial}{\text{Binomial}}
\newcommand{\logit}{\text{logit}}

\usepackage{adjustbox}
\newcolumntype{R}[2]{%
    >{\adjustbox{angle=#1,lap=\width-(#2)}\bgroup}%
    l%
    <{\egroup}%
}

\usepackage{array, ragged2e}
\newcolumntype{P}[1]{>{\raggedright\arraybackslash}p{#1}}
\graphicspath{{figures/}}

    \title{Comparison of new computational methods for geostatistical modelling of malaria}
    \author[1]{Spencer Wong}
    \author[1]{Jennifer A. Flegg}
    \author[2]{Nick Golding}
    \author[3]{Sevvandi Kandanaarachchi}
	\affil[1]{School of Mathematics and Statistics, The University of Melbourne, Parkville VIC 3010}
    \affil[2]{Telethon Kids Institute, Perth Children's Hospital, 15 Hospital Ave, Nedlands WA 6009}
    \affil[2]{Curtin University, Kent St, Bentley WA 6102}
    \affil[3]{CSIRO's Data61, Research Way, Clayton VIC 3168}

\date{}

\begin{document}

\maketitle
	
\begin{abstract}
\emph{Background}
Geostatistical analysis of health data is increasingly used to model spatial variation in malaria prevalence, burden, and other metrics. Traditional inference methods for geostatistical modelling are notoriously computationally intensive, motivating the development of newer, approximate methods. The appeal of faster methods is particularly great as the size of the region and number of spatial locations being modelled increases. 

\emph{Methods}
We present an applied comparison of four proposed `fast' geostatistical modelling methods and the software provided to implement them -- Integrated Nested Laplace Approximation (INLA), tree boosting with Gaussian processes and mixed effect models (GPBoost), Fixed Rank Kriging (FRK) and Spatial Random Forests (SpRF). We illustrate the four methods by estimating malaria prevalence on two different spatial scales -- country and continent. We compare the performance of the four methods on these data in terms of accuracy, computation time, and ease of implementation. 

\emph{Results}
Two of these methods -- SpRF and GPBoost -- do not scale well as the data size increases, and so are likely to be infeasible for larger-scale analysis problems. The two remaining methods -- INLA and FRK -- do scale well computationally, however the resulting model fits are very sensitive to the user's modelling assumptions and parameter choices. The binomial observation distribution commonly used for disease prevalence mapping with INLA fails to account for small-scale overdispersion present in the malaria prevalence data, which can lead to poor predictions. Selection of an appropriate alternative such as the Beta-binomial distribution is required to produce a reliable model fit. The small-scale random effect term in FRK overcomes this pitfall, but FRK model estimates are very reliant on providing a sufficient number and appropriate configuration of basis functions. Unfortunately the computation time for FRK increases rapidly with increasing basis resolution.

\emph{Conclusions}
INLA and FRK both enable scalable geostatistical modelling of malaria prevalence data. However care must be taken when using both methods to assess the fit of the model to data and plausibility of predictions, in order to select appropriate model assumptions and approximation parameters.   
\end{abstract}

\section{Introduction}


Spatial proximity often plays an important role in governing the spread of geographic processes, including fluctuations in the prevalence or incidence of infectious diseases. Thus, geostatistical modelling is a widely used method in the epidemiological mapping of infectious diseases and their impacts. This is particularly evident in the field of malaria mapping, where statistical models that explicitly account for space have been used to map important epidemiological metrics over broad spatial extents, to compensate for the spatial sparsity of data. For example, predictive maps created using geostatistical models have been published for malaria prevalence \citep{weiss2019mapping}, mortality \citep{gething2016mapping}, use of malaria interventions \citep{bertozzi2021maps} and antimalarial drug resistance \citep{flegg2013spatiotemporal, flegg2022spatiotemporal}. These metrics depend on various spatial processes including environmental factors (e.g., rainfall and temperature), variable access to health care, and human movement. In the absence of a full understanding of all these processes, spatial statistical modelling aims to describe the spatial variation in the metric of interest that is caused by the underlying spatial processes. 

In their book, \cite{Diggle2007} introduce a fundamental paradigm for modelling geospatial data that unites previous spatial modelling approaches with model-based statistical analysis. The quantity of interest or the response, $y_i$, is defined throughout a contiguous study region and each measurement at the sample location $\mathbf{x}_i$ is a realisation of the random variable $Y_i$ whose distribution is dependent on the location $\mathbf{x}_i$ as well as the random variables associated with the other data locations. That is, the random variables in space are dependent on each other based on their proximity.  Hence, the observed responses at $n$ locations are modelled as a joint $n$-dimensional vector of random variables where the dependency can be modelled using spatial random effects as part of a generalized linear geostatistical model. The spatially-correlated random variables are modelled as a Gaussian process (GP) and spatial covariates, such as bio-climatic and environmental layers, are often included as additional regressors to capture general trends. 


Gaussian processes are widely used in spatio-temporal modelling including in malaria prevalence mapping research \citep{Hay2006, Smith, Bhatt2017}. 
With the explosion of machine learning research, the popularity of GPs has remarkably increased in both theoretical and applied domains \citep{hensman2013gaussian}. \cite{Nickisch2010} made available a toolbox called \textit{GPML} for machine learning regression and classification tasks. GPs for large scale regression \citep{Park2018} and GPs for sparse approximations \citep{Quinonero-Candela2005} are examples of the use of GPs in machine learning. These new advances have made GPs a viable tool for modelling of very large datasets beyond the field of malaria mapping \citep{Datta2016}.
%


These modelling approaches vary in their inference procedures (e.g. Bayesian or frequentist) and computational techniques (e.g. simulation versus optimisation). One thing these newer methods have in common is that they have the potential to avoid calculations using a `full' (approximation-free) GP, due to the fact that full GP models scale cubicly with the number of unique locations in the training data. That is, a 10-fold increase in the number of unique spatial locations in the dataset results in a 1000-fold increase in computation time. Consequently, the full GP can become computationally infeasible for large datasets, such as those used in national- and continental-scale malaria mapping. For such large datasets, it may also be prohibitive to fit the model using asymptotically exact Bayesian methods such as Markov chain Monte Carlo (MCMC) methods, so deterministic approximations to such simulation approaches have also been explored. Newer approximation methods to both the spatial random effect and the inference method are often used to combat this limitation for example when using global-scale datasets \citep{Moraga2021, Pfeffer2018}.  


There are a multitude of approximation techniques available as alternatives to full Bayesian/frequentist inference, and the full Gaussian process, that still enable fitting of a spatially-explicit model as used in disease mapping. A review of all such techniques is beyond the scope of this paper. Here we present a comparison of four such approximation methods on a malaria prevalence mapping problem: Integrated Nested Laplace Approximation inference, with a Gaussian Markov Random Field approximation to the GP (INLA), Gaussian processes fitted via a boosting algorithm (GPBoost), Spatial Random Forests (SpRF), and Fixed Rank Kriging (FRK). These four methods are selected due to their different underlying models for the spatial correlation structure and varying approaches to inference.

The intended audience of this comparison are twofold.  We hope that it will be of interest to researchers interested in moving into the spatial (or spatio-temporal) mapping field who are looking for an introduction to currently available methods, as well as those who are already in the field with an interest in applying faster methods. We first present an analysis at a national scale, selecting Kenya as the country of interest. We then extend this to a continental-scale analysis over Africa. We introduce the four methods briefly in Section \ref{sec:methods} and specify the model used in each case. Due to their underlying mathematical differences in model specification and inference procedure, it is difficult to directly compare results. We mitigate this problem by comparing point and interval predictions against observed data in a cross validation scheme. In addition, we explore predictive spatial maps produced by each model. We discuss national and continent-scale results in Sections \ref{sec:resultsnational} and \ref{sec:resultcontinent}, and we compare the computation time taken by each of the methods at each scale in Section~\ref{sec:computational_results}. As concluding remarks, we briefly discuss nuances of the models uncovered by our analysis in Section \ref{sec:discussion}. The programming scripts for this work are available at \url{https://github.com/sevvandi/supplementary_material/tree/master/stcompare}.


\section{Methods}\label{sec:methods}

\cite{Diggle2007} start with the following basic geostatistical model that does not have any covariates.  They consider data given by $\left( \mathbf{x}_i, y_i \right)$ for $i \in \{1, \ldots, n\}$, where $\mathbf{x}_i$ denotes the spatial location (i.e. coordinates) and $y_i$ is the measured value for the quantity of interest at that location (e.g. the incidence of malaria at $\mathbf{x}_i$). They describe a model for normally-distributed response data with a stationary Gaussian process (one that tends back to the same average value, over the whole analysis region) as 
\begin{enumerate}
    \item $\left\{S(\mathbf{x}) : \mathbf{x} \in \mathbb{R}^2  \right\} $ where $S(\mathbf{x})$ is a Gaussian process with mean $\mu$ (the average value over the study region), variance or amplitude of the process at each location $\sigma^2 = \var\{ S(\mathbf{x}) \}$ and correlation function $\rho(u) = \text{cor}\left\{S(\mathbf{x}), S(\mathbf{x}')  \right\}$, where $u = \lVert \mathbf{x} - \mathbf{x}' \rVert$ and $\lVert \cdot \rVert$ denotes Euclidean distance (which controls the similarity of responses based on their distances apart);
    \item and $y_i$ are realisations of mutually independent Gaussian random variables $Y_i$ conditional on $\left\{S(\mathbf{x}) : \mathbf{x} \in \mathbb{R}^2  \right\} $ (i.e. after accounting for the spatial correlation, each $y_i$ is independent and normally-distributed).
\end{enumerate}
The model can be described by the equation
\begin{gather}\label{eq:diggle}
    Y_i \sim N(z_i,\tau^2),\\
    z_i = S(\mathbf{x}_i) \quad \text{for} \quad i = 1, \ldots, n \,.
\end{gather}
Common choices of correlation functions (termed covariance functions when they incorporate the variance term $\sigma^2$) include Mat\'ern, exponential and squared exponential functions.

INLA, GPBoost and FRK can be considered as different adaptations of the basic model described in equation \eqref{eq:diggle} by \cite{Diggle2007}, while SpRF is a spatial version of the random forest algorithm, which has its roots in machine learning and takes a fundamentally different approach. 

We use the following implementations of the four methods:
\begin{enumerate}
    \item INLA: Integrated Nested Laplace Approximations, implemented in the R package \texttt{INLA} \citep{rue2009approximate, lindgren2011explicit}. 
    \item GPBoost: Tree boosting with Gaussian processes and mixed effect models, implemented in the R package \texttt{gpboost} \citep{sigrist2020gaussian}.
    \item SpRF: Spatial Random Forests, implemented in the R package \texttt{ranger} \citep{hengl2018random}.
    \item FRK: Fixed Rank Kriging, implemented in the R package \texttt{FRK} \citep{zammit2021frk}. 
\end{enumerate}

As the national-scale dataset we have selected \textit{Plasmodium falciparum} prevalence data in Kenya from 2009, retrieved from the open-access portion of the Malaria Atlas Project malaria prevalence dataset. Kenya was selected as it had the highest number of surveys overall, with the most surveys occurring in 2009. Expanding to a continental scale, we have used the available surveys across Africa in 2009, keeping the same year between analyses.  We used the R package \texttt{malariaAtlas} \citep{Pfeffer2018} to download the malaria prevalence survey data.
Since our aim here was to evaluate the performance of statistical models for malaria mapping, rather than to produce reliable maps persay, we did not apply any additional validation, correction, or selection on these datapoints. We extract for each record only the spatial coordinates, the numbers of individuals screened, and the number of those individuals that were positive for \textit{P. falciparum}.

\subsection{INLA}
INLA (Integrated Nested Laplace Approximations) is a method for approximate Bayesian inference which offers an improvement in speed over asymptotically exact methods such as MCMC. Instead of estimating a high-dimensional joint posterior distribution by simulation, INLA obtains approximations to univariate posterior marginal distributions of the model parameters. INLA is restricted to the class of models that can be expressed as latent Gaussian Markov random fields. However, a multitude of commonly used models can be expressed in this form, including generalised linear geostatistical models. This approach to inference pairs well with an approximation to the spatial Gaussian process as a Gaussian Markov random field (GMRF) over a discrete `mesh' describing the study area, with piecewise linear interpolation to any locations that fall between nodes of this `mesh'. When the Gaussian process has a covariance function of the Matérn type, the stochastic partial differential equation (SPDE) representation of the GMRF can be used, which makes evaluation of the spatial process very fast for large spatial datasets, compared with the full GP approach. 

Over the years there have been many updates to INLA \citep{rue2017bayesian} to broaden its scope and facilitate diverse problem solving tasks. For more details we refer to their website \url{https://www.r-inla.org/home}.

Inference with INLA combines a series of assumptions and Laplace approximations to compute the marginal posteriors of model parameters and latent effects. INLA assumes that the response vector $\mathbf{y}$ depends on a vector of latent variables $\boldsymbol{\eta}$, and hyperparameters $\boldsymbol{\theta}_1$, with density $\pi(\mathbf{y}|\boldsymbol{\eta},\boldsymbol{\theta}_1)$. The latent variables for example may include the values of a linear predictor, an intercept, regression coefficients, and the values of any random effects. Importantly, $\boldsymbol{\eta}$ is assumed to be a mean $\mathbf{0}$ Gaussian Markov random field with precision matrix $\mathbf{Q}(\boldsymbol{\theta}_2)$ (the construction of $\mathbf{Q}$ for continuous spatial models is outlined in \citep{bakka2018spatial}) where $\boldsymbol{\theta}_2$ is a vector of hyperparameters. The hyperparameters are often combined into a single vector $\boldsymbol{\theta} = (\boldsymbol{\theta}_1, \boldsymbol{\theta}_2)$ with prior distribution $\pi(\boldsymbol{\theta})$. INLA then approximates the marginal posteriors $\pi(\eta_i|\mathbf{y})$ and $\pi(\theta_k|\mathbf{y})$ as follows. 

The first step is to write the joint posterior of the hyperparameters as
\begin{align*}
    \pi(\boldsymbol{\theta}|\mathbf{y}) &= \frac{\pi(\boldsymbol{\eta},\boldsymbol{\theta}|\mathbf{y})}{\pi(\boldsymbol{\eta}|\boldsymbol{\theta},\mathbf{y})}\,,\\
    &\propto \frac{\pi(\boldsymbol{\eta},\boldsymbol{\theta},\mathbf{y})}{\pi(\boldsymbol{\eta}|\boldsymbol{\theta},\mathbf{y})}\,.
\end{align*}
A Laplace approximation is applied to the denominator, replacing it with a Gaussian and giving the approximation 
\begin{equation}
    \tilde{\pi}(\boldsymbol{\theta}|\mathbf{y}) \propto \frac{\pi(\boldsymbol{\eta},\boldsymbol{\theta},\mathbf{y})}{\tilde{\pi}_G(\boldsymbol{\eta}|\boldsymbol{\theta},\mathbf{y})}\bigg|_{\;\boldsymbol{\eta} = \boldsymbol{\eta}^*(\boldsymbol{\theta})}\,, \label{eq:joint_posterior_approx}
\end{equation}
where $\boldsymbol{\eta}^*(\boldsymbol{\theta})$ is the mode of $\pi(\boldsymbol{\eta}|\boldsymbol{\theta},\mathbf{y})$, and $\tilde{\pi}_G(\boldsymbol{\eta}|\boldsymbol{\theta},\mathbf{y})$ is its Gaussian approximation. Approximate posterior marginals for the hyperparameters can then be obtained as
\begin{equation*}
    \tilde{\pi}(\theta_k|\mathbf{y}) =  \int \tilde{\pi}(\boldsymbol{\theta}|\mathbf{y}) d\boldsymbol{\theta}_{-k}\,.
\end{equation*}
The exact marginals for the latent effects
\begin{equation*}
    \pi(\eta_i|\mathbf{y}) = \int \pi(\eta_i|\boldsymbol{\theta}, \mathbf{y}) \pi(\boldsymbol{\theta}|\mathbf{y}) d\boldsymbol{\theta}\,,
\end{equation*}
are approximated using numerical integration as
\begin{equation} \label{eq:INLA_numerical_int}
     \tilde{\pi}(\eta_i|\mathbf{y}) = \sum_{k=1}^K \tilde{\pi}\left(\eta_i|\boldsymbol{\theta}^{(k)}, \mathbf{y}\right) \tilde{\pi}\left(\boldsymbol{\theta}^{(k)}|\mathbf{y}\right) \Delta_k\,, 
\end{equation}
where $\tilde{\pi}(\boldsymbol{\theta}|\mathbf{y})$ is as in \eqref{eq:joint_posterior_approx} and $\tilde{\pi}(\eta_i|\boldsymbol{\theta})$ is an approximation of $\pi(\eta_i|\boldsymbol{\theta})$. INLA provides three primary methods for computing $\tilde{\pi}(\eta_i|\boldsymbol{\theta})$, termed the \textit{Gaussian}, \textit{Laplace}, and \textit{Simplified Laplace} strategies, in addition to \textit{adaptive} and \textit{automatic} strategies. Each strategy applies Laplace approximations or series expansions to different conditional distributions, and has different trade offs for efficiency and accuracy. For full details on these methods, see for example \citep{rue2009approximate, GomezRubio_INLA_book, INLA_book_Wang}.
 
Predictions in INLA are carried out concurrently with model fitting, where the posterior predictive distribution of the response at each prediction location is computed \citep{GomezRubio_INLA_book}. The INLA software provides summary statistics including the mean, median, standard deviation and quantiles of the predictive distribution. 

\subsubsection{INLA model} \label{sec:INLA_model}
We formulate a model similar to \cite{kang2018spatio} and \cite{moraga2019geospatial} to predict malaria prevalence using INLA. 
Let $H_i$ denote the number of positive results (e.g., in our case, malaria infections) and $N_i$ the number of people screened at location $\mathbf{x}_i$  for $i = 1, \ldots, n$. Let $p_i$ denote the modelled prevalence at location $\mathbf{x}_i$, and $\mathbf{p}$ be the vector of modelled prevalences over all locations. Then we model $H_i$ using a binomial distribution as

\begin{equation}\label{eq:INLA_likelihood}
    H_i \sim \binomial\left(N_i, p_i \right).
\end{equation}

The standard link for the binomial distribution is the logit  function, which opens-up the probabilities in $[0,1]$ to real values in $(-\infty, \infty)$. Thus we obtain, 
\begin{equation}\label{eq:INLA_model}
    \logit\left(p_i \right)  = \beta_0 + S(\mathbf{x}_i) \, ,
\end{equation}
where $\beta_0$ denotes the intercept and $S$ is a spatial random effect that follows a zero-mean Gaussian process with Matérn covariance function
\begin{equation}\label{eq:matern_cov}
    \cov\left( S(\mathbf{x}_i), S(\mathbf{x}_j)  \right) = \frac{\sigma^2}{2^{\lambda - 1} \Gamma(\lambda)}\left(\kappa  \lVert \mathbf{x}_i - \mathbf{x}_j \rVert  \right)^\lambda K_\lambda \left(\kappa  \lVert \mathbf{x}_i - \mathbf{x}_j \rVert  \right) \, . 
\end{equation}
Here $\lambda$ is the smoothness parameter, $\sigma^2$ denotes the variance and $K_\lambda$ is the modified Bessel function of the second kind. The parameter $\kappa$ controls how fast the correlation decays with distance. 

We have based the implementation and parameter settings for our model on the examples available in \citep{moraga2019geospatial}. The first step in setting up a model is to construct a triangular mesh on which the SPDE will be solved. The software constructs this mesh based on restrictions provided by the user, and it usually contains a region of smaller triangles near the data surrounded by an extension of coarser triangles to avoid boundary effects \citep{lindgren2015bayesian}. When using the Kenya data, we have set the maximum triangle edge length to be $0.5$ for the inner region, and $4$ for extension. The \texttt{cutoff} parameter sets a distance, under which, points are grouped together when constructing the mesh vertices. We have set this to $0.01$, and additionally have left the \texttt{min.angle} and \texttt{offset} parameters, which determine the minimum allowed angles in the triangles and the size of the extension, to their default values of 21 degrees and $-0.1$ respectively. When using the Africa data we have used a mesh on the unit sphere, and have converted the above parameter values to radians. 

The smoothness parameter $\lambda$ in the Matérn covariance function \eqref{eq:matern_cov} must be chosen via the \texttt{alpha} parameter 
\begin{equation}
    \lambda = \alpha - \frac{d}{2}\,,
\end{equation}
where $d$ is the dimension of the space (ie. 2 for a spatial model). We have set \texttt{alpha} to its default value of $2$.

User settings additionally control the approximations during inference. We have used the default \texttt{auto} strategy for approximating $\tilde{\pi}(\eta_l|\boldsymbol{\theta},\mathbf{y})$. The \texttt{int.strategy} parameter then determines how the points $\boldsymbol{\theta}^{(k)}$ are selected for the numerical integration in \eqref{eq:INLA_numerical_int}, and we have used the faster \textit{empirical Bayes} strategy which selects a single point, namely the mode of $\tilde{\pi}(\boldsymbol{\theta}|\mathbf{y})$ and therefore does not average predictions over uncertainty in the hyperparameters, as would typically happen in an MCMC inference procedure. 
We have opted to use the median of the predictive distribution for point predictions, though other quantities such as the mean are available.
\subsection{GPBoost} \label{sec:GPBoost_intro}
GPBoost combines tree-boosting with Gaussian processes and mixed effects models. It aims to leverage the advantages of tree-boosting algorithms such as accounting for complex nonlinearities, discontinuities and higher order interactions with the versatility of Gaussian processes  \citep{sigrist2020gaussian}. It has the functionality to use mixed effects models, in particular models with grouped random effects.  

The general equation of a GPBoost model is given by
\begin{gather}\label{eq:gpboostgeneral}
    Y_i \sim N(z_i,\tau^2) \quad \text{for} \quad i = 1, ..., n, \\
    \mathbf{z} = F(X) + Z \mathbf{S}, \\
    \mathbf{S} \sim \mathscr{N}(0, \Sigma) \, ,
\end{gather}
where $Y_i$ is the response variable at location $\mathbf{x_i}$. The matrix $X \in \mathbb{R}^{n \times p}$ is the fixed effect predictor matrix, with the $i$th row containing covariates for location $\mathbf{x}_i$. The fixed effects function of the covariates $F$, is nonlinear and is learned with boosting. $\mathbf{S} \in \mathbb{R}^m$ contains the random effects with covariance matrix $\Sigma \in \mathbb{R}^{m \times m}$, while $Z \in \mathbb{R}^{n \times m}$ is the random effect predictor variable matrix, which is typically used to define grouped random effects.  

In a Gaussian process model the random effects $\mathbf{S} = \left(S(\mathbf{x}_1), S(\mathbf{x}_2), \ldots, S(\mathbf{x}_m)  \right)$ are a finite-dimensional version of a Gaussian process $S(\mathbf{x})$ with a covariance function
\begin{equation}
    \cov\left(S(\mathbf{x}), S(\mathbf{x}') \right) = c(\mathbf{x}, \mathbf{x}') \, , \,  \mathbf{x}, \mathbf{x}' \in \mathbb{R}^d \, .
\end{equation}
Here $c$ is a covariance function often parameterised as
\begin{equation}
    c(\mathbf{x}, \mathbf{x}') = \sigma_1^2 r\left( \lVert \mathbf{x} - \mathbf{x}' \rVert/\rho \right) \, , 
\end{equation}
where $r$ is an isotropic autocorrelation function with $\sigma_1^2 = \var(S(\mathbf{x}))$ and $\rho$ is the range parameter which determines how quickly $r$ decays with distance. GPBoost currently supports the exponential, Gaussian, Matérn, powered exponential, Wendland, and tapered exponential covariance functions. In a Gaussian process model, $Z$ is usually encoded as a diagonal matrix, so that each element of $\mathbf{S}$ contains the spatial random effect for that location.

With its default settings, GPBoost does not apply approximations to the Gaussian process. For increased efficiency, Vecchia approximations are available in the software, which assume conditional independence between responses based on distances, resulting in sparse matrices during computations \citep{sigrist2020gaussian}.

Inference with GPBoost is carried out by jointly optimising the nonlinear fixed effects function $F$, and the variance and covariance parameters $\boldsymbol{\theta}$ (i.e.$\tau^2$, $\sigma_1^2$, and $\rho$). In the Gaussian process case, the goal of the optimisation is to minimise the \textit{risk functional}
\begin{equation*}
    R(F, \boldsymbol{\theta}) = L(\mathbf{y},F(X),\boldsymbol{\theta})\,,
\end{equation*}
where $\mathbf{y} = (y_1,...,y_n)$ are the observed responses at locations $\mathbf{x}_1,..., \mathbf{x}_n$. 
Here, $L(\mathbf{y},F(X),\boldsymbol{\theta})$ is the negative log marginal likelihood for obtaining the observed responses $\mathbf{y}$, given the observed covariate matrix $X$, and model parameters $\boldsymbol{\theta}$,
\begin{equation*}
    L(\mathbf{y}, F(X), \boldsymbol{\theta}) = \frac{1}{2}(\mathbf{y}-F(X))^T\Psi^{-1}(\mathbf{y}-F(X)) + \frac{1}{2} \log\det(\Psi) + \frac{n}{2}\log(2\pi)\,,
\end{equation*}
where $\Psi = Z\Sigma Z^T + \tau^2 I$. The risk functional is minimised by iteratively updating $F$ and $\boldsymbol{\theta}$. At step $k$, $F_{k-1}$ is held fixed and $\boldsymbol{\theta}_k = \text{argmin}_{\boldsymbol{\theta}} (L(y, F_{k-1}(X), \boldsymbol{\theta}))$ is computed using a gradient or quasi-Newton method. With this value of $\boldsymbol{\theta}_k$, $F$ is updated via a single step of a boosting algorithm.

After optimisation, GPBoost produces predictions in a similar manner to Gaussian process regression. The joint distribution of the observed and predicted responses is formed, and conditioned on the observed responses. The mean of the resulting conditional distribution is used for the predicted value of the response. 

As we are not using covariates in our model, tree boosting is used only to find the intercept. While this does neglect GPBoost's functionality for learning nonlinear functions of covariates, we have included GPBoost in the analysis for users who may wish to apply it in more complicated scenarios that may benefit from tree boosting.

\subsubsection{GPBoost model} \label{sec:GPBoost_model}
GPBoost supports Gaussian, Bernoulli-probit, Bernoulli-logit, Poisson, and Gamma distributions for the response variable, however unlike INLA does not currently support a binomial response. We therefore model malaria prevalence by customising equation~\eqref{eq:gpboostgeneral} as follows:
\begin{gather}
    H_i/N_i \sim N(z_i, \tau^2)\,, \\
    z_i = \beta_0 + S(\mathbf{x}_i)\, , 
\end{gather}
where $\beta_0$ is the intercept, and $H_i$ and $N_i$ denote the number of positive results and the number of people tested at location $\mathbf{x}_i$. Note that for simplicity, we elected to use the direct proportion of positive tests rather than the empirical logit, and we clip predictions to lie within $[0,1]$ for the prevalence maps. We use the exponential covariance function $r\left( \lVert \mathbf{x} - \mathbf{x}' \rVert/\rho \right)  = \exp \left( -\lVert \mathbf{x} - \mathbf{x}' \rVert/\rho \right) $, which is the default choice in the software. We note that this model does not use GPBoost's full capability for learning nonlinear functions of the covariates, however it has been constructed to be consistent with our choice to not use covariates for any of the models.

The parameter settings for our model follow examples by the package author \citep{GPBoostGithub}. For the spatial random effect in our model we use a full Gaussian process without approximation, setting the \texttt{vecchia\_approx} parameter to \texttt{FALSE}. Other parameters in the software control the trees and boosting algorithm used to learn the fixed effects function $F$. We have set the number of boosting rounds to $247$ and the learning rate to $0.01$, using the parameters \texttt{nrounds} and \texttt{learning\_rate}. Other settings for our model include $\texttt{num\_leaves} = 1024$, $\texttt{max\_depth} = 6$, and $\texttt{min\_data\_in\_leaf} = 5$, each of which control the size of the trees.

\subsection{SpRF}
Spatial Random Forests (SpRF) \citep{hengl2018random} extend classical random forests to a spatial domain by using distances to observation points as explanatory variables, i.e. when fitting a model with SpRF, for each point $\mathbf{x}_i$, where $y_i$ is given, covariates are used that give the distance from each other observation point. That is, the design matrix for this part of the model is simply the distance matrix between all pairs of observation locations. In order to obtain uncertainty estimates, the SpRF authors use quantile regression forests which estimate specified quantiles of the conditional distribution $Y_i|X_i$ \citep{meinshausen2006quantile} where $X_i$ are the covariates for the $i$th response, in contrast to classical random forests which do not provide uncertainties. 

The generic equation of an SpRF model is given by
\begin{equation}\label{eq:sprf1}
    Y_i = f\left(X_{G_i},X_{R_i}, X_{P_i}\right) \, , 
\end{equation}
where $Y_i$ is the response at location $\mathbf{x}_i$,  $X_{G_i}$ denotes a vector of the distances to each of the observation locations from the querying point $\mathbf{x}_i$ (including a distance of 0 to itself, in the $i$th position of the vector) and $X_{R_i}$ and $X_{P_i}$ denote two types of covariates -- surface reflectance and process-based. The function $f$ is learned by the random forest. Unlike the other methods we discuss, SpRF does not use a covariance function.
 
SpRF is based on the \texttt{ranger} package for random forests, which provides an implementation of quantile regression forests with training procedure outlined in  \cite{meinshausen2006quantile}. Point predictions are given by the estimated medians from the quantile regression forests.

\subsubsection{SpRF model}
As in \citep{hengl2018random}, we include an additional normal assumption for the response to construct the simple SpRF model 
\begin{gather}
    H_i/N_i \sim N(z_i, \tau^2)\,, \\
    z_i = f(X_{G_i}) \quad \text{for} \quad i = 1,..., n, 
\end{gather}
where $H_i$ and $N_i$ are as defined above and $X_{G_i}$ contains the distances from each observation point to $\mathbf{x}_i$. 

The user parameters for SpRF determine the structure of the random forest and the rules for growing each tree, including the number of trees and the number of variables to split on at each node via the \texttt{num.trees} and \texttt{mtry} parameters. We have left each parameter at its default value, resulting in a forest with 500 trees where each node splits at $\sqrt{n_v}$ variables ($n_v$ is the total number of variables input into the random forest). Other parameters which further tune the structure of the trees and forest have been left at their default values, and our code for SpRF is based on a tutorial from the method's authors \citep{SpRFGithub}.

\subsection{FRK}
Fixed Rank Kriging (FRK) \citep{zammit2021frk} is a spatio-temporal modelling framework built for large datasets. It uses a spatial random effects (SRE) model, which decomposes a spatially correlated mean-zero random process using a linear combination of spatial basis functions. This dimensionality reduction 
using a relatively small number of basis functions ensures FRK's computational efficiency. The spatial domain $D$ is partitioned into $M$ subsets, $A_1, ..., A_M$, called basic areal units (BAUs) with centroids $\mathbf{x}_1, ..., \mathbf{x}_M$. The SRE model is constructed on these BAUs which determine the granularity of the model, and the process is assumed to be piecewise continuous over the BAUs. 

The general equation for FRK with a Gaussian response can be written as
\begin{gather}\label{eq:frk1}
    Y_i \sim N(z_i, \tau^2) \quad \text{for} \quad i = 1,...,n\,,\\
    \mathbf{z} = C_{Z} \boldsymbol{\zeta}\,,\\
    \zeta_j = \mathbf{t}(\mathbf{x}_j)^T \boldsymbol{\beta} + v(\mathbf{x}_j) +\xi(\mathbf{x}_j) \quad \text{for}\quad  j = 1, ..., M\,.
\end{gather}
Here, $Y_i$, $i=1,..., n$ are the responses at the observation locations, $\boldsymbol{\zeta} = (\zeta_1,...,\zeta_M)^T$ is the value of a latent spatial process evaluated at each of the BAUs with centroids $\mathbf{x}_1, ..., \mathbf{x}_M$, and $C_Z$ is an $n$ by $M$ matrix connecting the observation locations to the BAU locations. The vector $\mathbf{t}(\mathbf{x}_j)$ is a collection of covariates at BAU $j$ and $\boldsymbol{\beta}$ is a vector of regression coefficients, while $v(\mathbf{x}_j)$ is the value of a small-scale, spatially correlated random effect. Lastly, $\xi(\mathbf{x}_j)$ is a fine-scale random effect, which is treated as uncorrelated across the BAUs \citep{zammit2021frk}.

FRK introduces non-Gaussian data to the model by replacing the observation distribution with a member of the exponential family and using a link function to transform the latent process into a mean process. The general structure of such a model is 
\begin{gather}\label{eq:frk2}
Y_i | \mu_i, \psi \sim \text{EF}(\mu_i,\psi)\quad \text{for  } i=1,...,n,\\
\boldsymbol{\mu} = C_{Z}\boldsymbol{\mu}'\,,\\
g(\boldsymbol{\mu}') = \boldsymbol{\zeta}\,,\\
\zeta_j = \mathbf{t}(\mathbf{x}_j)^T\boldsymbol{\beta} + v(\mathbf{x}_j)+ \xi(\mathbf{x_j})\quad \text{for  } j=1,...,M\,,
\end{gather}
where $\psi$ is a dispersion parameter for the context dependent member of the exponential family $\text{EF}$, $\boldsymbol{\mu}$ is called the \textit{mean process}, and $g(\cdot)$ is the link function. We represent the mean process at the observation locations by $\mathbf{\mu}$, while $\mathbf{\mu}'$ represents the mean process at the BAUs. 


The spatially correlated random effect $\nu(\mathbf{x})$ is decomposed as
\begin{equation}
\nu(\mathbf{x}) = \sum_{l=1}^r \phi_l(\mathbf{x})\eta_l \,,\label{eq:basis_fn_frk}
\end{equation}
where $\phi_1,...,\phi_r$ are a fixed collection of basis functions on the spatial domain, and $\boldsymbol{\eta} = (\eta_1, ..., \eta_r)^T$ is an r-variate Gaussian random variable with covariance matrix $\mathbf{K}$. To estimate model parameters including the coefficients $\alpha$, variance parameters for the fine scale random effect $\xi$, and covariance parameters for the covariance matrix $\mathbf{K}$, FRK carries out maximum likelihood estimation. When working with non-Gaussian data, a Laplace approximation is used to approximate the marginal likelihood, which is then maximised via a quasi-Newton method.

By default, FRK produces a prediction for the mean process $\mu(\cdot)$ at each of the BAUs. Predictions and uncertainties are generated via a Monte Carlo sampling approach, and the predicted value of $\mu$ in each BAU is taken to be the average of the samples. 

\subsubsection{FRK Model}\label{sec:FRK_model}
As with INLA, we model the number of positive tests $H_i$ using a binomial distribution
\begin{gather}
    H_i  \sim \binomial\left(N_i, p_i \right) \, , \\
    \mathbf{p} = C_Z\mathbf{p}'\,,
\end{gather}
where $p_i$ is the prevalence at the $i$th observation location. The vector $\mathbf{p}'$ gives the prevalence at the BAUs, and is transformed into the prevalence at the observation locations via the $C_Z$ matrix, which has construction detailed in \citep{frk2}. The logit function is then used as the link function $g$ in equation~\eqref{eq:frk2}, i.e.
\begin{equation}
    \logit(p'_j) = \zeta_j\,.
\end{equation}
 As we are not using any covariates, the latent process over the BAUs $\zeta_j$ can be written as  
\begin{equation}
    \zeta_j = \beta_0 + v(\mathbf{x}_j) + \xi(\mathbf{x}_j)\, ,
\end{equation}
where $\beta_0$ denotes the intercept. 

Our model decomposes the spatial random effect, $\nu(\mathbf{x})$, using Gaussian basis functions of two different scales placed regularly across the spatial domain, as controlled by the \texttt{type}, \texttt{nres}, and \texttt{regular} parameters respectively. The spatial scale of these basis functions is determined jointly by the \texttt{regular} parameter and the \texttt{scale\_aperture}, which we left at their default values of $1$ and $1.25$ respectively. The assumed correlation structure of the random coefficients $\boldsymbol{\eta}$ is controlled by the \texttt{K\_type} parameter. When using a non-Gaussian model, this takes a default value of \texttt{precision}, which models the coefficient dependence using a precision matrix $\mathbf{Q}$ based on the Leroux model \citep{frk2}. During prediction, the user can specify the number of Monte Carlo samples to be drawn, which we have left at the default value of $400$. Code and parameter choices for our FRK model are based on examples from the package authors in \citep{zammit2021frk, frk2}.

\subsection{Methods for the country scale analysis}

\begin{figure}
    \centering

    \subfloat[][]{
        \includegraphics[width=0.4\textwidth]{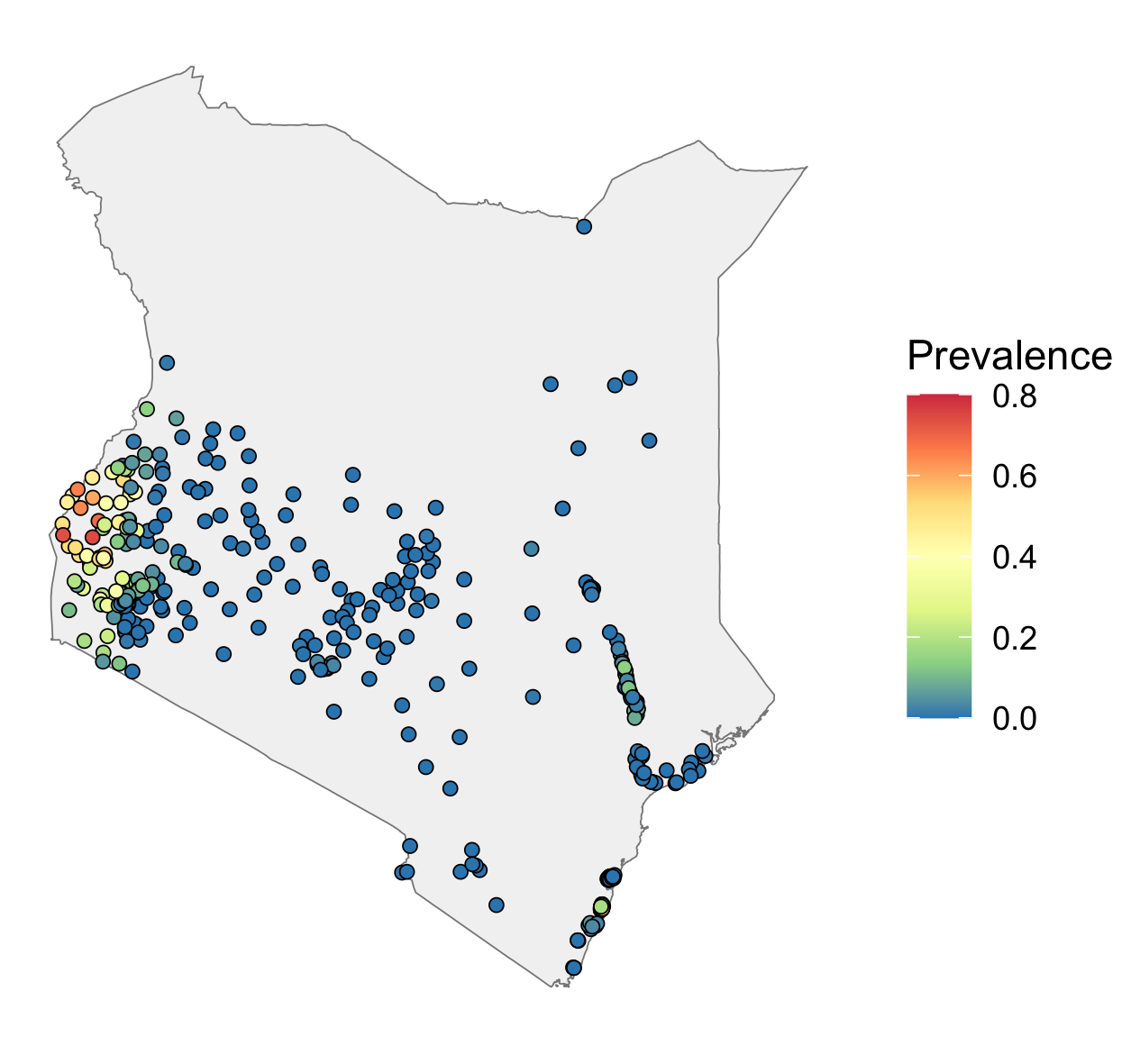}
        \label{fig:Kenya_2009_prevalence}
    }\hspace{30px}
     \subfloat[][]{
        \includegraphics[width=0.4\textwidth]{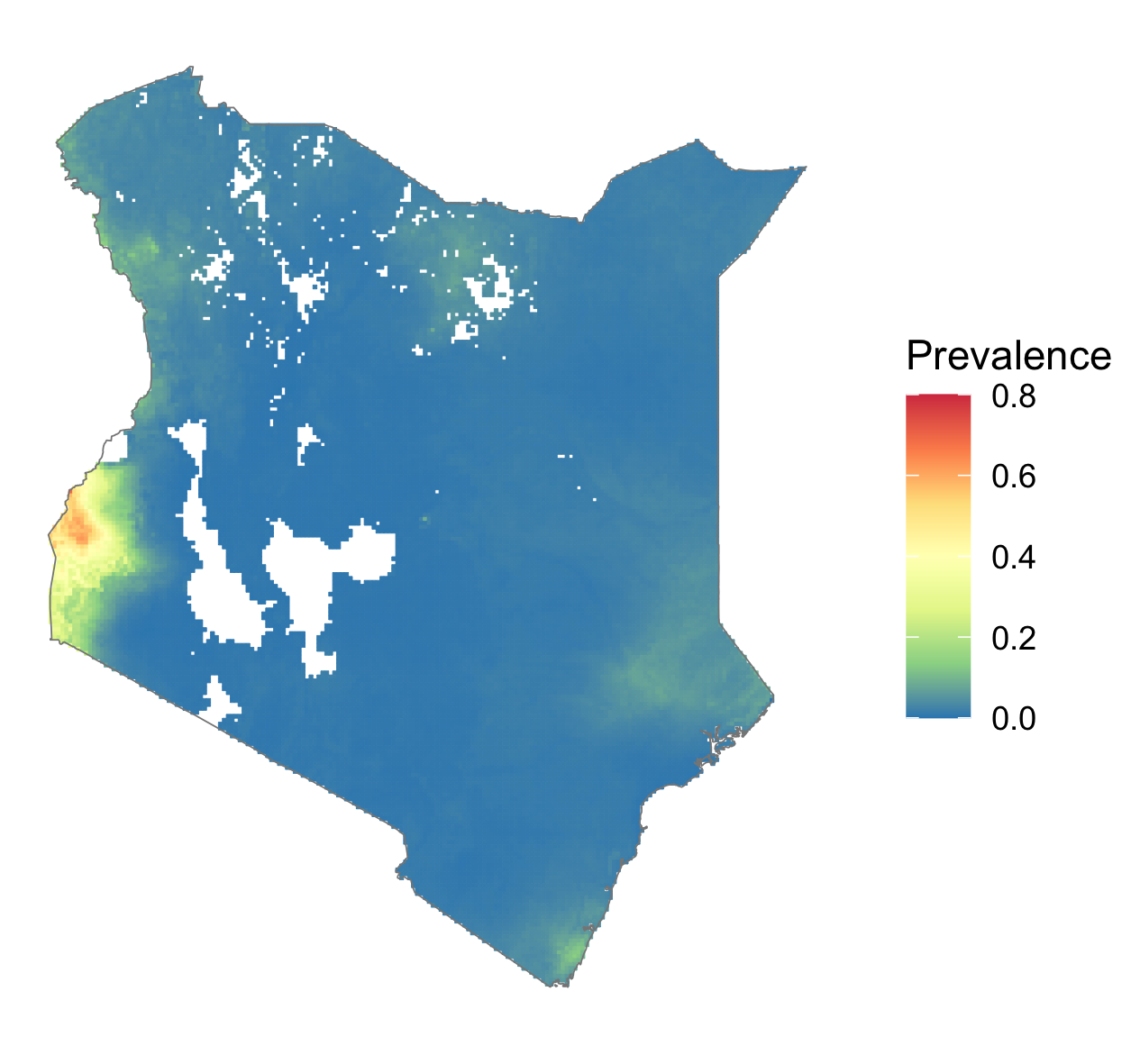}
        \label{fig:Kenya_MAP_surface}
    } 
    \caption{2009 \textit{P. falciparum} prevalence data in Kenya. (a) shows prevalence survey results, while (b) shows the Malaria Atlas Project predicted prevalence.}
    \label{fig:Kenya_data}
    
\end{figure}

At the country scale, the four models were compared qualitatively using their predictive maps, while cross validation was used to compare their predictive performance. To produce maps of predicted prevalence, we fit each model on all available \textit{P. falciparum} prevalence surveys from Kenya in 2009 from the malariaAtlas R package. This consisted of 382 surveys at points across the country which are shown in Figure~\subref*{fig:Kenya_2009_prevalence}. Point estimates of prevalence and uncertainties were produced by the fitted models on a grid over Kenya, with each cell covering a nominal 0.1 degrees (approximately 11x11 km at the equator) in longitude and latitude. 

Model fitting and prediction were run on a 2014 MacBook Pro with a two core, 2.8GHz Intel Core i5 processor running macOS 10.13.6. We ran each model using a single thread to obtain a baseline performance comparison to accompany our main focus on the model predictions; we note that parallelisation options are available for each model which may provide performance improvements. Recorded times were measured as the total time to run a model's R script, including both fitting and prediction.

\begin{figure}[!ht]
\centering
    \captionsetup{width=0.3\textwidth}
    \subfloat[][10-fold CV locations]{
        \includegraphics[width=0.3\textwidth]{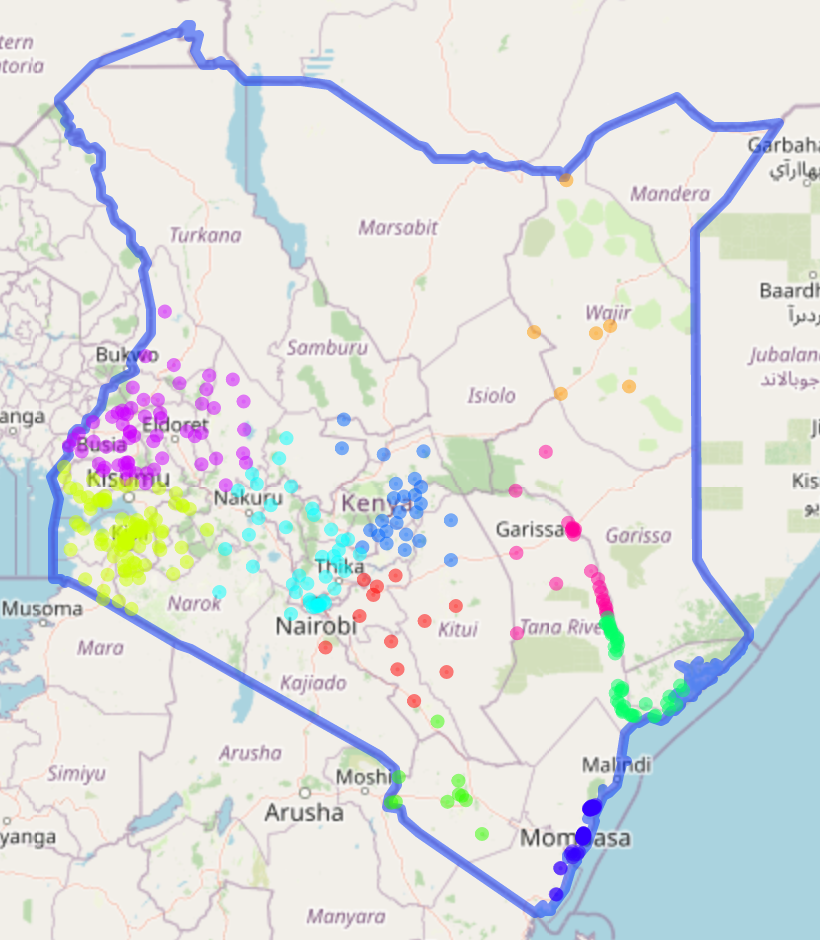}
        \label{fig:10foldspos}
    }\hspace{40px}
     \subfloat[][50-fold CV locations]{
        \includegraphics[width=0.3\textwidth]{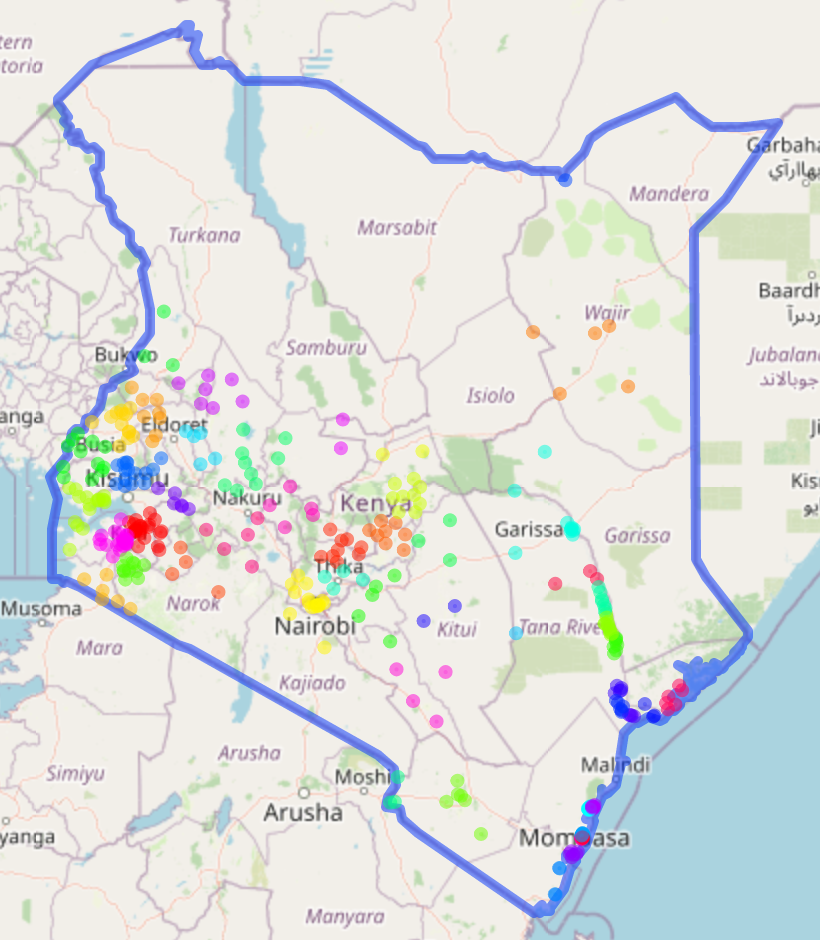}
        \label{fig:50foldspos}
    } 
    \captionsetup{width = 0.8\textwidth}
    \caption{\textit{P. falciparum} prevalence survey locations in Kenya for 2009. Colours represent different cross validation folds for 10-fold CV (a) and 50-fold CV (b).  }
    \label{fig:cvpositions}
    
\end{figure}

To evaluate the models we use spatial block cross validation (CV) \citep{Roberts2017} using 10 and 50 folds. In a spatial setting, randomly allocating points to cross validation folds is not effective because close by points can act as proxies. The folds were selected using $k$-means clustering \citep{Likas2003} on the spatial coordinates of the prevalence surveys - resulting in a series of `blocks' of spatially-adjacent points.  Figure~\ref{fig:cvpositions} shows the location of points for the two sets of CV folds, where each colour represents a fold.  The 10 and 50 CV folds measure different abilities of the methods. The 50-fold CV quantifies short-scale interpolation ability, while the 10-fold quantifies the ability to interpolate over longer distances.  

Using 10 and  50-fold cross validation, we investigate the following: 
\begin{enumerate}
    \item analysis of the point predictions including a comparison between the predictions and out-of-sample prevalence values using multiple measures, 
    \item analysis of the uncertainty bounds for each model, and
    \item analysis of the predictions with respect to density of the sampled locations (Appendix~\ref{sec:detailedCV}).
\end{enumerate}

Analysis of uncertainties is complicated by the differing measures of uncertainty output by the models. INLA contains information on the summaries of the posterior marginal densities of the fitted model, and can compute the standard deviation and different quantiles of the predictions.  GPBoost provides the variance of each prediction. FRK predicts the standard deviation of each 
prediction in the linear, Gaussian setting. For the non-Gaussian case, it provides the predictions using a Monte Carlo approach (Sainsbury-Dale et al. 2021). SpRF uses quantile regression and the quantiles can be specified in the ranger package. To compare SpRF with the other methods, we assume a normally distributed response as in \citep{hengl2018random}, and estimate the standard deviation for SpRF’s predictions as 
\begin{equation}
    \text{SD} \approx \text{IQR}/1.34898\,. \label{eq:normal_approx}
\end{equation}
Hengl et al. \citeyearpar{hengl2018random} note that this assumption may not always be valid, and hence we are only able to roughly compare the SpRF model's uncertainty with the other three models.

For each model we measure how many observed prevalence values lie within the predicted uncertainty intervals. Let $\hat{y}_i$ denote the mean of the predicted response for observation $y_i$. We define 
\begin{align}
    \text{Within 1SD}(\hat{y}_i)  & = \text{TRUE if}  \, \,   \vert y_i - \hat{y}_i \vert \leq \text{SD}(\hat{y}_i) \\
    \text{Within 2SD}(\hat{y}_i)  & = \text{TRUE if Within 1SD}(\hat{y}_i) = \text{FALSE and} \, \,   \vert y_i - \hat{y}_i \vert \leq 2\text{SD}(\hat{y}_i)     
\end{align}
where SD denotes the standard deviation.  As prevalence values are between 0 and 1, we trim the bounds if they exceed these limits. Note that $\hat{y}_i$ corresponds to the predicted prevalence for our GPBoost and FRK models, but not for our INLA and SpRF models which use the median for predictions.

\subsection{Methods for the continent scale analysis}

\begin{figure}[!ht]
    \centering
    \subfloat[][]{
        \includegraphics[width = 0.45\textwidth]{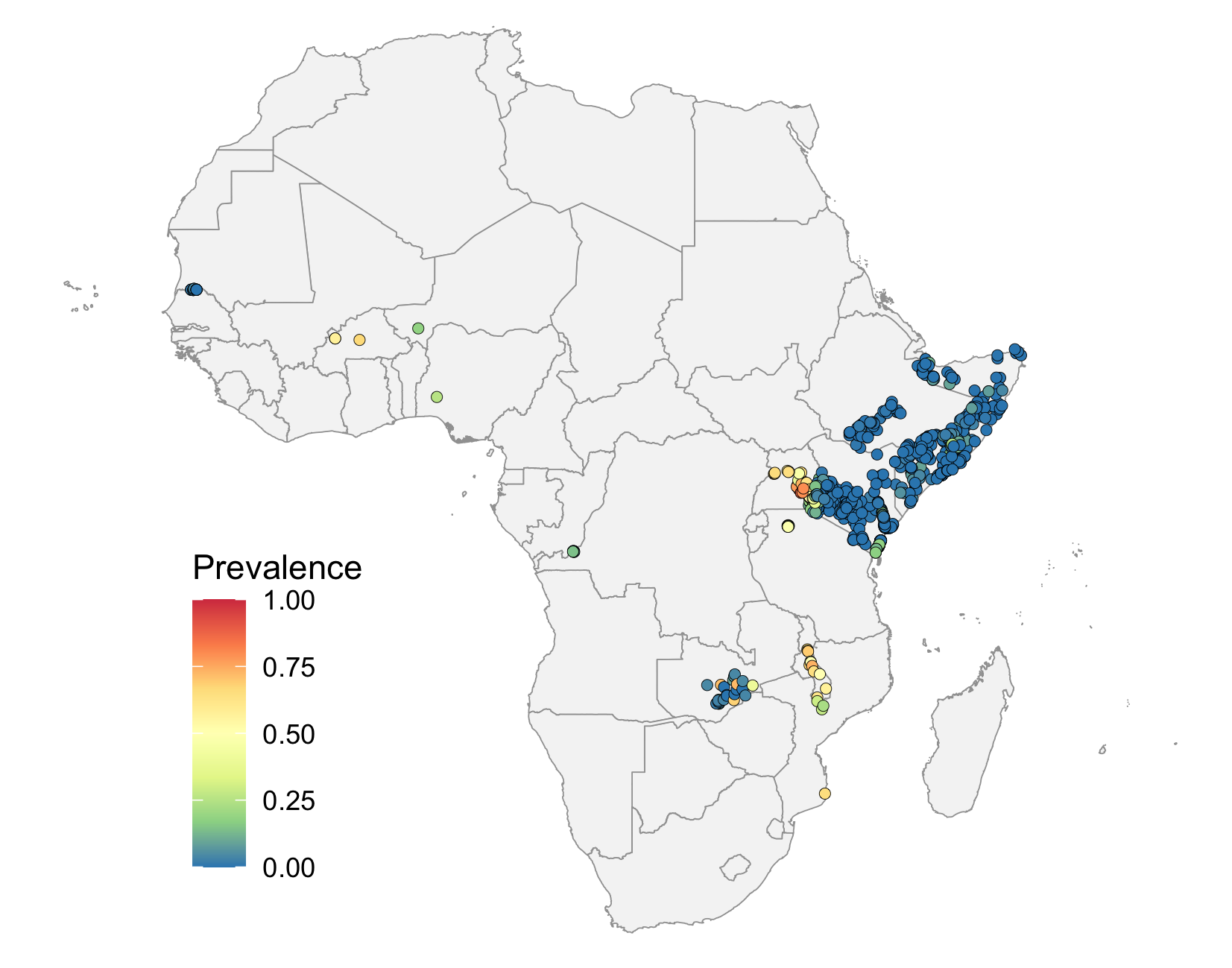}
        \label{fig:Africa_observation_points}
    }
    \subfloat[][]{
        \includegraphics[width = 0.45\textwidth]{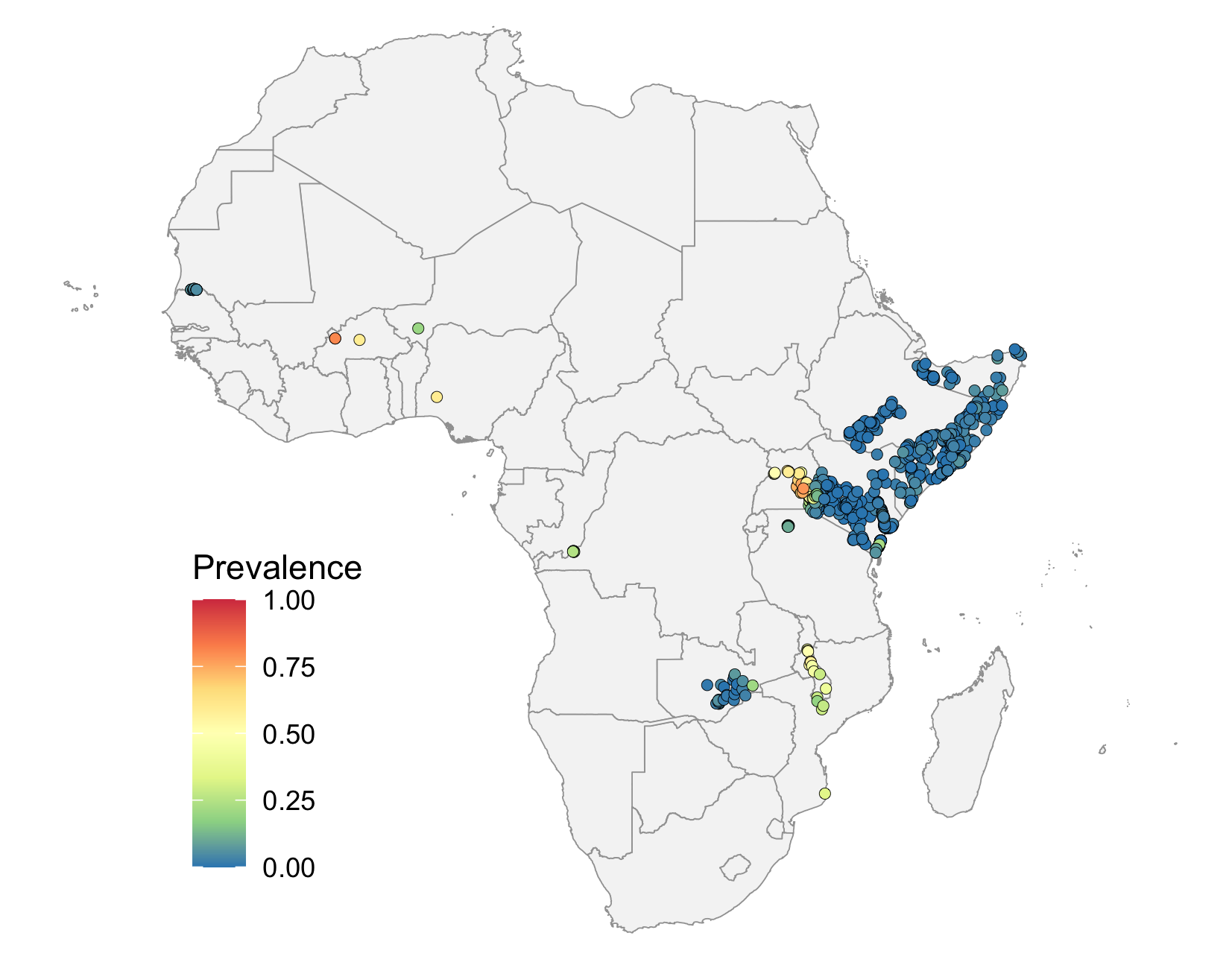}
        \label{fig:Africa_observation_binomial_points}
    }
    \\
    \subfloat[][]{
        \includegraphics[width = 0.45\textwidth]{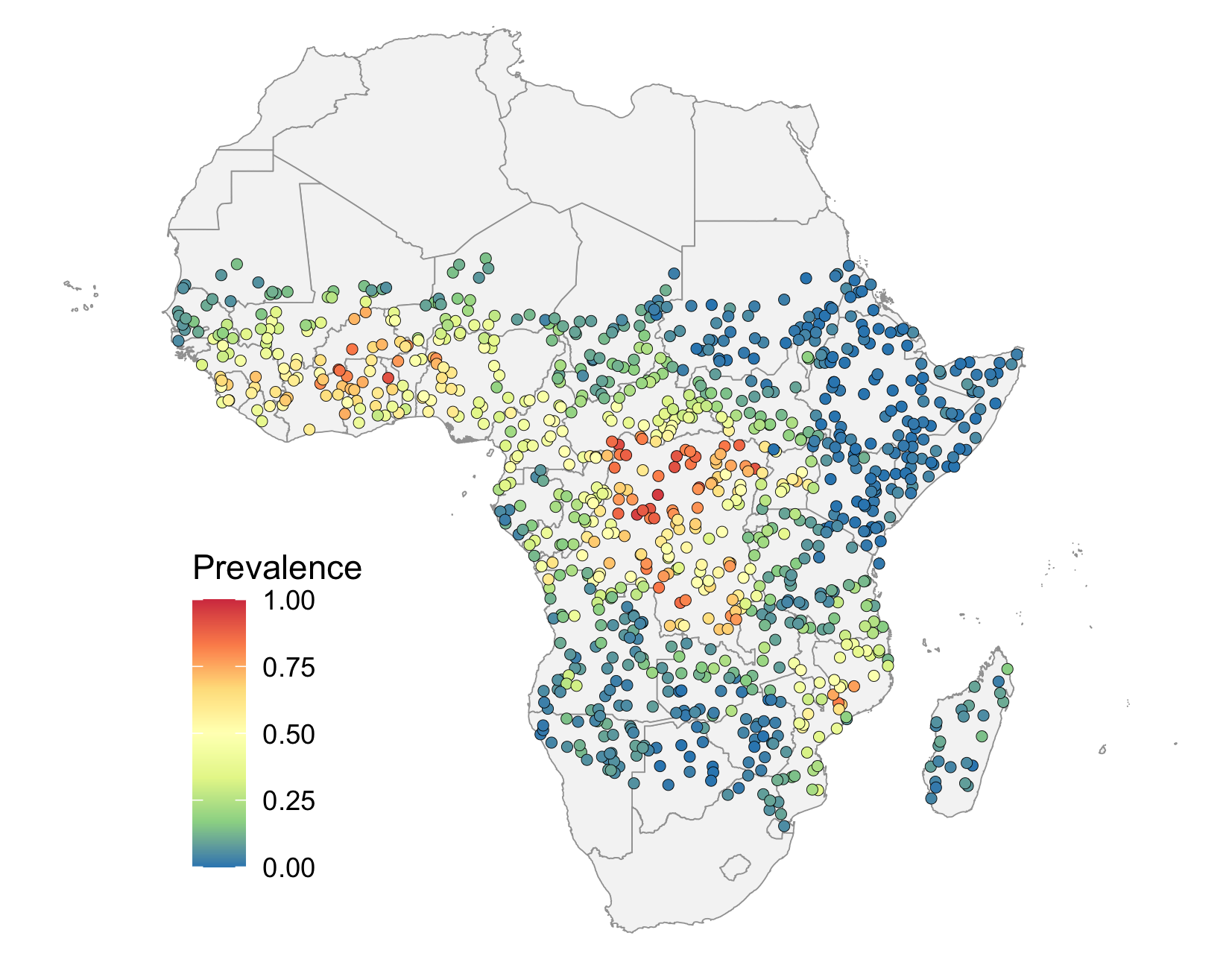}
        \label{fig:Africa_uniform_binomial_points}
    }
    \subfloat[][]{
        \includegraphics[width = 0.45\textwidth]{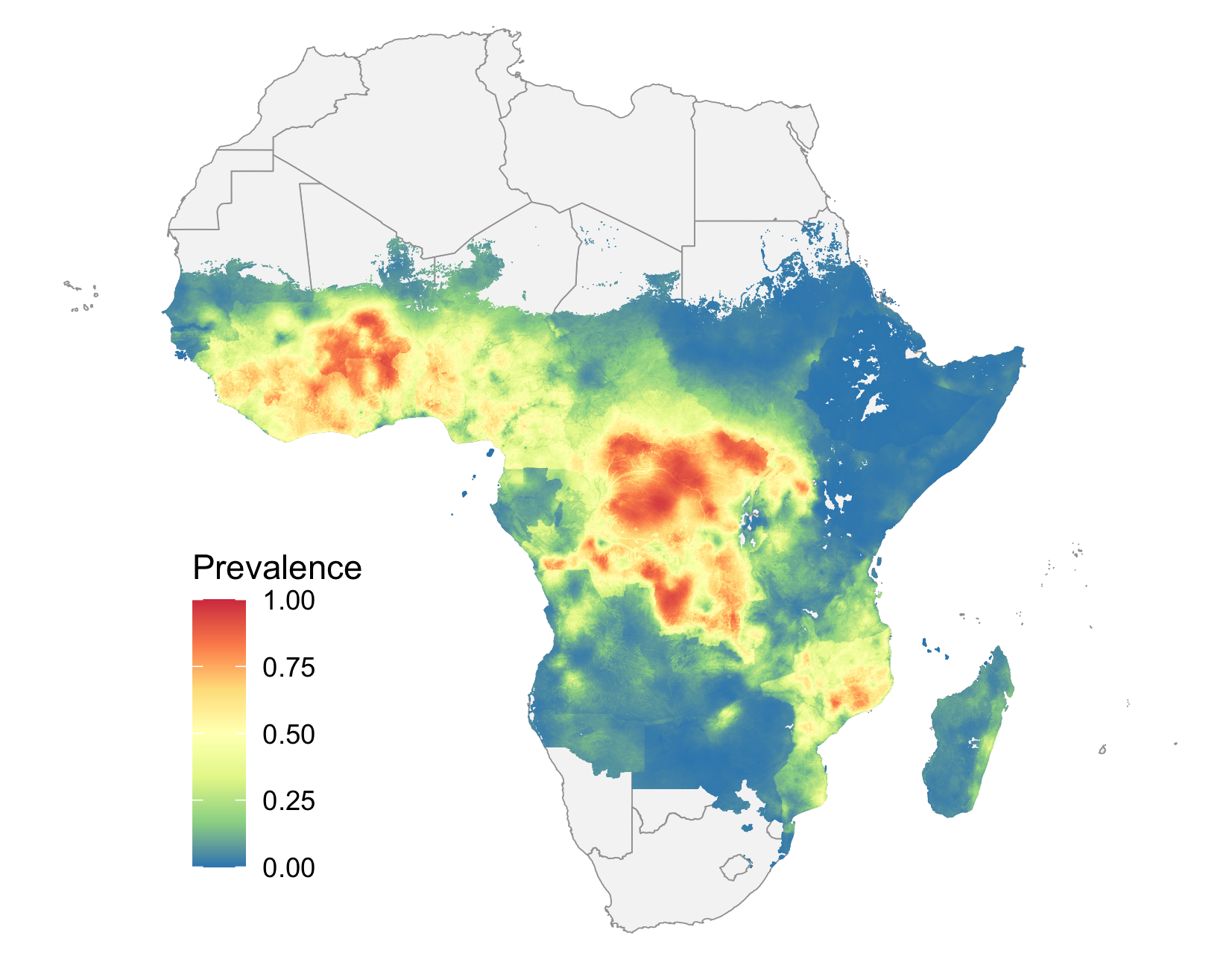}
        \label{fig:MAP_raster}
    }
    
   \caption{\textit{P. falciparum} prevalence data used to fit the four models at the continental scale. \protect\subref*{fig:Africa_observation_points} shows the 2009 observed data at 868 locations. \protect\subref*{fig:Africa_observation_binomial_points} shows the prevalence generated from binomial samples at the observation locations. \protect\subref*{fig:Africa_uniform_binomial_points} shows the prevalence generated by binomial samples at 1000 uniformly random locations. \protect\subref*{fig:MAP_raster} is the Malaria Atlas Project predicted prevalence raster from 2009 used to generate the samples in \protect\subref*{fig:Africa_observation_binomial_points} and \protect\subref*{fig:Africa_uniform_binomial_points}.}
   \label{fig:Africa_input_data}
\end{figure}

At the continent scale we focused on prediction maps, fitting each model to three sets of prevalence data over Africa. The first set consists of 868 \textit{P. falciparum} prevalence surveys from 2009, available via the malariaAtlas R package. This data is shown in Figure~\subref*{fig:Africa_observation_points}, with survey points concentrated in Kenya and Somalia. Each model was additionally fit using two types of simulated data to allow comparison of the predictions with a known truth and to compare model performances on both interpolation and extrapolation tasks, and lastly to assess how properties of the data such as spatial sparsity and noise impact model predictions.

Simulated data was generated using the 2009 \textit{P. falciparum} prevalence raster created by the Malaria Atlas Project (MAP), shown in Figure~\subref*{fig:MAP_raster} \citep{weiss2019mapping}. Prevalence was sampled from the raster at the locations of the 2009 surveys and combined with the number of tests at each location to generate a binomial sample for the number of people testing positive. Of the 868 observation locations, 28 points lie on gaps in the prevalence raster and were excluded, leaving 840 points in this second dataset, which is shown in Figure~\subref*{fig:Africa_observation_binomial_points}. This spatially clustered simulated data allows us to evaluate each model's ability to extrapolate over regions with little or no data. While this dataset shares locations with the observation data, its prevalence notably contains less noise.

The second set of simulated data was generated by selecting 1000 points at random on the MAP raster, allowing for comparison of the models' interpolation performance when trained on data with good spatial coverage. A binomial sample for the number of positive tests was generated at each location, where the number of people tested was set to 85, approximately the average number in the surveys from 2009. The prevalences from this simulated dataset are shown in Figure~\subref*{fig:Africa_uniform_binomial_points}. 

We have used the same parameter settings as in the country scale analysis, though whenever possible, we have used settings that compute an appropriate spherical distance between points, due to the larger spatial extent of the data. This was possible SpRF which uses great circle distances, and for INLA which allows for meshes to be constructed on the unit sphere. GPBoost does not have this functionality at the time of writing this article, however correspondence with the package authors reveals that they hope to add this functionality in future.
While FRK does support using great circle distances for some models, this feature is not currently well supported for models with a binomial response and did not work in our tests. Hence we have used Euclidean distances between coordinates for both GPBoost and FRK.

Using INLA with a spherical geometry requires a mesh to be built on a subset of the sphere. Although several methods for constructing this mesh are used in the literature \citep{lindgren2015bayesian, bakka2018spatial, HUMPHREYS2017192}, each produced similar results and we have followed the method outlined by Lindgren and Rue (\citeyear{lindgren2015bayesian}).

Model fitting and prediction were carried out on a single 3.00GHz Intel(R) Xeon(R) Gold 6154 CPU in the Physical partition of the University of Melbourne’s high performance computing cluster, Spartan, and each model run was allocated 32 GB of RAM. As with the country scale data, each model was run using a single thread. Predictions were produced on a grid with cell side length 0.15 degrees (approximately 16.7 km at the equator).

\section{Results} \label{sec:results}
This section presents the analysis on Kenya, which includes the predictive maps and cross validation results, and the continent scale analysis including models trained on three different input datasets discussed above.

\subsection{Case study: Kenya}
\label{sec:resultsnational}
At a national scale we have used two means of verifying our models: \begin{inparaenum}\item predictive maps, and \item 10-fold and 50-fold cross-validated predictions.  \end{inparaenum} To produce the maps we have used all the data while for cross validation we have left out some data in each fold. 

\subsubsection{Predictive maps}
\begin{figure}[!p]
    \captionsetup{position = top}
    \centering
    \subfloat[][INLA]{
       \includegraphics[width = 0.7\textwidth]{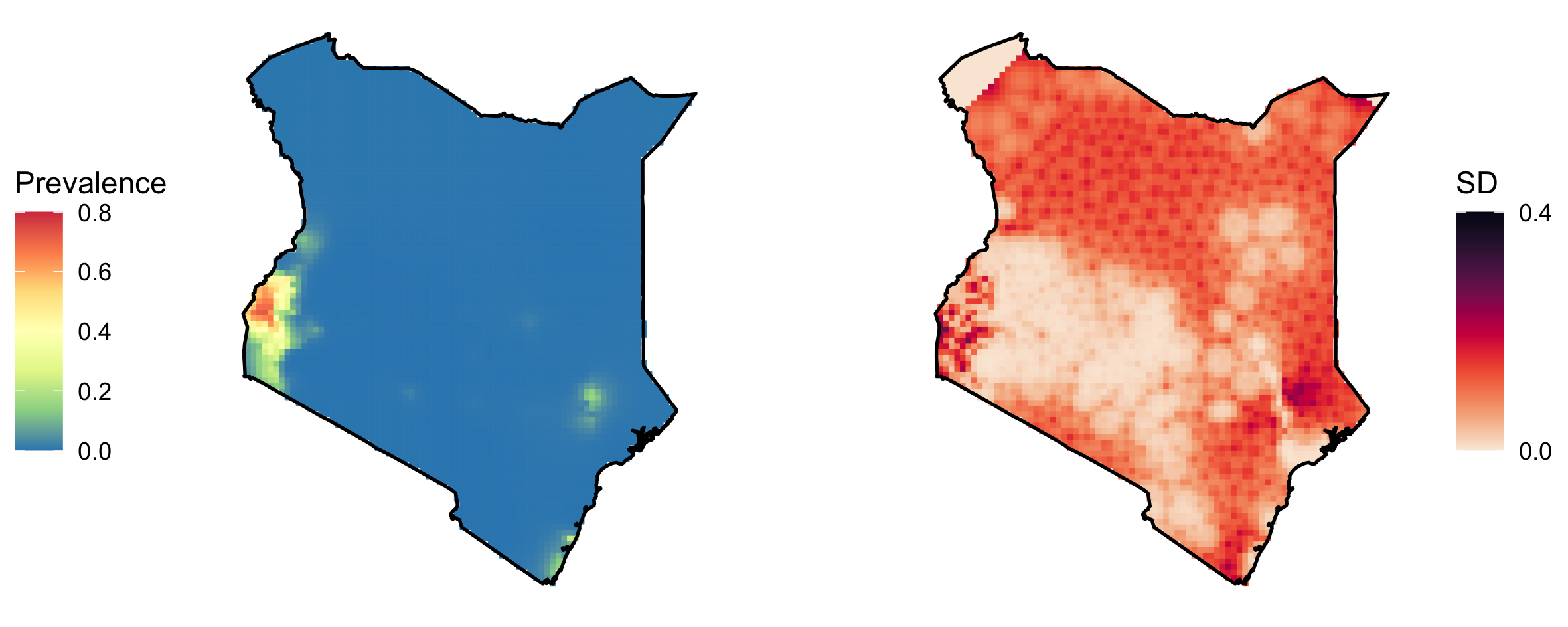}
       \label{fig:INLA_Kenya_map}
    }\vspace{-10px}\\
    \subfloat[][GPBoost]{
       \includegraphics[width = 0.7\textwidth]{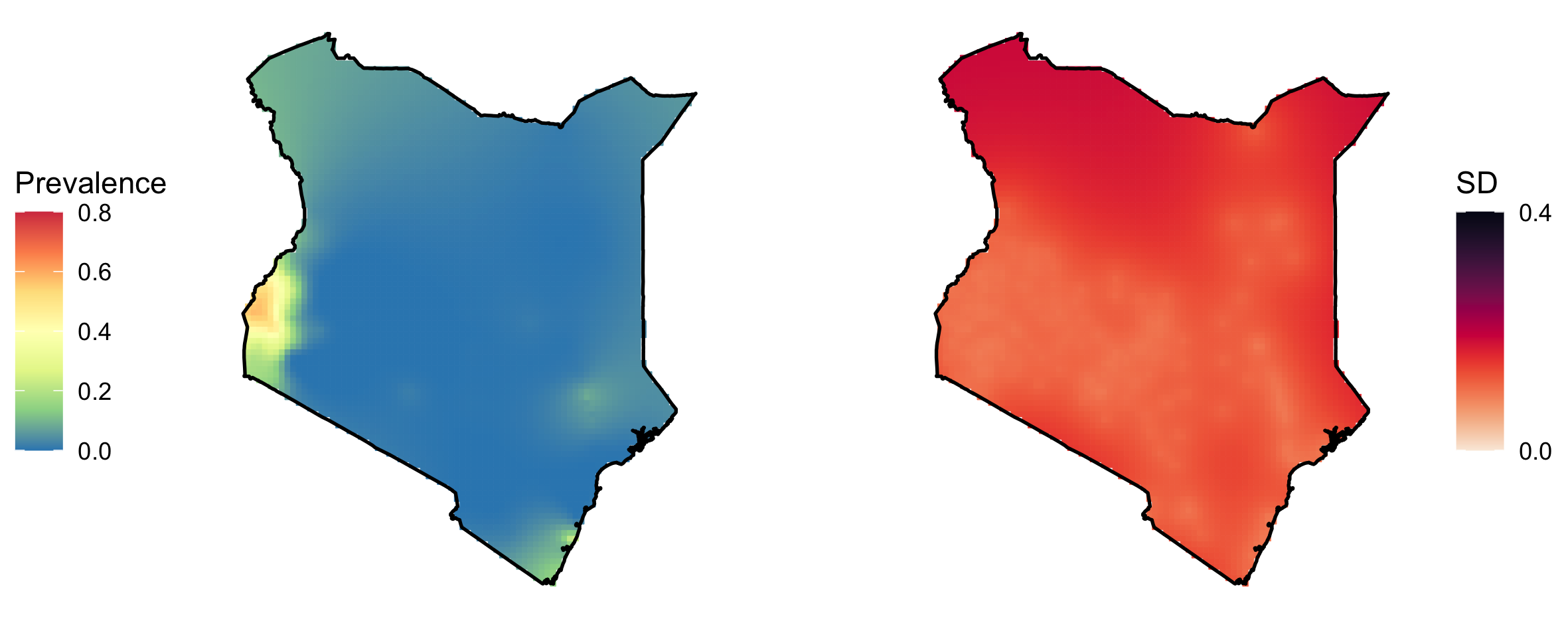}
       \label{fig:GPBoost_Kenya_map}
    }\vspace{-10px}\\
    \subfloat[][SpRF]{
       \includegraphics[width = 0.7\textwidth]{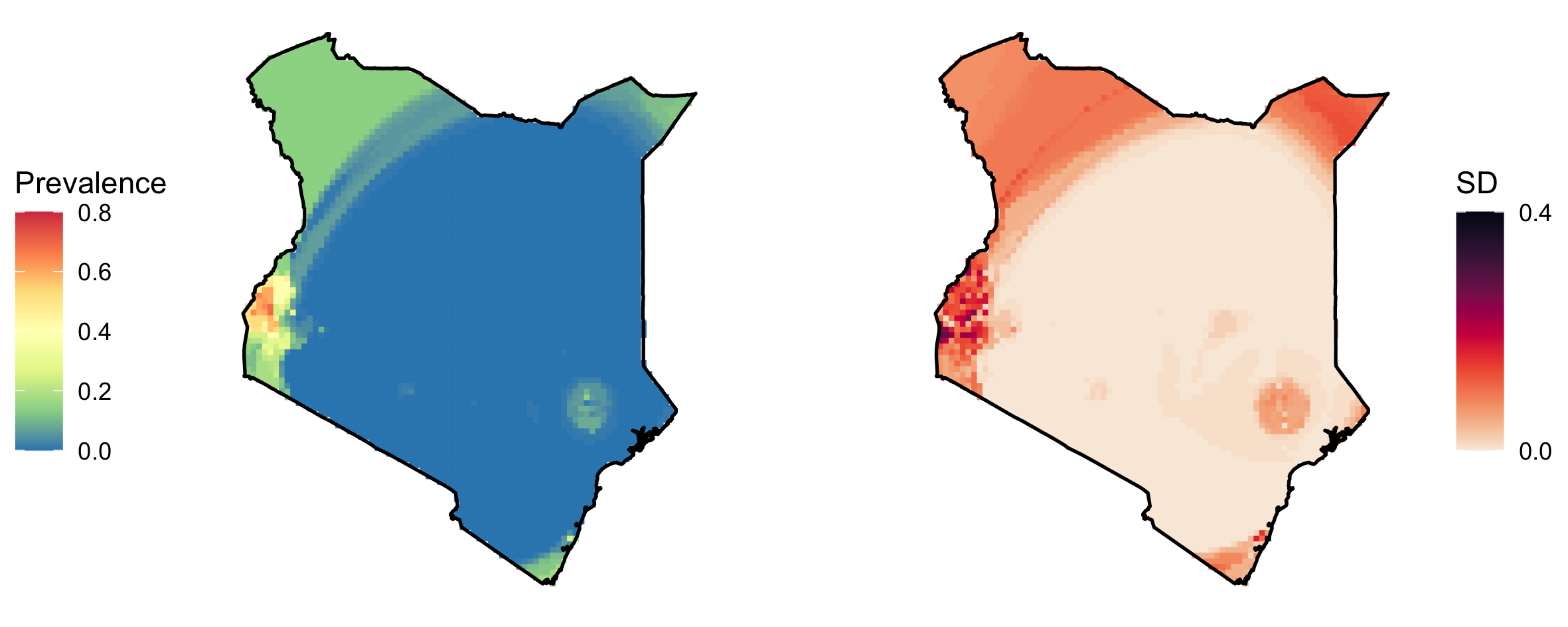}
       \label{fig:SpRF_Kenya_map}
    }\vspace{-10px}\\
    \subfloat[][FRK]{
       \includegraphics[width = 0.7\textwidth]{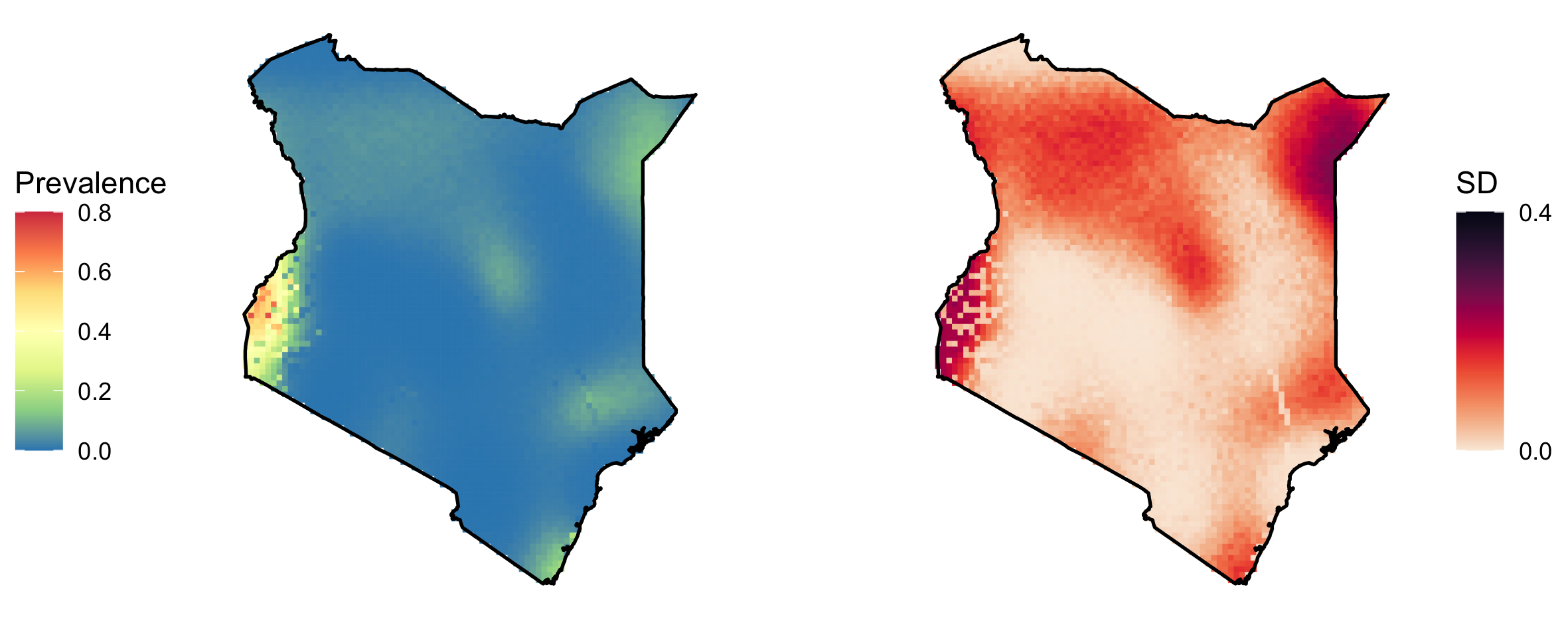}
       \label{fig:FRK_Kenya_map}
    }
    \captionsetup{position = bottom}
    \caption{Predicted prevalences and uncertainties for the four models when trained on \textit{P. falciparum} prevalence data from Kenya in 2009. Note that these maps are intended only to illustrate differences in model predictions when fitted to a small data sample, and are not likely to accurately represent malaria prevalence across the country in this year. }
    \label{fig:Kenya_maps}
\end{figure}

The predictions and uncertainties produced by the four models when trained on the 2009 Kenya prevalence data are shown in Figure~\ref{fig:Kenya_maps}. At the broadest scale, each model is similar in predicting a region of high prevalence in Western Kenya, with clusters of higher prevalence in the East, but low prevalence over much of the rest of the country. For each model, the predicted prevalence drops to zero quite quickly away from the data, indicative of a smaller spatial range than might be expected. This is especially prominent with INLA, and may be indicative of overdispersion in the data. 

A notable feature is the arc like band of higher prevalence in the north west of Kenya in SpRF's predictions in Figure~\subref*{fig:SpRF_Kenya_map}, which is further discussed in Section~\ref{sec:resultcontinent}. Higher prevalence in this region is also somewhat apparent in GPBoost predictions and, to a lesser extent, FRK. This area of predicted higher prevalence falls in a broad region with no prevalence data and so represents different approaches to extrapolation in the four models.

\subsubsection{Cross-validation results} \label{sec: CV}
\begin{table}[!ht]
	\centering
	\caption{Cross validation results of the four models with best results in each category in boldface.}
	{
	\begin{tabular}{cp{1.5cm}p{1.5cm}p{1.5cm}p{1.5cm}p{1.5cm}p{2cm}}
		\toprule
  Model & 10-fold-RMSE &  50-fold RMSE & Training Correlation & 10-fold Correlation & 50-fold Correlation & \text{\% points} \text{within 1SD} (10-fold)
   \\
        \midrule
       INLA    & 0.181 & 0.124  & 0.909  & 0.235   &  0.683 & 75\\
     GPBoost   & 0.127 & \textbf{0.11} & 0.873 & 0.646  & \textbf{0.751} & 84.211 \\
     SpRF      & 0.132 &  0.121  & \textbf{0.912}  & 0.641  & 0.702 & 37.105 \\
     FRK       & \textbf{0.125}  &  0.123  & 0.902 & \textbf{0.661 }& 0.702  & 83.421\\
    \bottomrule
	\end{tabular} }
	\label{tab:predactualscv}
\end{table}

Table \ref{tab:predactualscv} gives the cross validation results. In terms of cross validation RMSE and correlation, FRK performs the best for 10-fold CV, and  GPBoost performs the best for 50-fold CV. SpRF predictions had the highest correlations to the data used to train the model, but poorer correlation to out-of-sample data, indicating that this model may be overfitting to the training data. INLA performs poorly with respect to the 10 fold RMSE and correlation.  

From Table \ref{tab:predactualscv} we see SpRF has only 37.105\% of the points within 1SD for 10-fold cross validation, which is much lower than for the other models, and we discuss this further in Appendix \ref{sec:detailedCV}. GPBoost performs the best in terms of the percentage of points within 1SD. However these results need to be taken in context, because a higher standard deviation can increase this percentage. 

\begin{figure}[!ht]
    \centering
    \captionsetup{width=.3\linewidth}
     \subfloat[][INLA]{
        \includegraphics[width=0.3\textwidth]{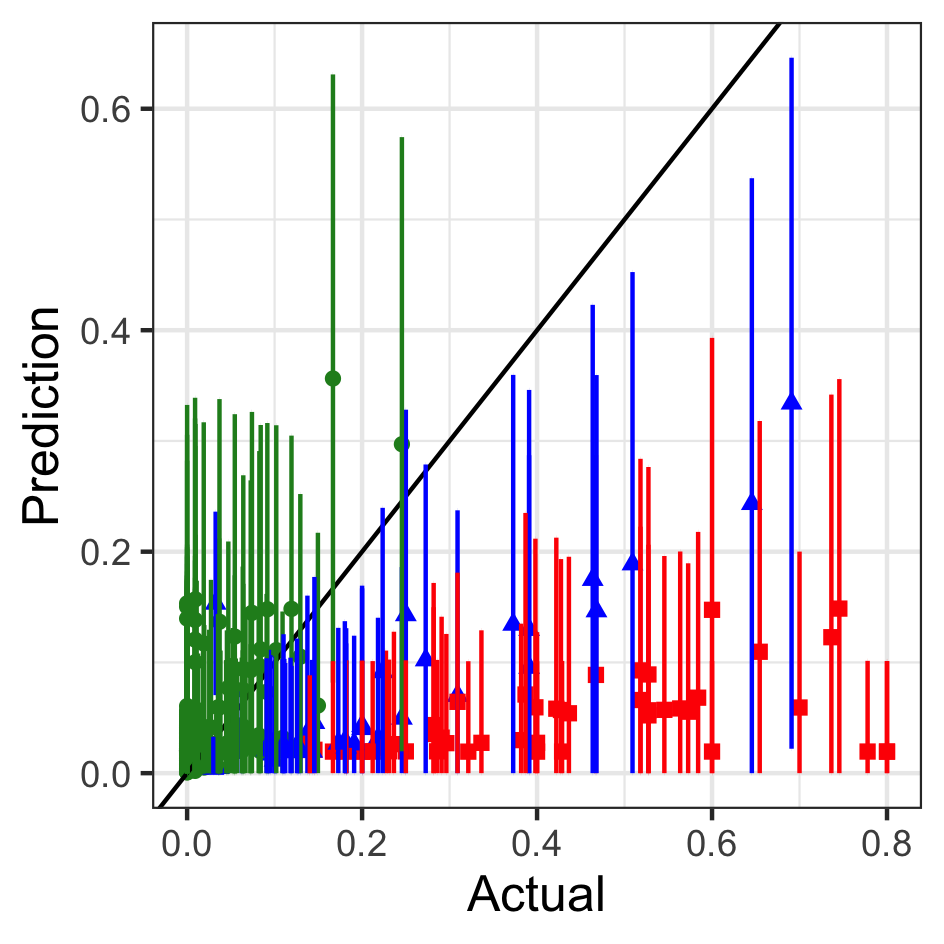}
        \label{fig:INLA10}
    } 
     \subfloat[][GPBoost]{
        \includegraphics[width=0.3\textwidth]{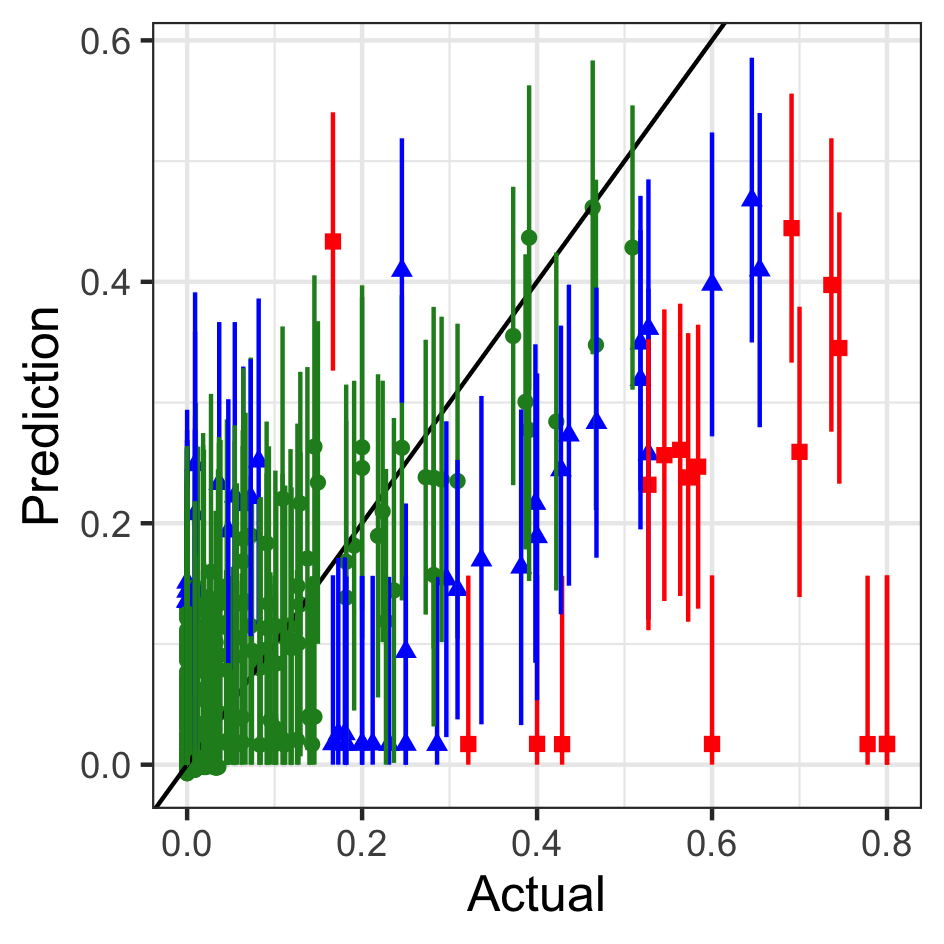}
        \label{fig:GPBoost10}
    } 
     \\
   \subfloat[][SpRF]{
        \includegraphics[width=0.3\textwidth]{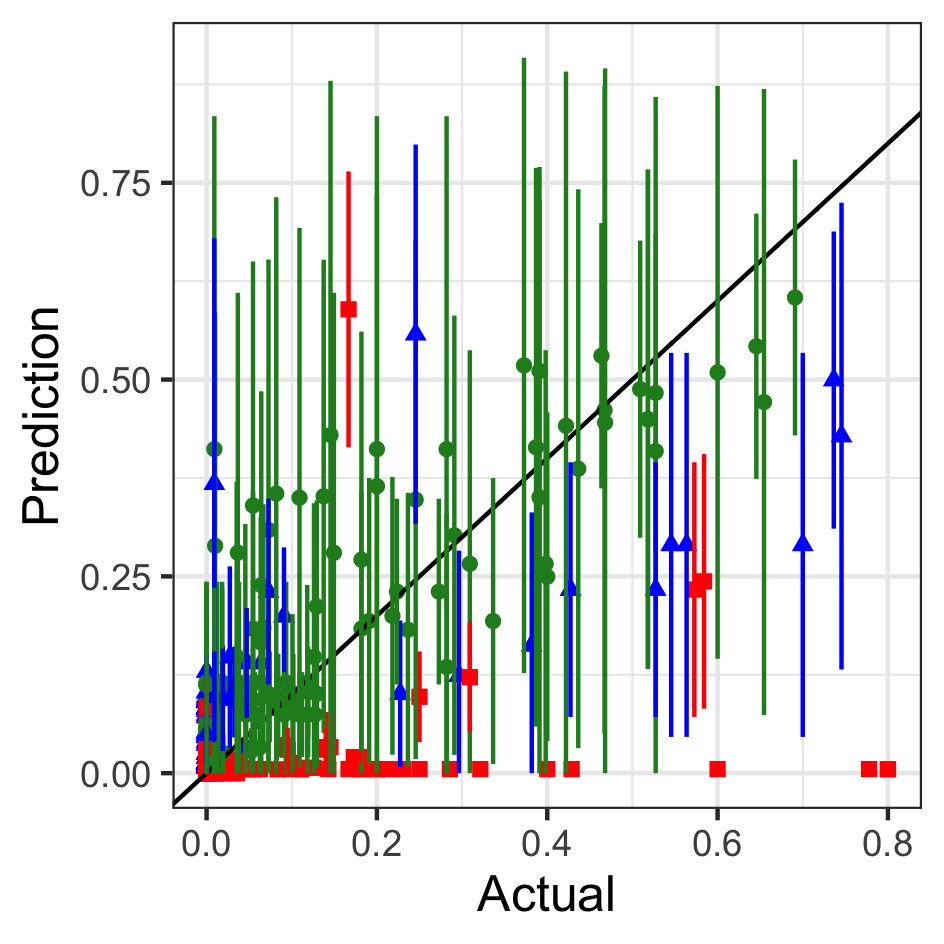}
        \label{fig:SpRF10}
    }
    \subfloat[][FRK]{
        \includegraphics[width=0.3\textwidth]{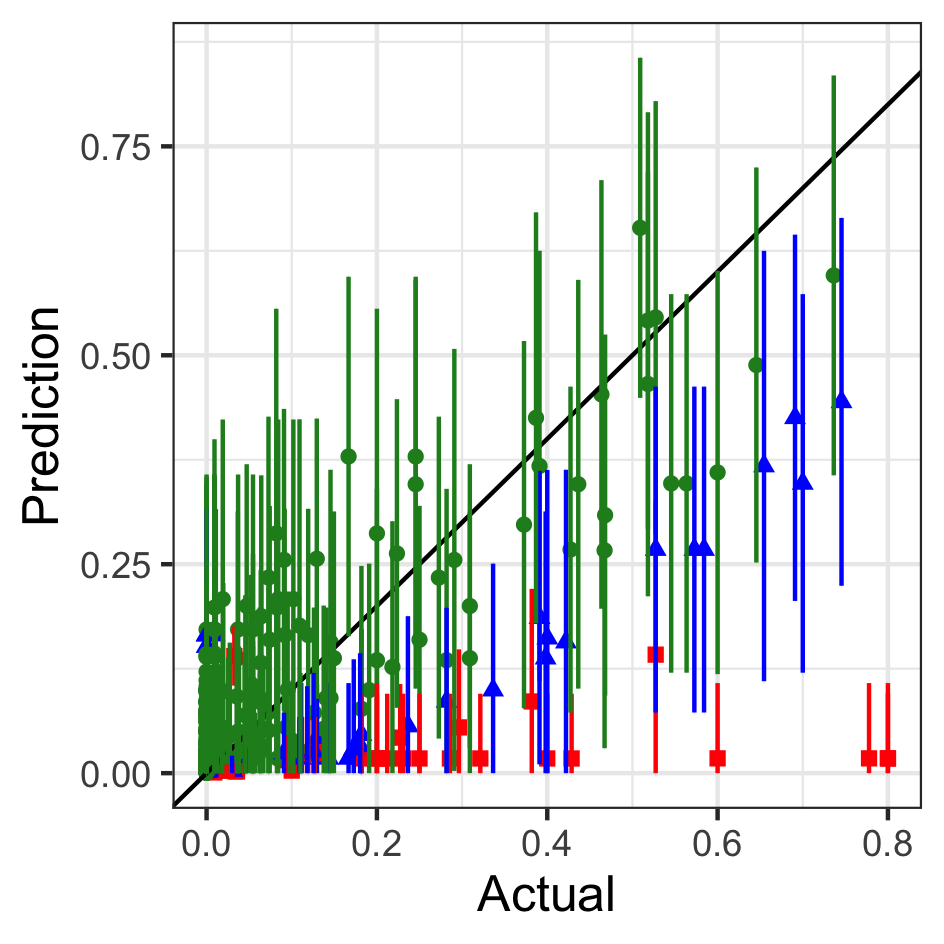}
        \label{fig:FRK10}
    } 
     \captionsetup{width=0.8\textwidth}
    \caption{Interval predictions for 10-fold cross validation for national level Kenya data. Points show the predicted mean from each model, and intervals show one standard deviation above and below the mean. }
    \label{fig:interval10fold1}
\end{figure}

Figure \ref{fig:interval10fold1} shows the interval and point predictions for the four methods for 10-fold cross validation. Points within 1SD are shown in green, points within 2SD are shown in blue, and the rest are shown in red. We see that many of INLA's predictions are close to zero and we investigate this issue in Appendix \ref{sec:input_noise}. Additional cross validation metrics that consider the prediction error divided by its standard deviation are detailed in \cite{cressie2015statistics}. However, we do not consider these metrics as some methods produce very small standard deviations and thus will result in very large values. More details on cross validation results in terms of the clusters and density of locations are given in Appendix \ref{sec:detailedCV}.
\subsection{Continent scale results}\label{sec:resultcontinent}
Geostatistical mapping is often carried out at a continent or global scale and frequently uses large datasets of observations. As computation time for inference and prediction can scale poorly with the amount of data and size of the domain, it is important to assess the performance, both in terms of predictive power and time, of recent methods. To examine how each of the four methods perform at larger scales, and to understand how predictions are affected by potential violations of model assumptions, and by clustering and sparsity of observation points, we expanded the study area to the whole continent of Africa in 2009, and fit each model to the three prevalence datasets shown in Figure~\ref{fig:Africa_input_data}. Figure~\ref{fig:Africa_prevalence_maps} shows the prevalence predicted by the models when trained on each input dataset. The corresponding uncertainties appear in Appendix~\ref{sec:prediction_uncertainty}.

\newcolumntype{M}[1]{>{\centering\arraybackslash}m{#1}}
\begin{center}
\begin{figure}
\begin{tabular}{p{2mm}M{0.31\textwidth}M{0.31\textwidth}M{0.31\textwidth}}
        &  (i) Observation data& (ii) Binomial sampling at observation locations & (iii) Binomial sampling with uniform coverage \\
    \rotatebox[origin = c]{90}{(a) INLA} 
    & \includegraphics[width=0.31\textwidth]{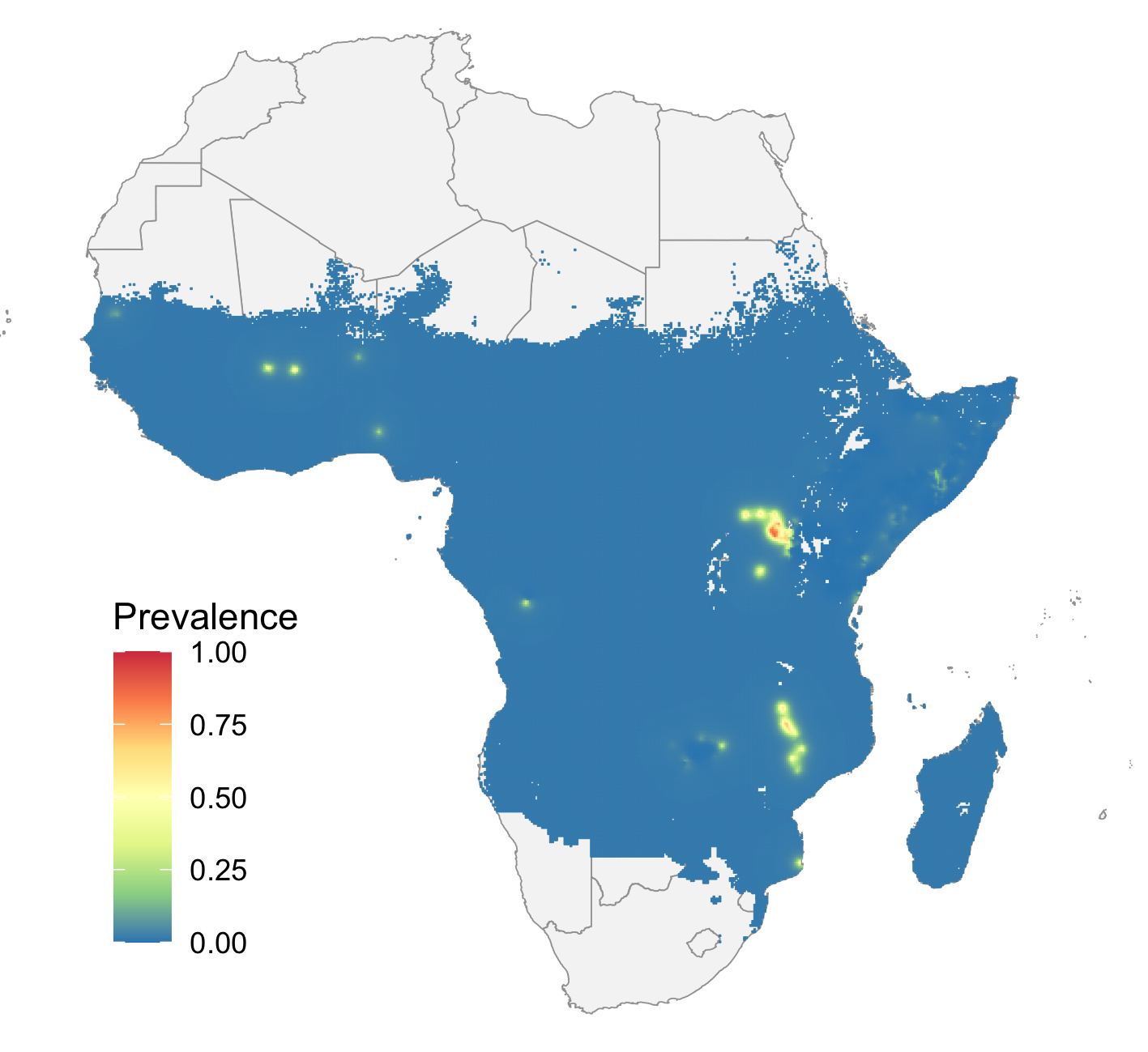}
    & \includegraphics[width=0.31\textwidth]{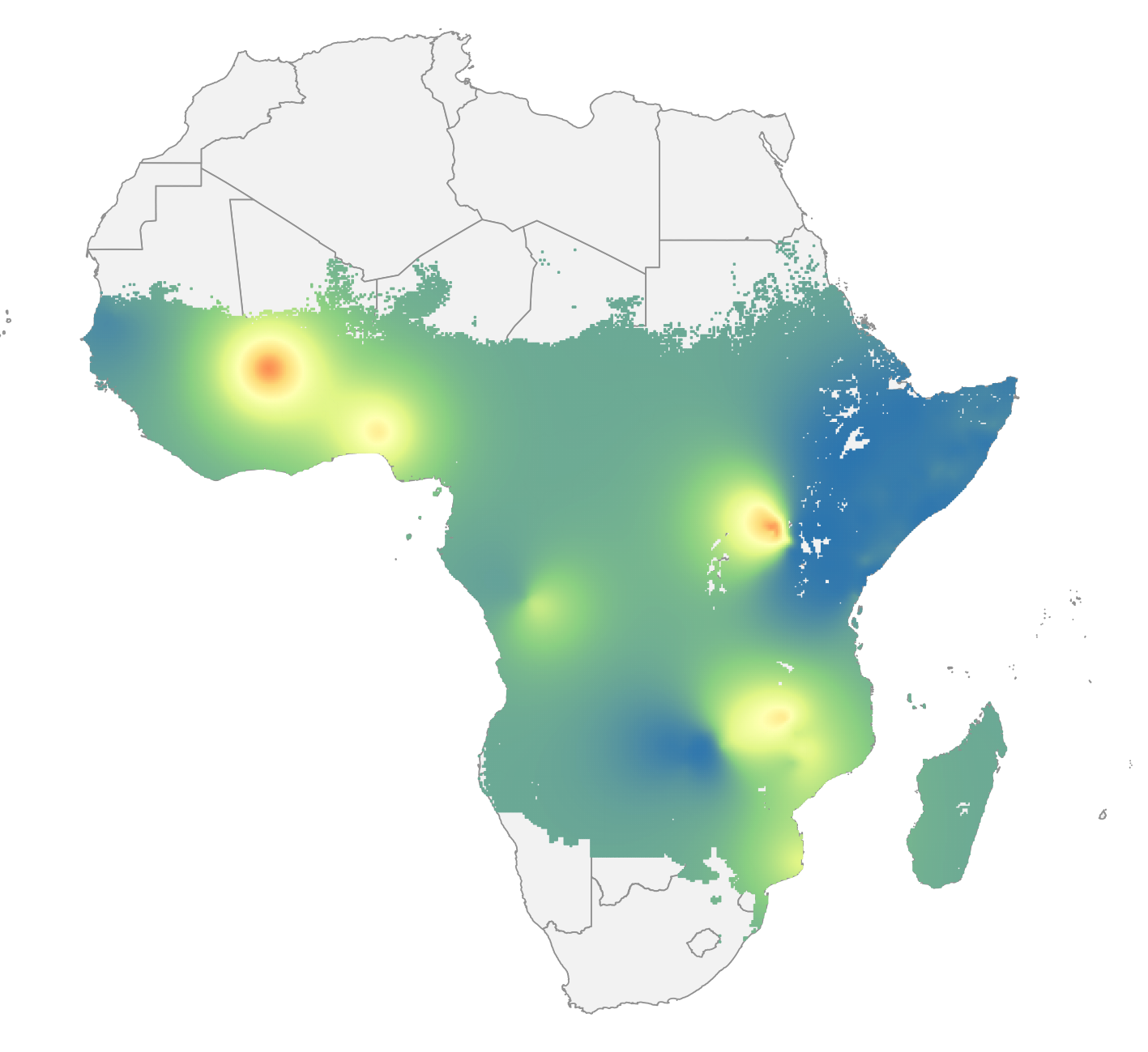}
    & \includegraphics[width=0.31\textwidth]{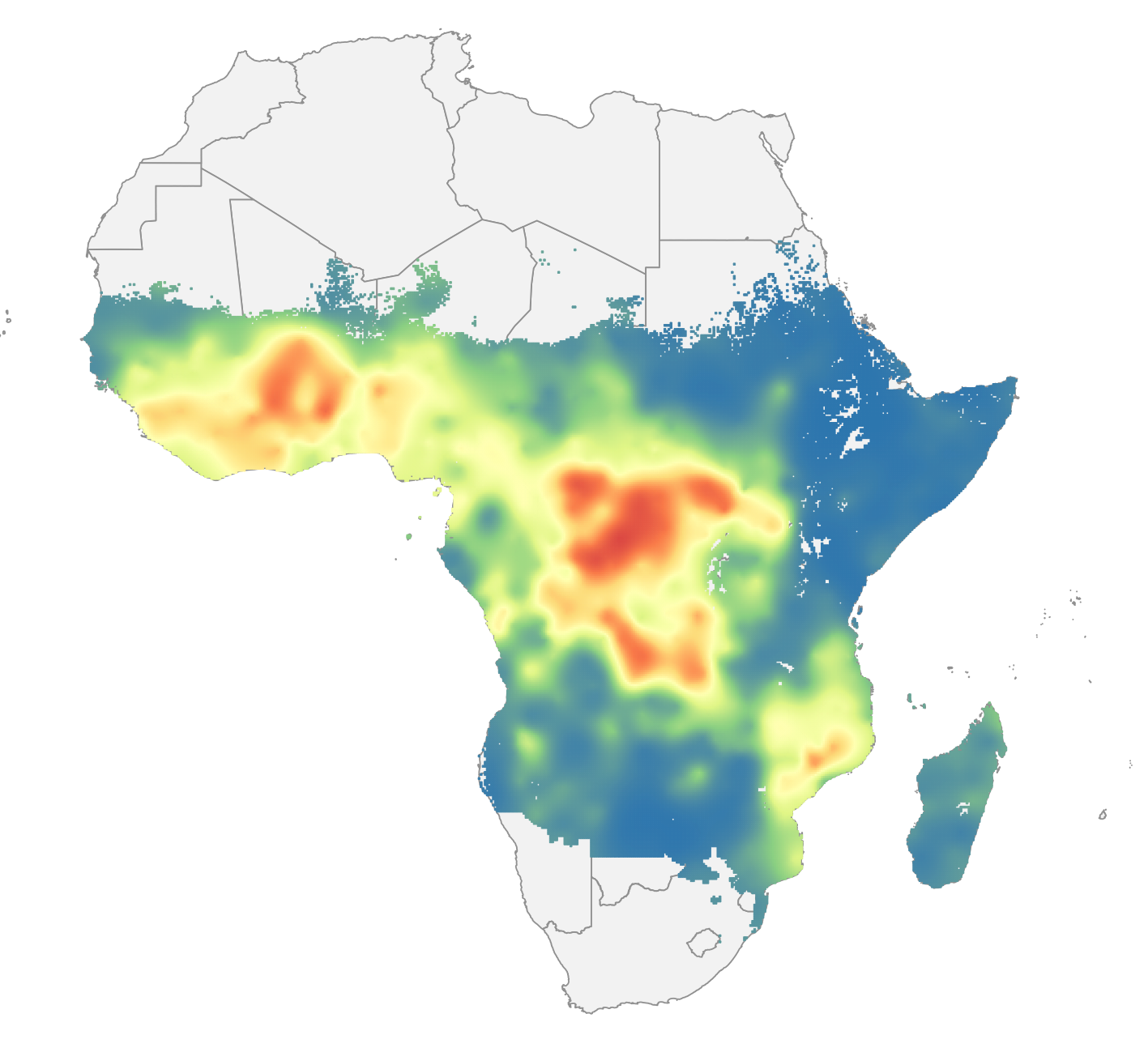} \\
    \rotatebox[origin = c]{90}{(c) GPBoost} 
    & \includegraphics[width=0.31\textwidth]{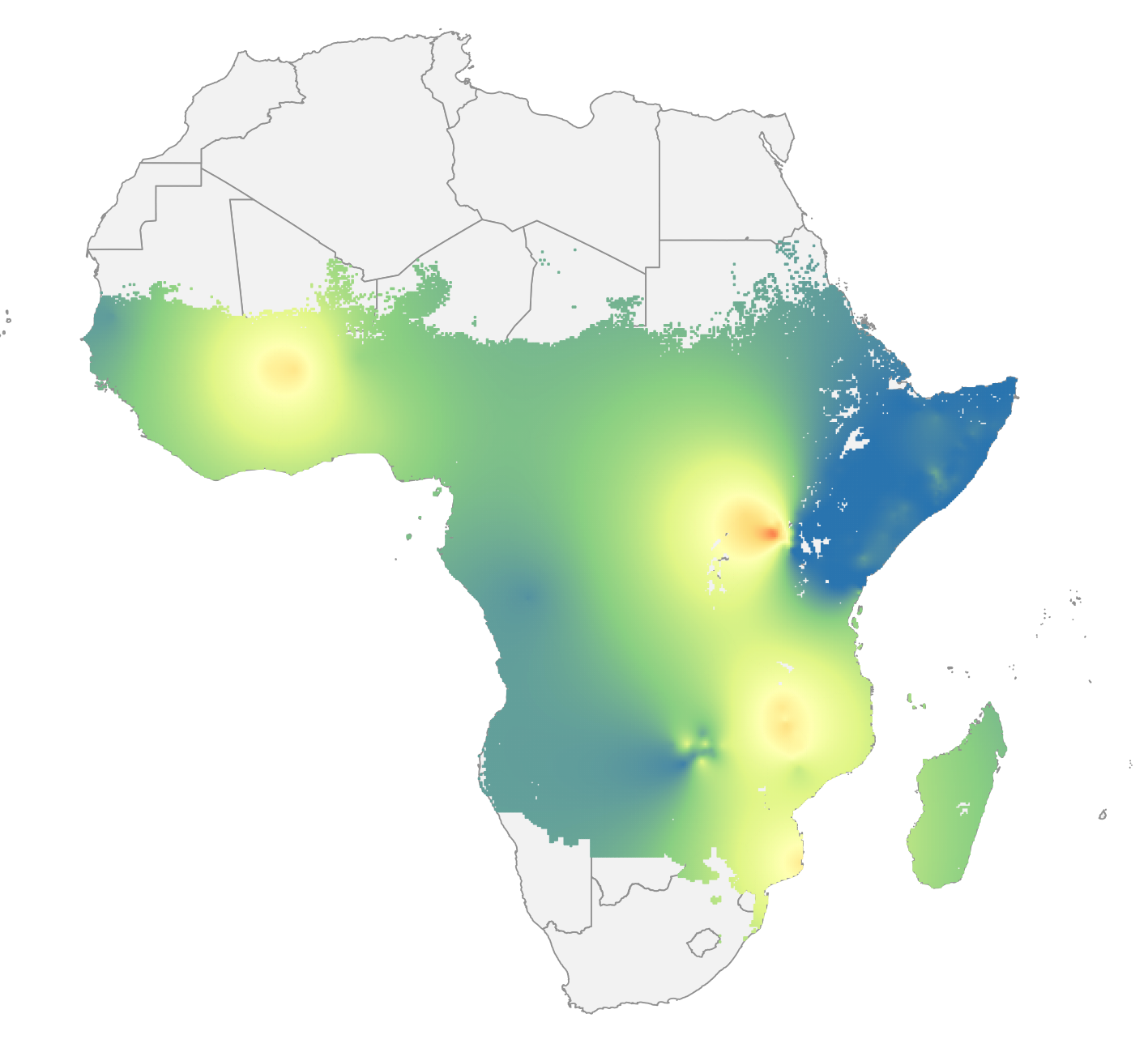} 
    & \includegraphics[width=0.31\textwidth]{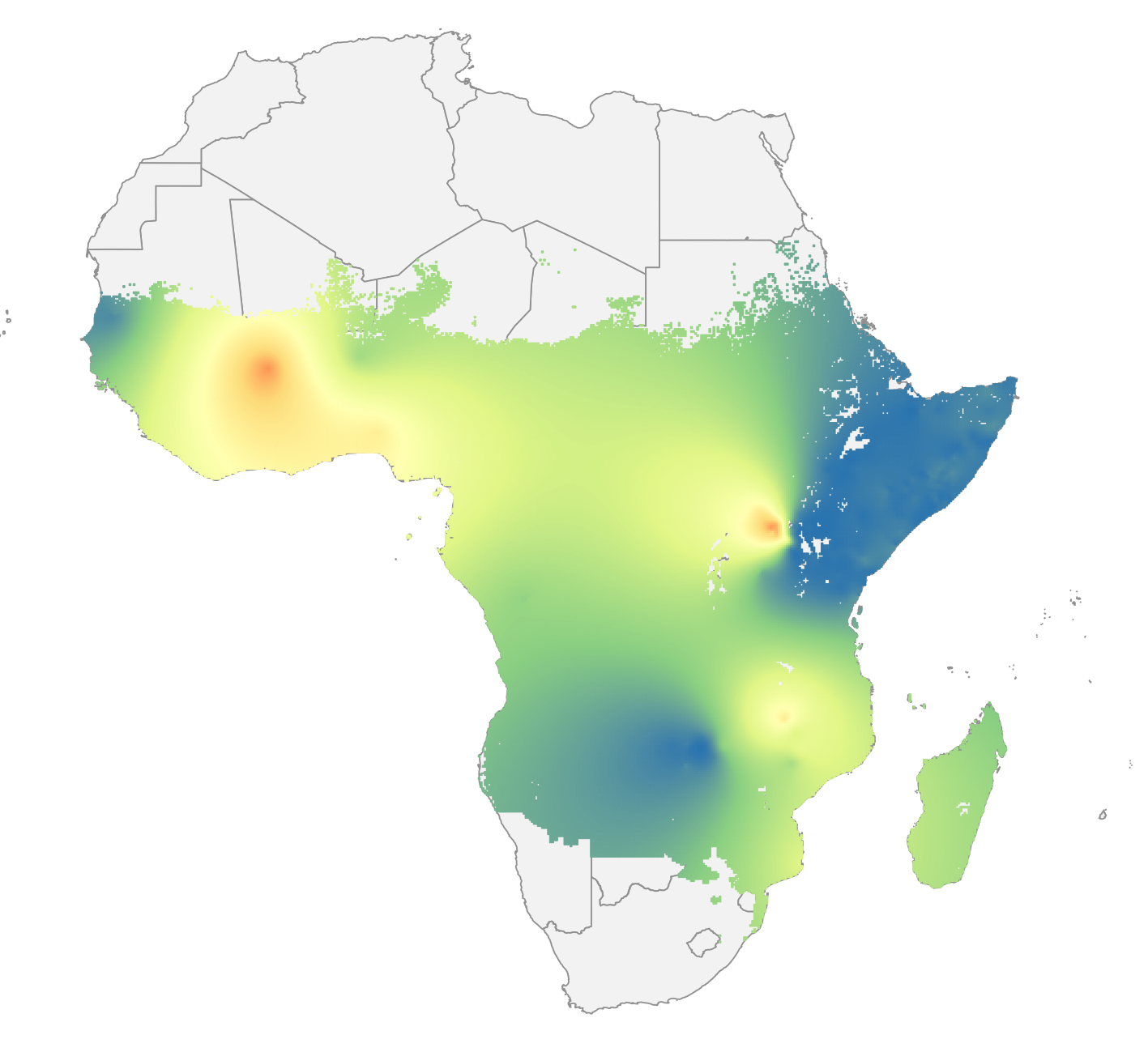} 
    & \includegraphics[width=0.31\textwidth]{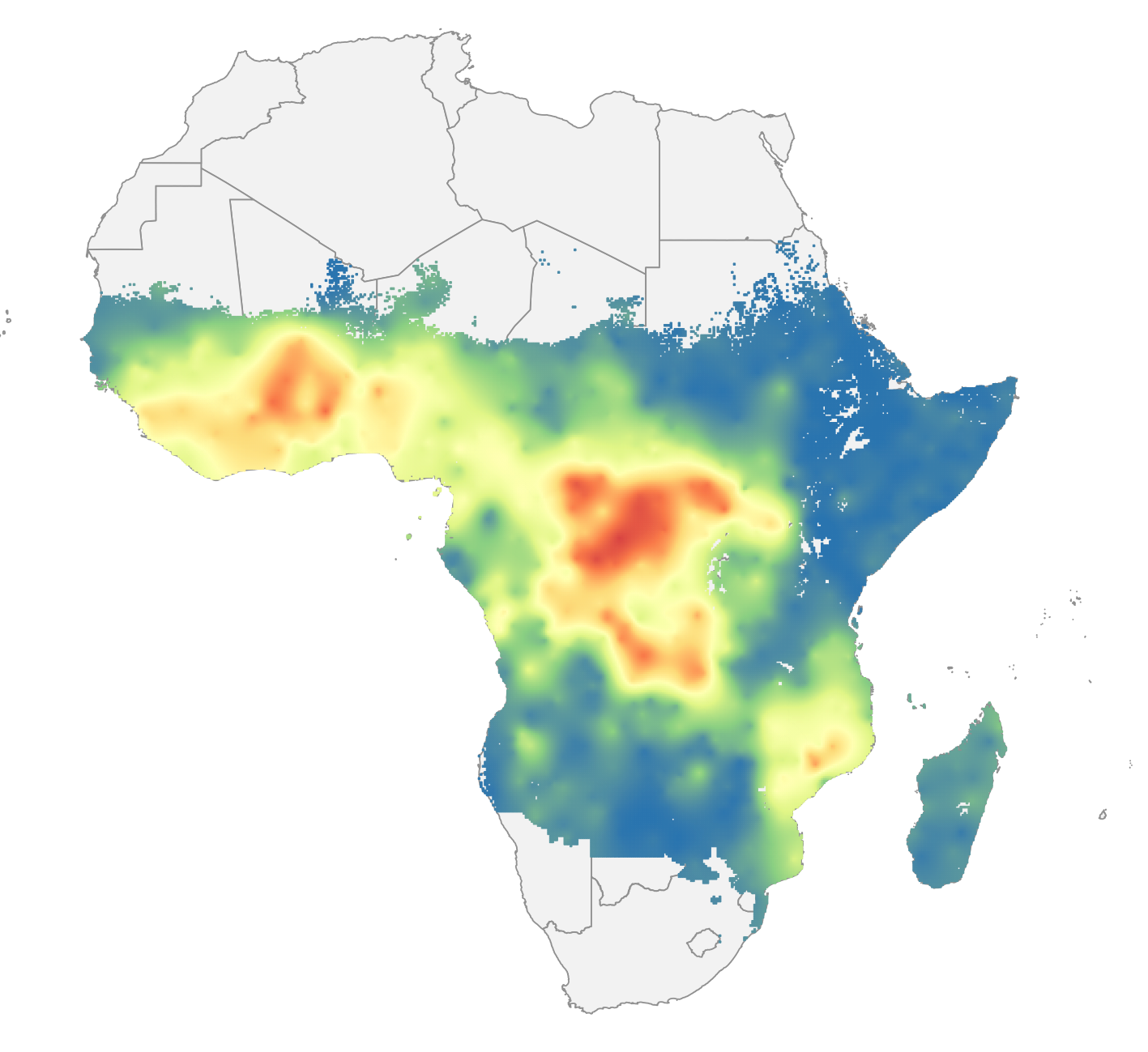}
    \\
     \rotatebox[origin = c]{90}{(b) SpRF} 
    & \includegraphics[width=0.31\textwidth]{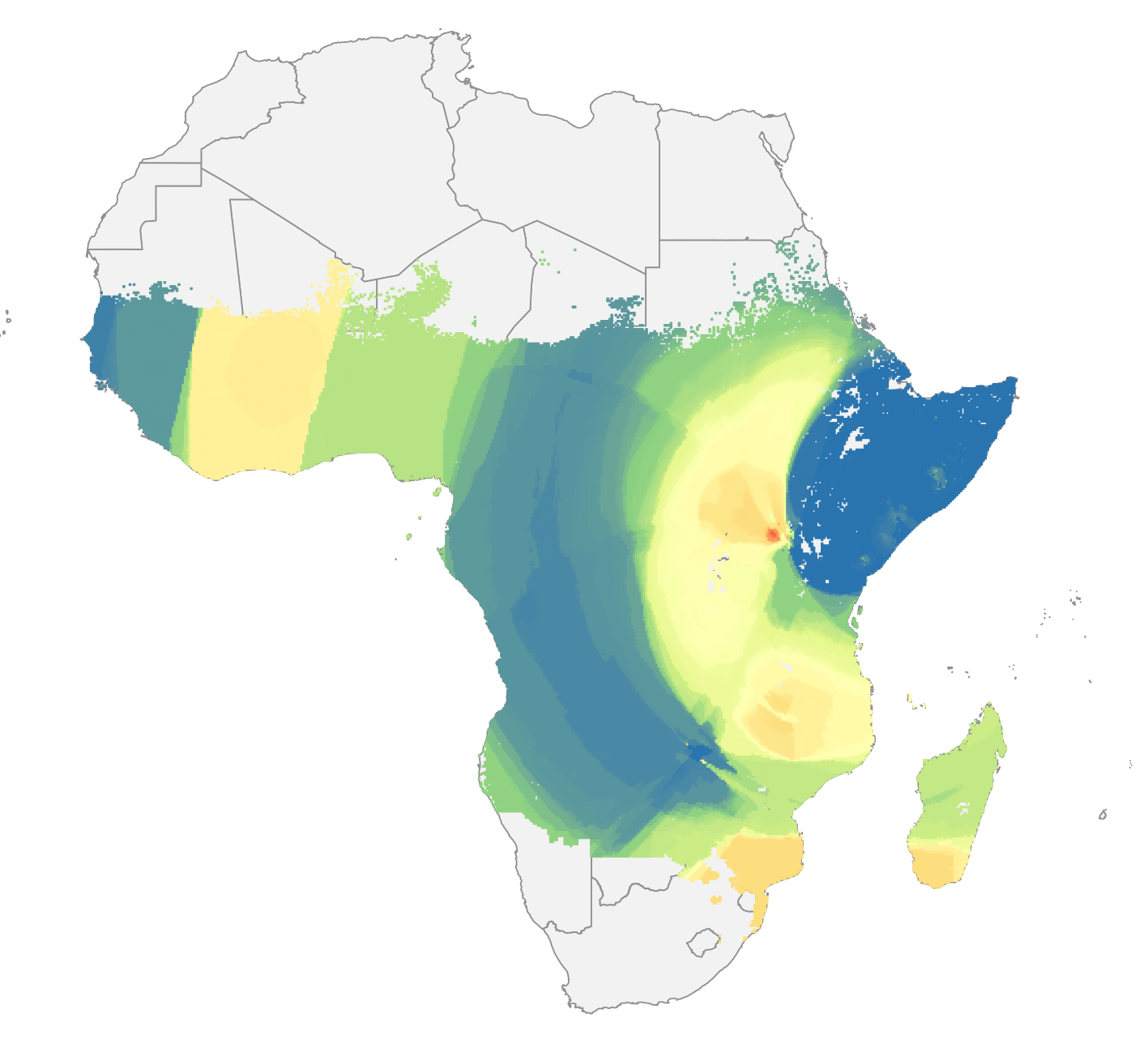}
    & \includegraphics[width=0.31\textwidth]{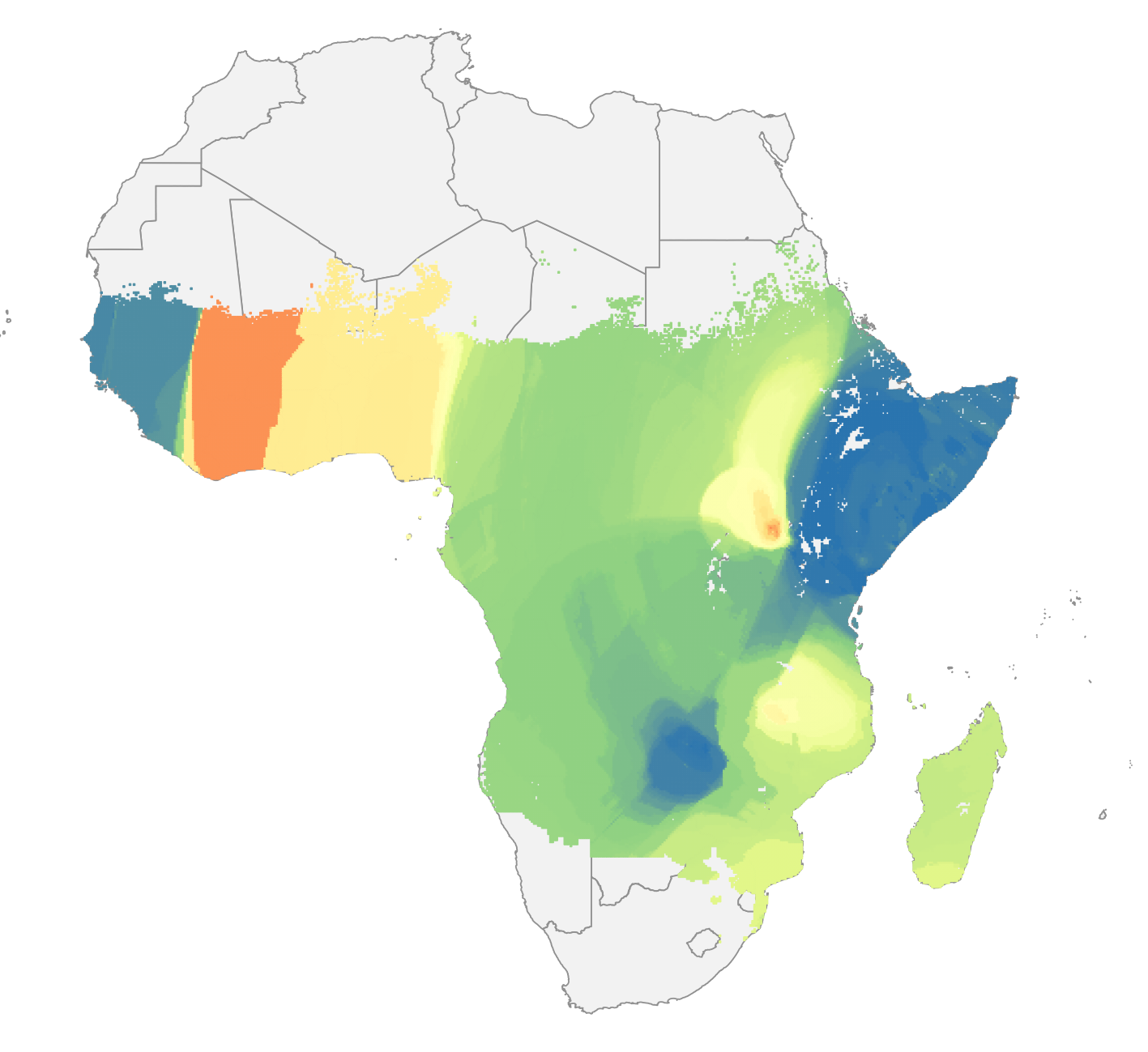} 
    & \includegraphics[width=0.31\textwidth]{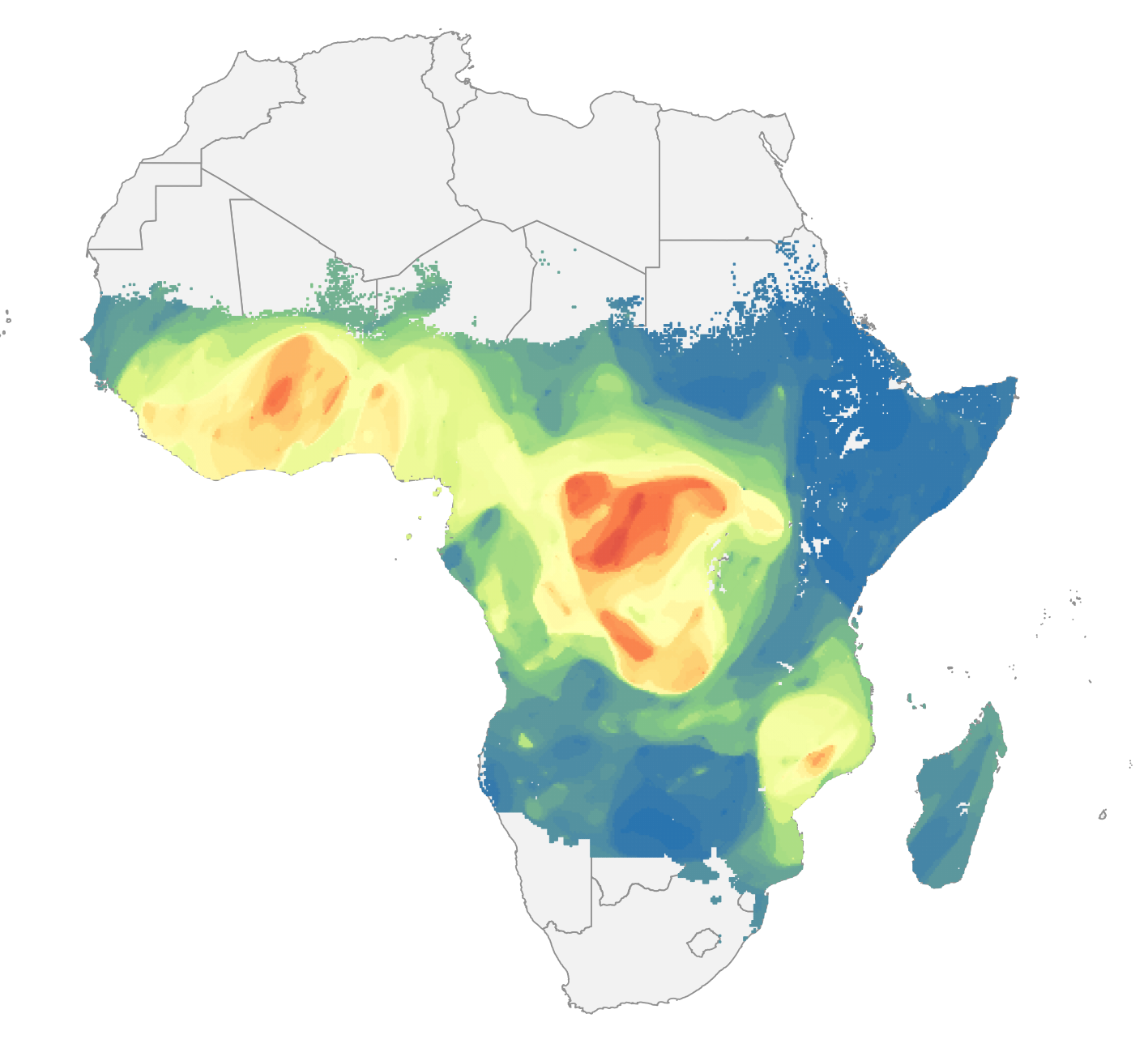}
     \\
    \rotatebox[origin = c]{90}{(d) FRK} 
    & \includegraphics[width=0.31\textwidth]{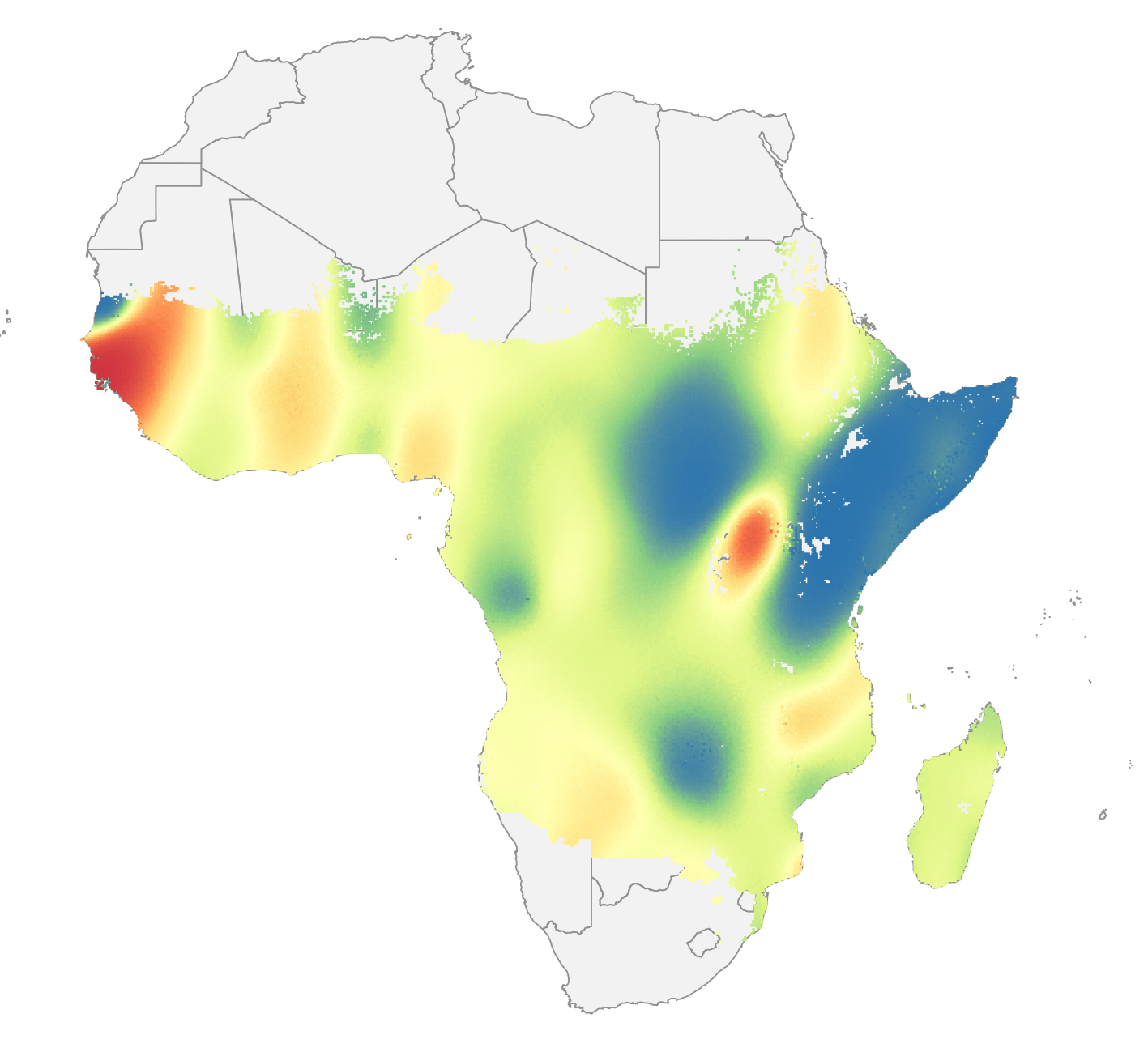}
    & \includegraphics[width=0.31\textwidth]{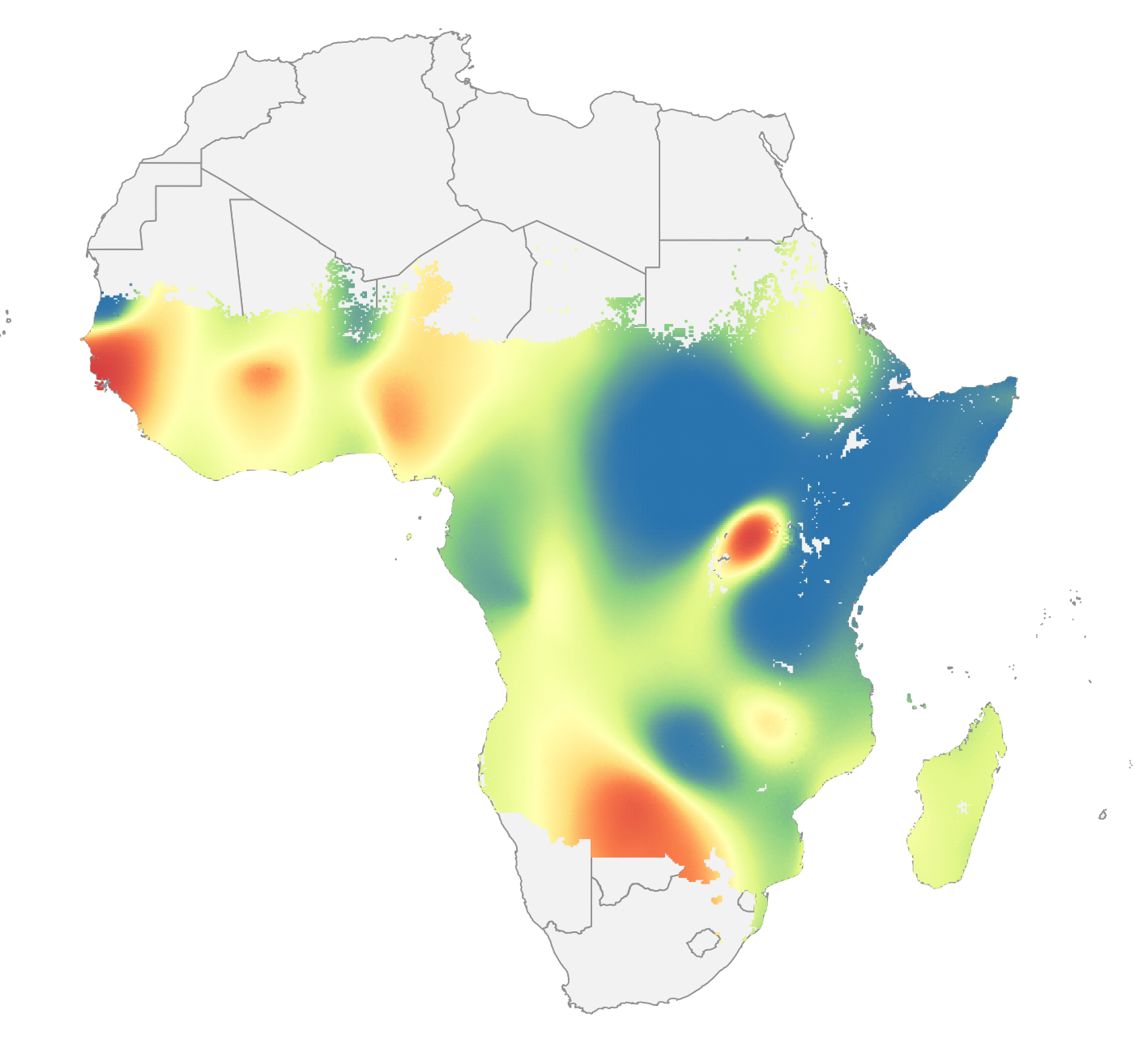} &
    \includegraphics[width=0.31\textwidth]{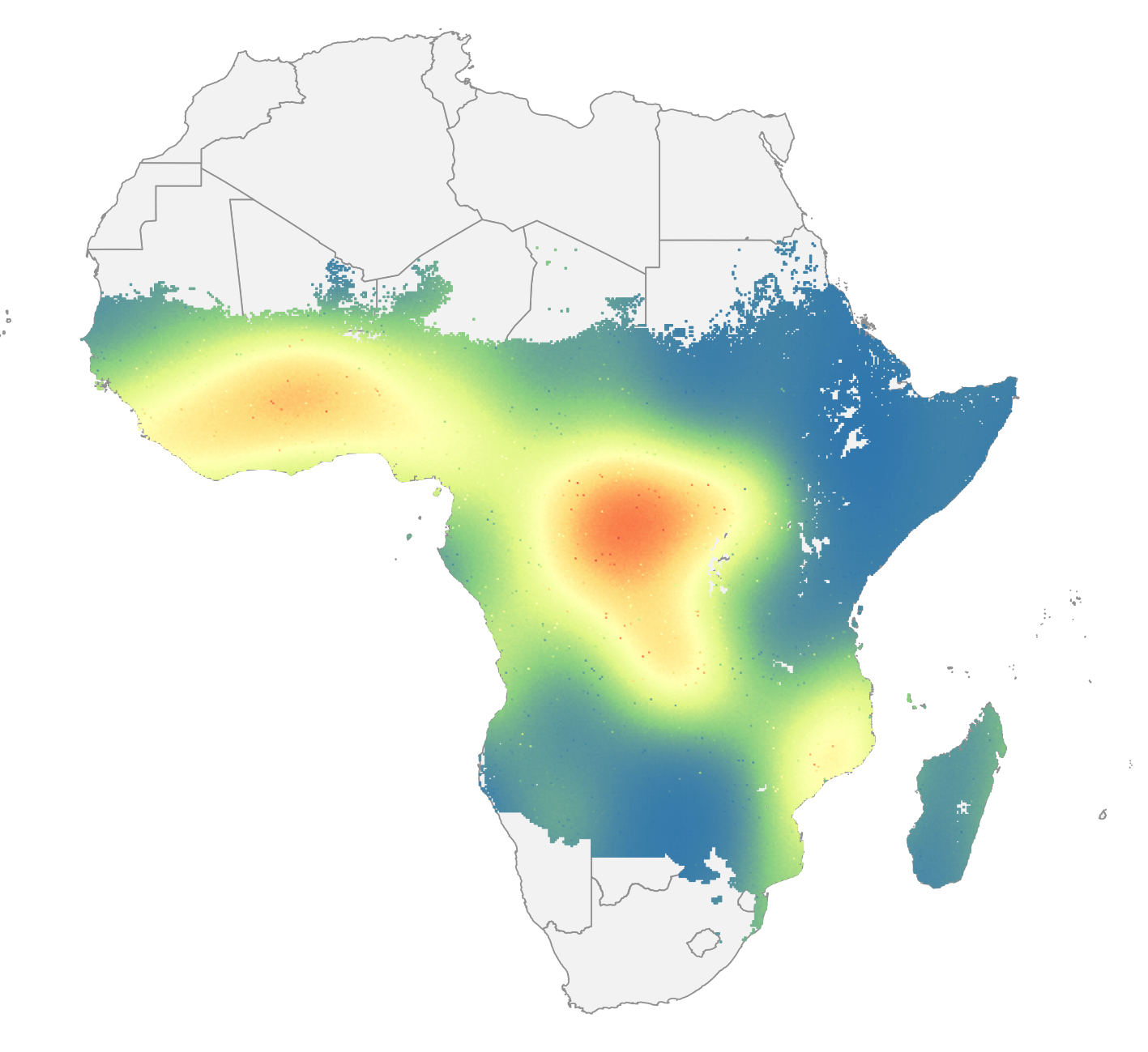}
  
\end{tabular}
  
    \caption{\textit{P. falciparum} prevalence predictions when fit using three different datasets. In column (i), models are fit using the survey data from Africa in 2009, shown in Figure~\protect\subref*{fig:Africa_observation_points}. In column (ii), the models are fit to binomial samples drawn from the Malaria Atlas prevalence raster at the same survey locations, shown in Figure~\protect\subref*{fig:Africa_observation_binomial_points}. In column (iii), they are fit to binomial samples drawn from the raster at 1000 uniformly selected locations across the continent, shown in Figure~\protect\subref*{fig:Africa_uniform_binomial_points}. Outputs have been masked by the Malaria Atlas Project raster in Figure~\protect\subref*{fig:MAP_raster}.}
    \label{fig:Africa_prevalence_maps}
\end{figure}
\end{center}

Scaling up to the continent level reveals differences between the models that are not apparent at smaller scales. While the national scale prevalence maps in Figure \ref{fig:Kenya_maps} are largely similar apart from the slight banding effect seen in SpRF's predictions, the prevalence maps in Figure~\ref{fig:Africa_prevalence_maps} differ significantly, with artifacts appearing in several of the maps.

Overall, the four models are better at local interpolation than extrapolation over large regions without data. The predictions in Figure~\ref{fig:Africa_prevalence_maps}(iii) generated using the randomly distributed data recover the prevalence structure of the MAP raster in Figure \subref*{fig:MAP_raster} much more faithfully than the predictions in Figure~\ref{fig:Africa_prevalence_maps}(ii) from the sparser non-uniform data. This behaviour is expected as malaria prevalence is known to be highly heterogeneous and our models do not use covariate data.

SpRF's predictions display a prominent banding effect, visible in both the country and continent scale maps where contiguous arc-like bands of high prevalence appear in both point and uncertainty estimates. This may be explained by the fact that SpRF models the quantity of interest -- malaria prevalence in our case -- based on distances to points with known values. Thus we observe bands of high or low prevalence at different radii from clusters of observations, and the piecewise constant nature of random forests would contribute to the sharp steps between each band. The banding effect is particularly prominent in Figures \ref{fig:Africa_prevalence_maps}(ci) and (cii), where the points were clustered into smaller regions, while it is less obvious when the datapoints have good spatial coverage, as in Figure \ref{fig:Africa_prevalence_maps}(ciii), which does not show bands spanning the continent. Further increasing the number of simulated points was found to further reduce the prominence of these bands (results not shown). 

Even though SpRF produces maps with this unwelcome feature, the cross validated point estimates are quite accurate. From Table~\ref{tab:predactuals} in Appendix \ref{sec:detailedCV} we see that SpRF has the highest proportion of points with absolute errors less than 0.05 and 0.1 for 10-fold cross validation, which is a harder task for the algorithms than 50-fold cross validation. Thus, we can argue that SpRF gives reliable predictions at points even though it may produce a predictive map that can be misleading in regions where there are no sample points.   

Figure~\ref{fig:Africa_prevalence_maps}(ai), produced by INLA with the observation data, displays a sudden drop in prevalence away from observations, resulting in flat near-zero predictions covering most of the continent. This appears to result from a combination of both the sparsity and noise present in the data, rather than clustered nature of the data alone. Figure~\ref{fig:Africa_prevalence_maps}(aii) uses nearly the exact same locations, yet shows higher values of prevalence spreading much further from the observations. In Appendix~\ref{sec:input_noise} we outline evidence that this effect arises from unaccounted-for overdispersion in the observation data. In particular, increased noise in the data appears to reduce the estimated range for the spatial random effect, resulting in the model reverting to constant predictions away from observation locations. This behavior is consistent with INLA's poor performance in the 10-fold cross validation analysis in Section~\ref{sec: CV}, where the model predicted near-zero malaria prevalence for each of the held out folds.   

We observe that FRK's predictions depend strongly on the arrangement of the basis functions, which are generally placed by the software based on the data locations and the user parameters introduced in Section~\ref{sec:FRK_model}. For example, FRK's predictions in Figures~\ref{fig:Africa_prevalence_maps}(di) and (dii) display spurious oscillations in regions with little or no data, however these oscillations correspond to periodic placement of the basis functions. Other arrangements we tested led to flat predictions over the whole continent (results not shown). These types of artifacts are not present in Figure~\ref{fig:Africa_prevalence_maps}(diii), where the input data has good spatial coverage. However, this map appears as a smoothed version of the input data, and does not resolve the finer structure in the MAP surface. The impact of the arrangement and number of basis functions on the prediction maps is detailed further in Section~\ref{sec:FRK_bases}.

For both Kenya and Africa, GPBoost produces prevalence maps without the artifacts appearing in the other models' outputs. However, the uncertainty maps in Figures \ref{fig:Africa_uncertainty_maps}(ci)-(cii) in Appendix~\ref{sec:prediction_uncertainty} exhibit a high level of overall uncertainty regardless of whether the regions have more survey points or not. This is further confirmed by the near-constant interval widths that rarely fluctuate with the density of the survey points in Figure \ref{fig:widthdensity} (Appendix \ref{sec:detailedCV}) . Even though GPBoost currently computes only Euclidean distances between coordinates, both the prevalence maps for Africa and Kenya appear to be reasonable. However, it is sub-optimal to use Euclidean distances between longitude and latitude coordinates for a global model.  

\subsection{Computational results} \label{sec:computational_results}
Times taken to run each model on each of the datasets are shown in Table~\ref{tab:Map_times}.  While FRK is consistently the fastest, INLA shows great variation among the African datasets, ranging from less than 10 minutes with the uniform simulated data to 69.11 minutes with the observation data. Further analysis of this variation for INLA is given in Appendix B.

\begin{table}[!ht]
    \centering
    \caption{Times taken in minutes to train the models on each dataset and generate the prediction maps.  The \textit{Kenya: Observations} column corresponds to the maps in Figure~\ref{fig:Kenya_maps}. The \textit{Africa: Observations}, \textit{Africa: Simulated observations}, and \textit{Africa: Simulated uniform} columns correspond to columns (i), (ii), and (iii) of Figure~\ref{fig:Africa_prevalence_maps} respectively. Note that different machines were used to run the models for the Kenya and Africa datasets.}
    \begin{tabular}{cp{2cm}p{2cm}p{1.7cm}p{3cm}p{3cm}}
            \cmidrule{1-6}
           & & \multicolumn{4}{c}{Dataset} \\
           \cmidrule{3-6}
           & & Kenya: \mbox{Observation} &Africa: \mbox{Observation} & Africa: \text{Simulated observation}\ & \text{Africa: Simulated} \mbox{uniform} \\
           \cmidrule{1-6} \vspace{-5px}
           \\ 
           \multirow{4}{3em}{Model} & INLA & 0.34 & 69.11 & 11.49 & 7.05 \\
           & GPBoost & 0.99 & 11.27813 & 6.56 & 13.85\\
           & SpRF & 0.44 & 24.54 & 24.64 & 27.60\\
           & FRK & 0.35 & 3.28 & 3.41 & 3.09
           \vspace{5px}\\ 
           \cmidrule{1-6}
    \end{tabular}
    \label{tab:Map_times}
\end{table}

\begin{figure}
    \centering
    \includegraphics[width = 0.6\textwidth]{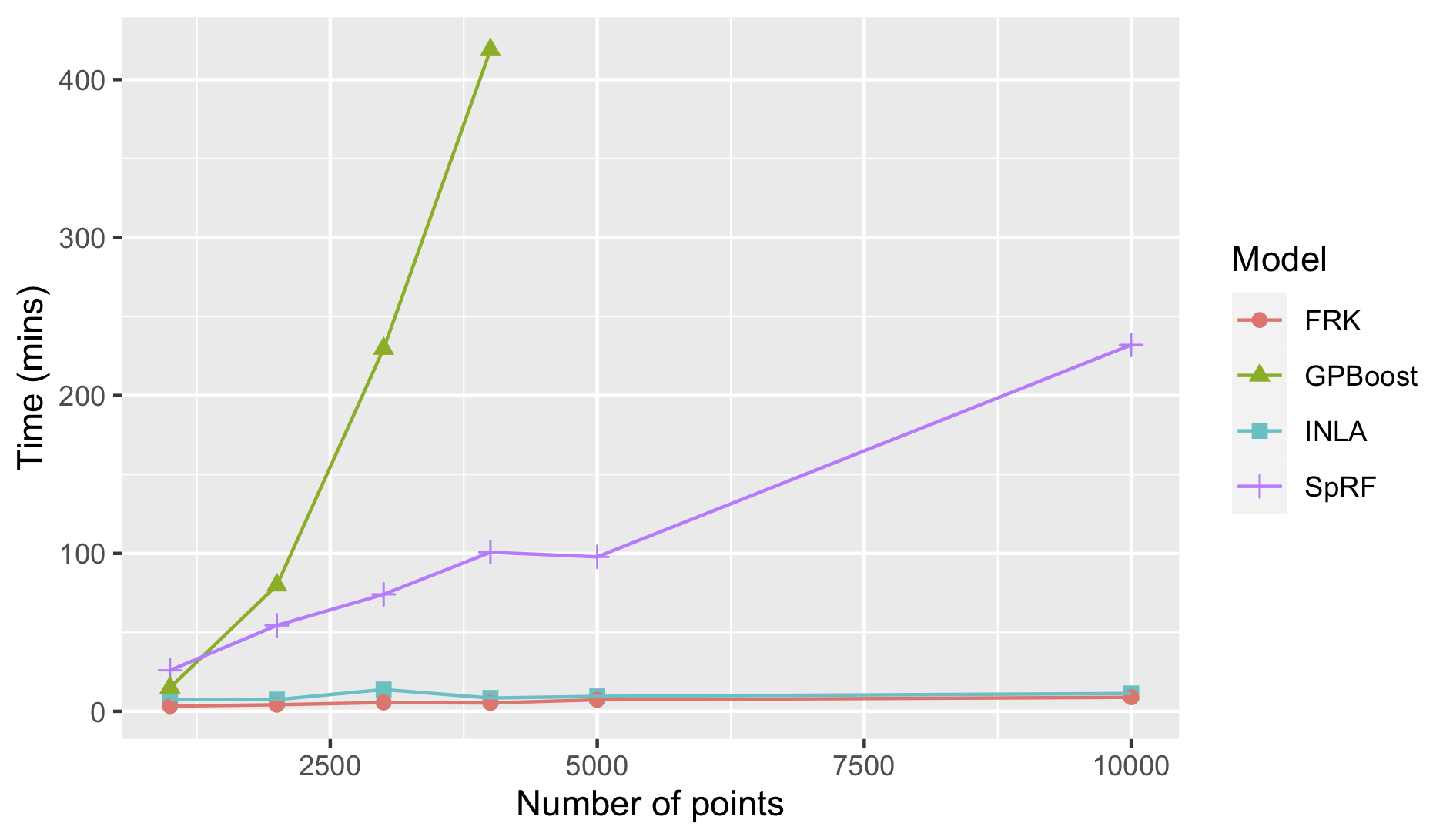}
    \caption{Times taken by each model on uniformly distributed simulated datasets. GPBoost was not run with 5000 or 10000 points due to the likely long computation time.}
    \label{fig:time_vs_n}
\end{figure}

Each model was additionally trained on simulated prevalence datasets with 1000 - 10000 points selected at random. Figure~\ref{fig:time_vs_n} shows times taken to fit each model and produce predictions as the dataset size varies. Both INLA and FRK remained very fast on larger datasets, showing little variation in their times. In contrast SpRF's time appears to increase linearly with dataset size, and GPBoost rapidly slows down on larger datasets, reflecting the computational requirements of using an unapproximated Gaussian process. As noted in Section~\ref{sec:GPBoost_intro}, a Vecchia approximation is available with GPBoost to improve the computational efficiency. In Appendix~\ref{sec:Vecchia}, we examine the effects on computation and model predictions of applying this approximation.

\subsection{Sensitivity of FRK and INLA on parameter choices}
Of the four methods, INLA and FRK show promise in their computational efficiency, displaying favourable scaling compared to SpRF and GPBoost in Figure~\ref{fig:time_vs_n}. Additionally, while the artifacts in SpRF's output appear to stem from the way it uses distances as an input, it is less clear whether the artifacts in INLA and FRK's prevalence maps are due to specific model parameters, or if they are fundamentally caused by the approximations used by each method. For this reason we examine these two methods more closely and test the sensitivity of their predictions to the model parameters.

\subsubsection{INLA} \label{sec:INLA_betabinomial}
The primary artifact visible in INLA's prediction maps is the flat, near zero, predictions when the model is fit to the observation data, as shown in Figure~\ref{fig:Africa_prevalence_maps}(ai). Appendix~\ref{sec:input_noise} outlines evidence that this feature is due to overdispersion, suggesting that the binomial response is unsuited for modelling the variability in the observed malaria data, despite commonly being used in tutorials on the application of INLA to disease mapping problems.

\begin{figure}
    \centering
    \includegraphics[width = 0.4\textwidth]{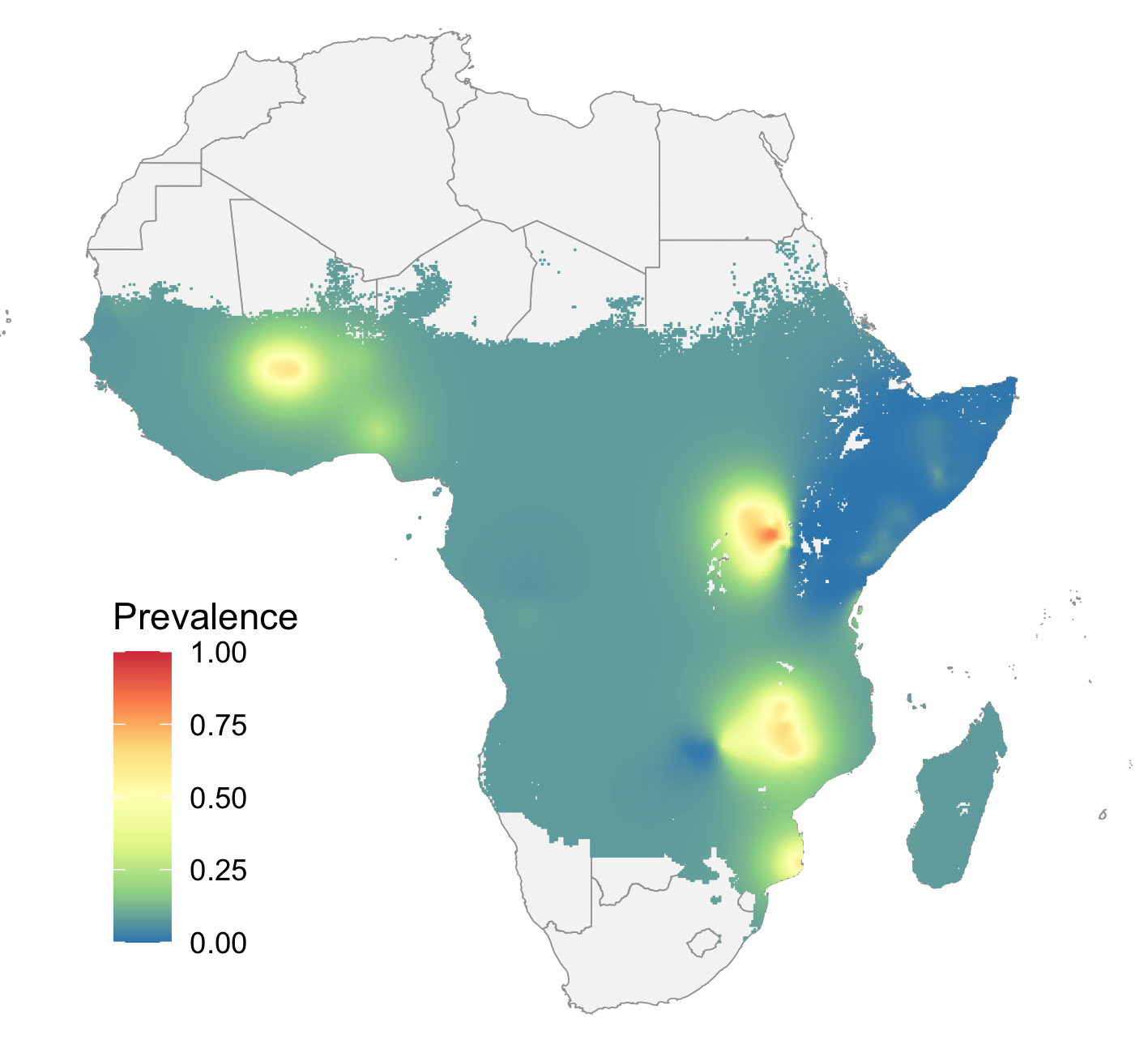}
    \caption{Prevalence predictions from an INLA model with a Beta-binomial response, fit to the observation data in Figure~\protect\subref*{fig:Africa_observation_points}.}
    \label{fig:beta-binomial_prediction}
\end{figure}

Several adjustments can be made to the model to address this overdispersion, such as the use of a Beta-binomial or Gaussian response (either directly on the proportion positive, or its empirical logit transform), or the inclusion of an independent error term in the linear predictor. All of these options include an additional parameter in the model to capture error variance at the level of the observation. Figure~\ref{fig:beta-binomial_prediction} shows predictions from an INLA model with a Beta-binomial response which has been fit to the observation data. The flat predictions of Figure~\ref{fig:Africa_prevalence_maps}(ai) are notably absent, suggesting that the Beta-binomial is effective in resolving the overdispersion. A Gaussian response was additionally tested and was found to also handle the variability in the observation data, with results shown in Appendix~\ref{sec:INLA_gaussian}. These results highlight a need for caution when applying INLA with a binomial response to disease mapping problems, and the importance of checking for overdispersion.

\subsubsection{FRK} \label{sec:FRK_bases}
While the fastest of the four methods, FRK's continent-scale predictions display a spurious ``spotty'' pattern when fit to either of the spatially sparse datasets and a much less detailed map when fit to the simulated data at randomly selected locations (Figure  \ref{fig:Africa_prevalence_maps}). These features appear to stem from FRK's use of a small number of basis functions in approximating the Gaussian process. In this section, we examine whether increasing the number of these functions can resolve the artifacts in FRK's outputs.

The number of basis functions used in FRK's approximation is primarily controlled by the \texttt{nres} and \texttt{regular} parameters. Increasing the \texttt{nres} parameter adds an additional ``resolution'' or layer of basis functions with a finer spatial scale, while increasing the value of \texttt{regular} reduces the scale of each basis function and adds additional rows and columns to their arrangement. Details on the effects of these parameters are available in the software documentation \citep{FRK_documentation}. The model used throughout Section~\ref{sec:resultsnational} - \ref{sec:computational_results} had these parameters set to \texttt{nres}$=2$ and \texttt{regular}$=1$.

\begin{figure}
    \centering

    \subfloat[][]{
        \includegraphics[width=0.4\textwidth]{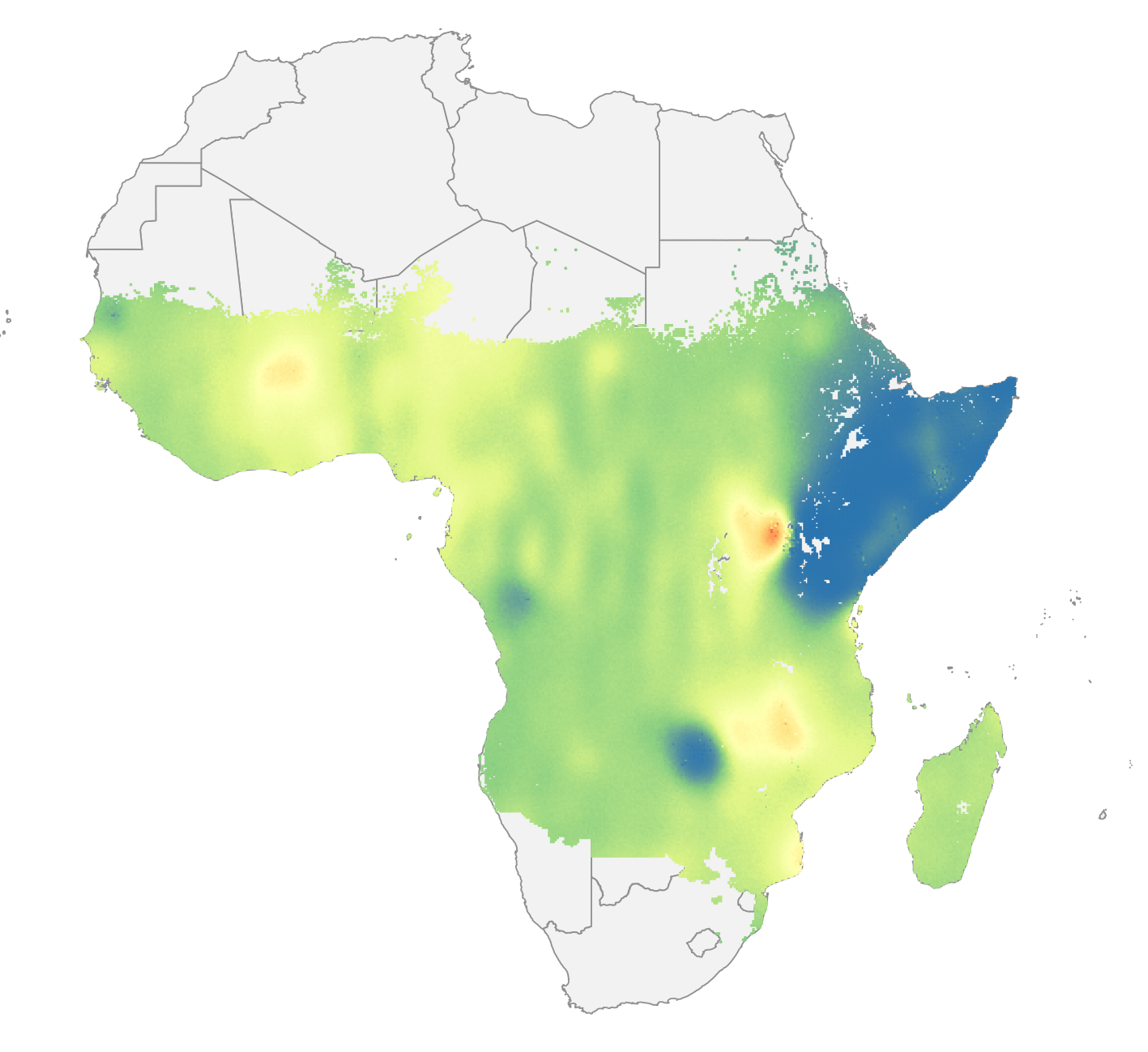}
        \label{fig:FRK_nres_3}
    }\hspace{30px}
     \subfloat[][]{
        \includegraphics[width=0.4\textwidth]{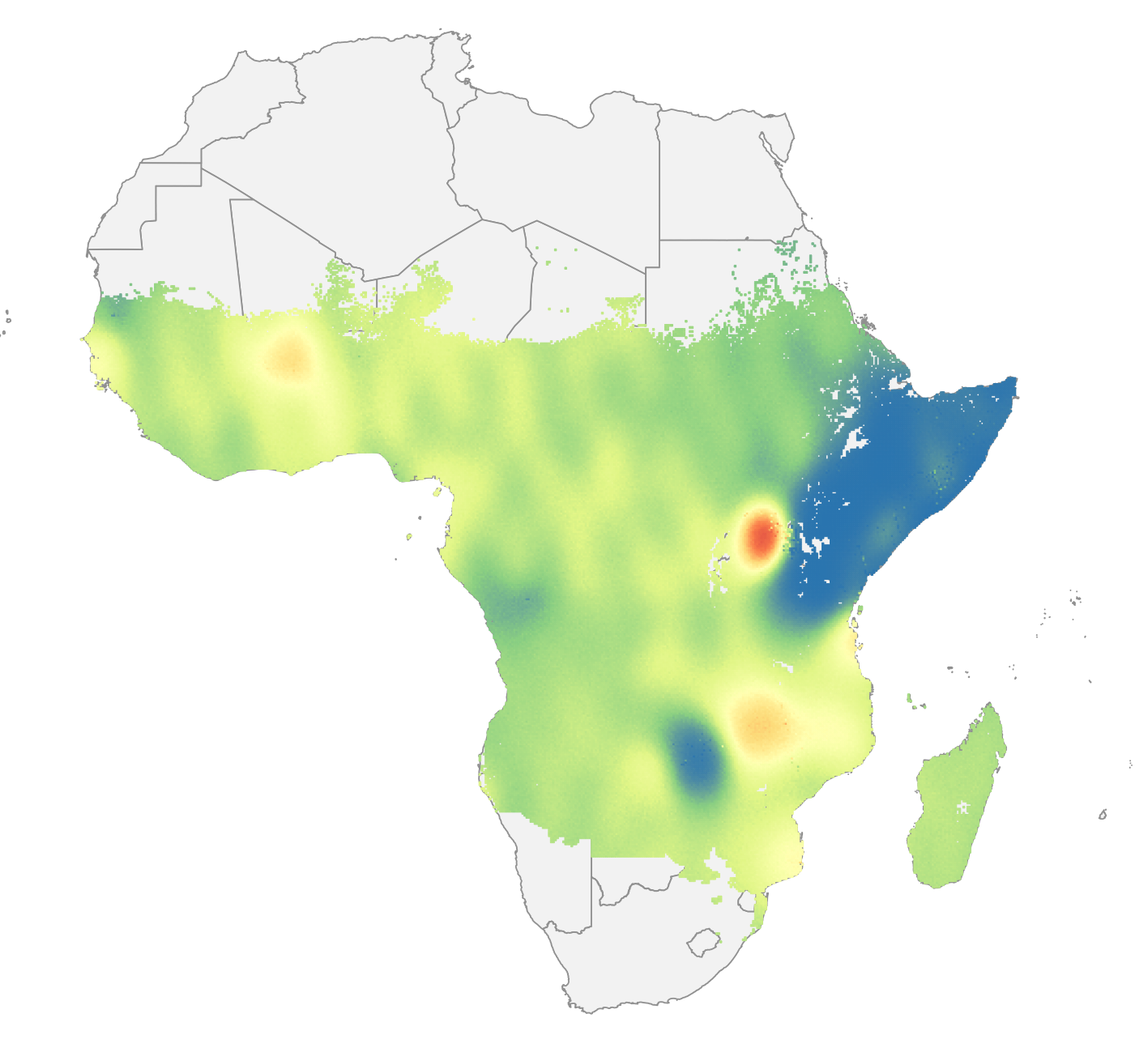}
        \label{fig:FRK_regular_2}
    } 
    \caption{\textit{P. falciparum} prevalence predictions from the FRK model with (a) \texttt{nres = 3} and \texttt{regular = 1}, and (b) \texttt{nres = 2} and \texttt{regular = 2}.}
    \label{fig:FRK_parameter_tuning}
    
\end{figure}

Figure~\subref*{fig:FRK_nres_3} shows FRK's predictions when fit to the observation data with \texttt{nres} increased to 3, and \texttt{regular} left at 1, which resulted in a model with 1338 basis functions. Whilst the broad-scale spottiness is less prominent in this model, finer-scale oscillation is quite visible in regions of Central Africa. This modest improvement came at a significant computational cost, as the model took over 55 minutes and required 106GB of RAM, compared to the 4.77GB of RAM and 3.28 minutes required when \texttt{nres} was set to 2, and \texttt{regular} was set to $1$. Figure \subref*{fig:FRK_regular_2} shows the predictions when \texttt{nres} is kept at 2 and \texttt{regular} is increased to 2. These settings resulted in 600 basis functions, and required 22.74GB of memory and 9 minutes to run, significantly less than when increasing the \texttt{nres} parameter. However the fine-scale oscillation is noticably more pronounced in areas with little or no data.

While the key to FRK's computational efficiency is its decomposition of the spatial random effect into a small number of basis functions, these results suggest that it is challenging in practice to balance this efficiency with the risk of artifacts appearing in the model output, especially on large scale mapping problems, with sparse data. 

\section{Discussion}\label{sec:discussion}
Our applied comparison of four geostatistical modelling methods found that two of them (SpRF and GPBoost) are not sufficiently scalable or accurate to be applicable to large-scale malaria prevalence modelling problems.  SpRF's spatial predictions displayed a prominent `banding' artifact, and at first glance SpRF models appeared to be overfitted (matching closely to training data, but making poor predictions to hold-out data). However, on closer inspection (Appendix \ref{sec:detailedCV} we see that SpRF's tighter uncertainty intervals in low density regions may result in this perception. Unlike the other methods, SpRF does not incorporate a covariance function, but  instead treats the columns of the distance matrix between coordinates as covariates for inclusion in the Random Forest. We note that a covariance function could in fact be applied to the distance matrix before inclusion in the model. This would not have the same interpretation as in the Gaussian-process based models we consider, but would enable SpRF to consider a distance-based decay in the unobserved spatial effects being modelled. However, the RandomForest inference machinery in SpRF would have no means to estimate the parameters of such a function (such as the rate of decay with distance), and it seems unlikely they could be reasonably specified in advance. Due to these issues of fit, and the fact that the computation time of SpRF scaled approximately linearly with the size of the data, this approach is unlikely to be useful for applied geostatistical modelling of malaria data.

GPBoost made reasonably good predictions to hold-out data, being the best-performing model at 50-fold spatially-blocked cross-validation in the national-scale comparison (implying a good ability to extrapolate over short distances) and the second-best, behind FRK, at 10-fold cross validation (ability to extrapolate over longer distances). However the computation time using the default GPBoost specification scaled very poorly with increasing data size. This is because by default GPBoost performs inference on the full (unapproximated) Gaussian process, with each step of the inference procedure requiring an $\mathcal{O}(n^3)$ inversion of the covariance matrix. Neither the maximum-likelihood inference of GP hyperparameters, and boosting inference on the intercept (and covariate effects if used) reduce this computational burden. However, even deploying the Vecchia approximation to the Gaussian process provided by GPBoost did not resolve these issues, as shown in Appendix~\ref{sec:Vecchia}. The Vecchia approximation resulted in faster, but linearly increasing computation times, but also resulted in severe artifacts in the model predictions. Increasing the complexity (number of neighbouring points to consider) in the approximation resolved these artifacts, but at the cost of a substantial increase in the required computation time and RAM usage. We were unable to determine a combination of these and the boosting parameters that would reduce computation time to a comparable level to INLA and FRK.  We also note that GPBoost is a relatively new technique, and future versions may include faster approximations.  

Both INLA and FRK offered substantially better scalability to increasing data size than SpRF and GPBoost, taking only minutes to fit to 10,000 datapoints.  Whilst it was computationally scalable, and is a widely established method and software for geostatistical modelling of malaria data, implementing INLA using the commonly suggested binomial distribution for prevalence data (e.g. as suggested in \citep{moraga2019geospatial, Moraga2021}) resulted in spurious predictions and poor ability to extrapolate in both the 10-fold and 50-fold cross-validation tests. We have demonstrated that this is due to the fact that the malaria prevalence data being modelled are overdispersed relative to the binomial sampling assumption and spatial-only model. That is, the assumption is violated that the infection status of each individual in a given sample is independent of the others, given the estimated prevalence estimate at that location. This should not be surprising from an epidemiological perspective, given that the infections in a given place do not arise independently - each infection is caused by another. This gives rise to local noise, either at the level of a pixel or group of pixels (that particular location may have some risk factor not accounted for by the smooth spatial model), or at the level of the observation (on the day of sampling, that population may have had a higher or lower than usual prevalence). INLA's behaviour in this case is an attempt to capture these small-scale variations with a very `wiggly' spatial random effect, i.e. one with rapid decay with increasing distance. It favours this parameter configuration on overdispersed data because the observation variance is fixed when using a binomial likelihood, and the variance is not sufficiently large to explain the data. This issue of poor identifiability between the observation-level variance and the lengthscale of a Gaussian processes has previously been described (\cite{Rasmussen2018}, see Figure 5.4), and can be resolved in classical (and model-based) geostatistics with the use of an independent `nugget' effect either on each observation or each observed location (\cite{Diggle2007}). Despite also using a binomial observation distribution, FRK does not suffer the same pitfall because it includes a type of spatial nugget effect in its `small-scale' effect parameter.

For malaria prevalence modelling with INLA, we suggest that a more reliable `default' model than the standard binomial observation model would be one which includes additional observation-level random noise. This can be achieved by using a Beta-binomial or Gaussian (on the observed prevalences or on the empirical-logit scale). Both of these options have an additional observation-level variance parameter that can be used to explain the overdispersion relative to the binomial. Of these, the Beta-binomial is most likely to be generally applicable to malaria prevalence data, since it is able to accurately account for observation errors in the common situation where only very few the individuals tested are infected. Though we note that fitting with a Gaussian response is substantially more computationally efficient in INLA, and so may be preferable if computation time is a major constraint. An alternative approach would be to include an independent observation-level random effect in the model specification.

Whilst FRK scaled well to large datasets (generally taking slightly less time than INLA) and performed well in both the 10-fold and 50-fold extrapolation comparisons, for continental-scale modelling, we were unable (with modest model modification) to specify the model in such a way that it was both computationally scalable and avoided the spurious oscillating effect of the basis functions. Whilst less noticeable, similar patterns are visible in the national-scale analysis in parts of North-Western and far North-Eastern Kenya where no data are available to inform such a prediction. Given these issues, we believe significant care must be taken when applying FRK to mapping of sparse malariometric data, to avoid these spurious predictions that are driven by computationally convenient approximations rather than data.

Comparing four methodologically different techniques has its limitations. One such limitation is that the inherent differences of the methods make a comparison somewhat difficult. For example, the likelihoods are different as well as the underlying model structure and/or covariance functions. Thus, each method has its own measures and we cannot compare an INLA goodness of fit measure with that of SpRF and vice versa.  We have mitigated this problem by focusing on the outputs -- predictive maps and cross validation results. 

Another aspect of interest is the parameter settings. There are many different parameter settings for each method. We have selected the commonly used (default) parameter settings in this work and even though we have explored several different parameter settings, we have not conducted a comprehensive exploration of the parameter space of these algorithms.  While the default parameter settings were acceptable for Kenya, we expect that algorithms can benefit from customised parameters when running the model on the scale of Africa. The limitations of the choice of parameters is brought to light by the extent of the geographical region. Exploring optimal parameter selection is another avenue of research.  Furthermore, there might be other parameter settings that can make the inference approximations of the different models more comparable.   

From a practitioner's point of view, it is challenging to adopt a new method for spatial modelling mostly because it takes a long time to learn the methodology and write code to produce meaningful output. This is a significant barrier to entry. If the methods discussed provide tuning functions that explore the parameter space and select a set of parameters that enables the practitioner to build a good model, it would increase the usability of these methods.    

An in-depth investigation of strengths and weaknesses of the models would be another avenue of interest. One option is to construct a meta-model that can predict the best model based on features of different locations \citep{Wang2009}. Such a meta-model could combine the strengths of the  diverse models to make a stronger prediction. The findings of this paper should be of use for those creating, interpreting or working with spatial data, as a baseline comparison of new computational geostatistical models.  %


\section{Acknowledgements}
We would like to thank Håvard Rue, Andrew Zammit-Mangion, Matthew Sainsbury-Dale, Fabio Sigrist and Noel Cressie for their correspondence and help with setting up and troubleshooting models. This research was supported by The University of Melbourne’s Research Computing Services and the Petascale Campus Initiative. J.A. Flegg’s research is supported by the Australian Research Council (DP200100747, FT210100034) and the National Health and Medical Research Council (APP2019093). 
\bibliography{references}
\bibliographystyle{agsm}

\appendix

\section{Detailed cross-validation results} \label{sec:detailedCV}

\subsection{Point predictions}

\begin{table}[!ht]
	\centering
	\caption{Cross validation results of the four models with best results in each category in boldface.}
	{
	\begin{tabular}{ccp{2cm}p{2cm}p{2cm}p{2cm}p{2cm}}
		\toprule
   \multirow{2}{*}{Fold} & \multirow{2}{*}{Model} & \multirow{2}{*}{RMSE}  & \multirow{2}{*}{Correlation} & \multicolumn{3}{c}{ \% points with absolute error less than } \\  
   \cmidrule{5-7} & & &  & $ 0.05$ & $ 0.1$  & $  0.2$  \\
        \midrule
\multirow{4}{*}{10-fold}  
    &   INLA    & 0.181    &  0.235    & 69.211    & 76.316    & 86.053 \\
    & GPBoost   & 0.127  &  0.646 & 52.632    & 73.947 & \textbf{93.158} \\
    & SpRF      & 0.132     & 0.641     & \textbf{69.474}       & \textbf{80.263}  & 91.053 \\
    & FRK       & \textbf{0.125}     &  \textbf{0.661}    & 57.105    & 76.579   &  92.105\\
        \midrule
\multirow{4}{*}{50-fold}  
    &   INLA    & 0.124     &  0.683 & \textbf{69.474}   & \textbf{80.789}   & 87.895 \\
& GPBoost & \textbf{0.11} & \textbf{0.751} & 65.789   & \textbf{80.789}   &  90.0\\
    & SpRF      & 0.121     &  0.702    & 67.632        & 78.947            & \textbf{90.526} \\
    & FRK       & 0.123     & 0.702     & 66.053        & 79.211           & 90.263\\        
		 \bottomrule
	\end{tabular} }
	\label{tab:predactuals}
\end{table}

\begin{figure}[!ht]
    \centering
    \includegraphics[width=0.6\textwidth]{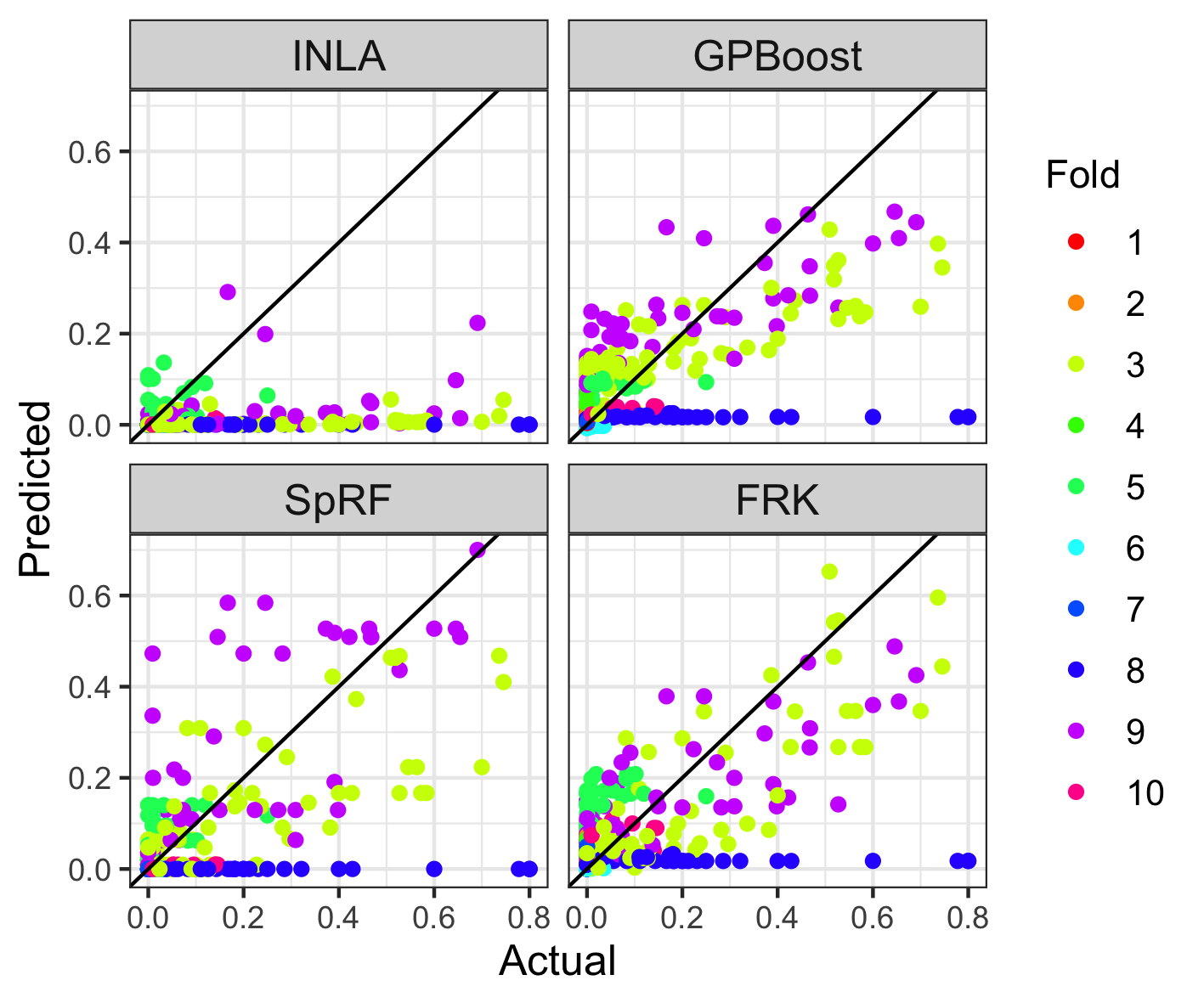}
    \caption{Model predictions of the four models vs actual prevalences using 10-fold CV.}
    \label{fig:actpreds10fold}
\end{figure}

Table~\ref{tab:predactuals} gives the results for 10 and 50-fold cross validation. For each set of folds it gives the Root Mean Square Error (RMSE), correlation coefficient between the predicted and actual values and the percentage of observations that have an absolute error $(\vert \text{predicted} - \text{actual}\vert)$ less than a specified threshold (thresholds of 0.05, 0.1 and 0.2 considered).  As noted earlier, FRK and GPBoost have the best RMSE and correlation values for 10-fold and 50-fold cross validation respectively. SpRF gives the best performance in terms of the percentage of observations with absolute error less than 0.05 and 0.1. Compared to the other models, INLA performs poorly for the 10-fold cross validation, with a higher RMSE and significantly lower correlation coefficient. However, it gets a high percentage of observations with absolute error less than the three thresholds. This is because a large number of observations have low prevalence values. This is further illustrated in Figure~\ref{fig:actpreds10fold}, which shows the actual and predicted values using 10-fold cross validation for each model.  

Figure~\ref{fig:actpreds10fold} shows the points by cross validation fold as determined in Figure~\subref*{fig:10foldspos}. As the folds are determined by $k$-means clustering, observations in each fold lie close together.  We see that data points in most folds have similar prevalence values. However, data points in Folds 3, 8 and 9 have a broad range of values. 
The points assigned to Fold 8 are difficult to predict for all four models. These points are along the coast near the city of Mombasa and are somewhat isolated from other clusters, which might be a contributing reason.  

\begin{figure}[!ht]
    \centering
    \includegraphics[width = 0.5\textwidth]{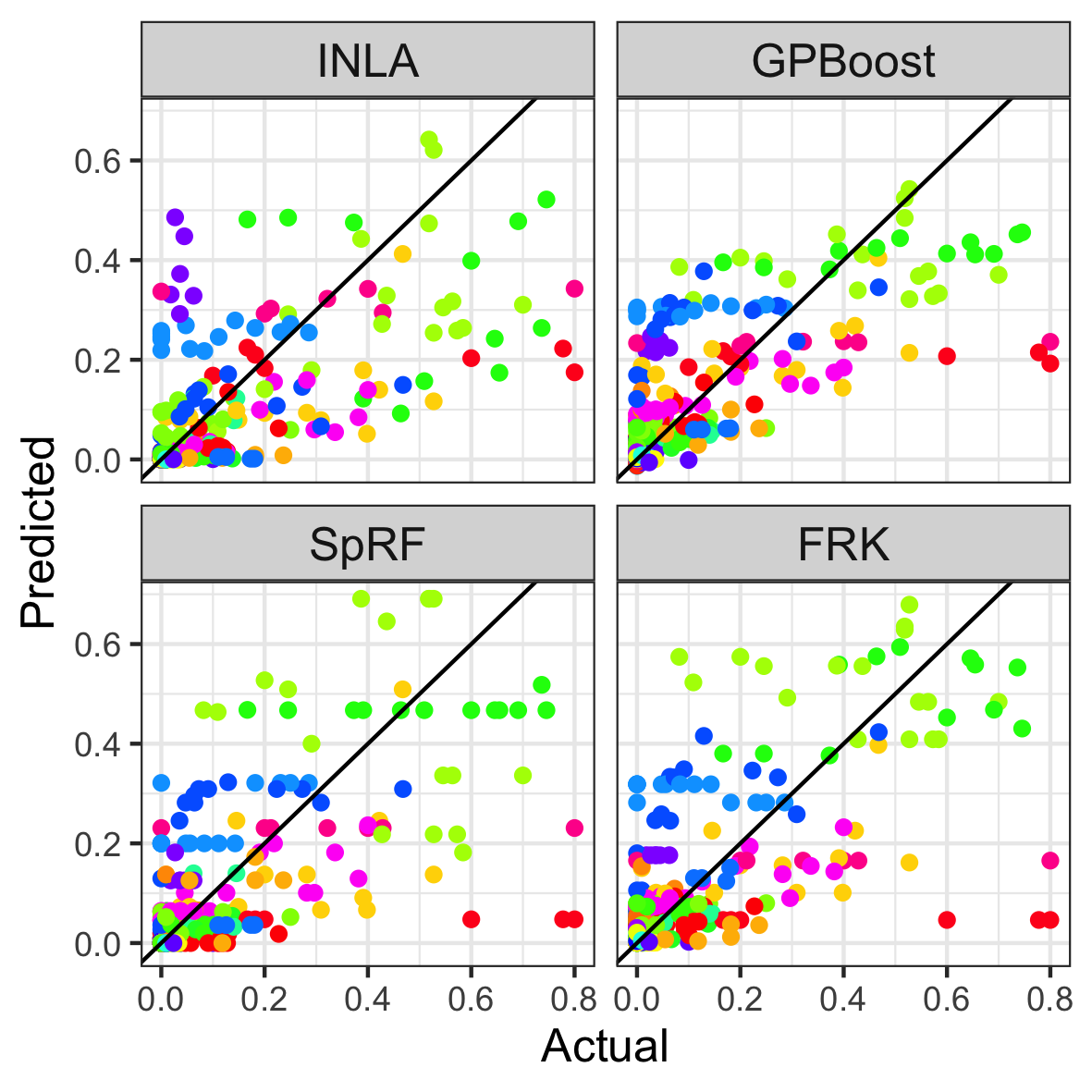}
    \caption{Model predictions of the four models vs actual prevalences using 50-fold CV with folds in different colours.}
    \label{fig:actpreds50fold}
\end{figure}

\begin{figure}[!ht]
    \centering
    \captionsetup{width = 0.3\textwidth}
     \subfloat[][INLA]{
        \includegraphics[width=0.3\textwidth]{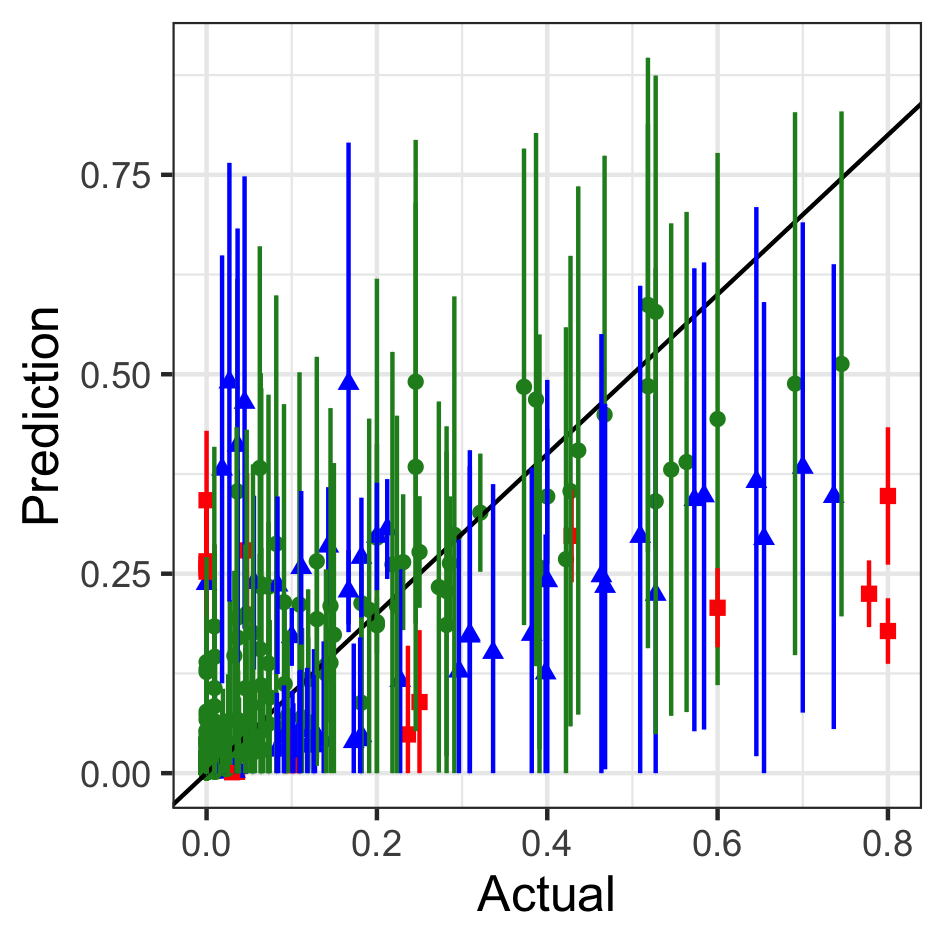}
        \label{fig:INLA50}
    } 
    \subfloat[][GPBoost]{
        \includegraphics[width=0.3\textwidth]{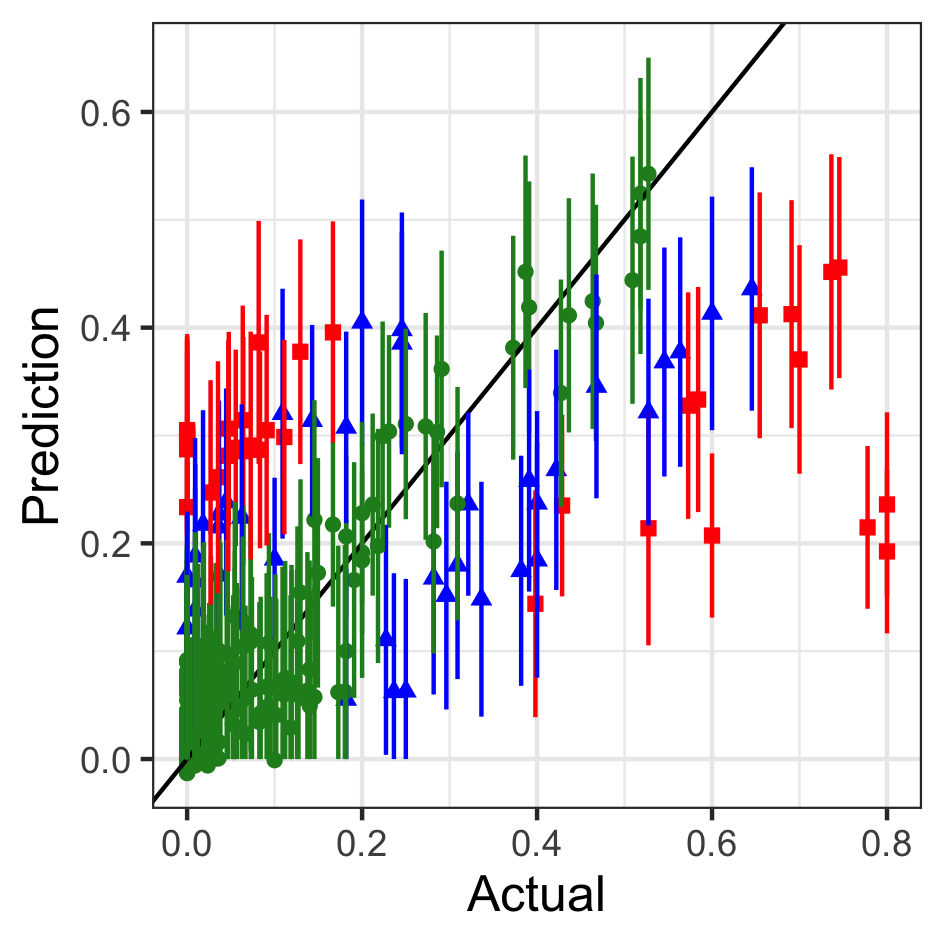}
        \label{fig:GPBoost50}
    } \\
    
    \subfloat[][SpRF]{
        \includegraphics[width=0.3\textwidth]{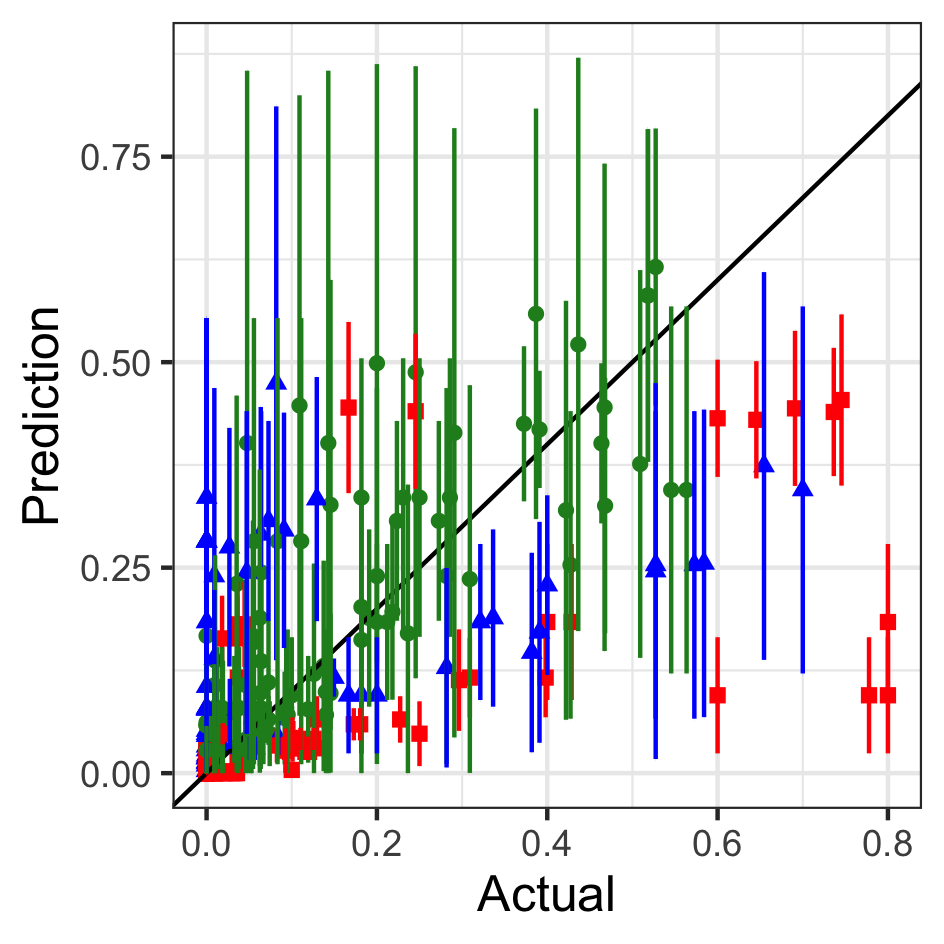}
        \label{fig:SpRF50}
    } 
    \subfloat[][FRK]{
        \includegraphics[width=0.3\textwidth]{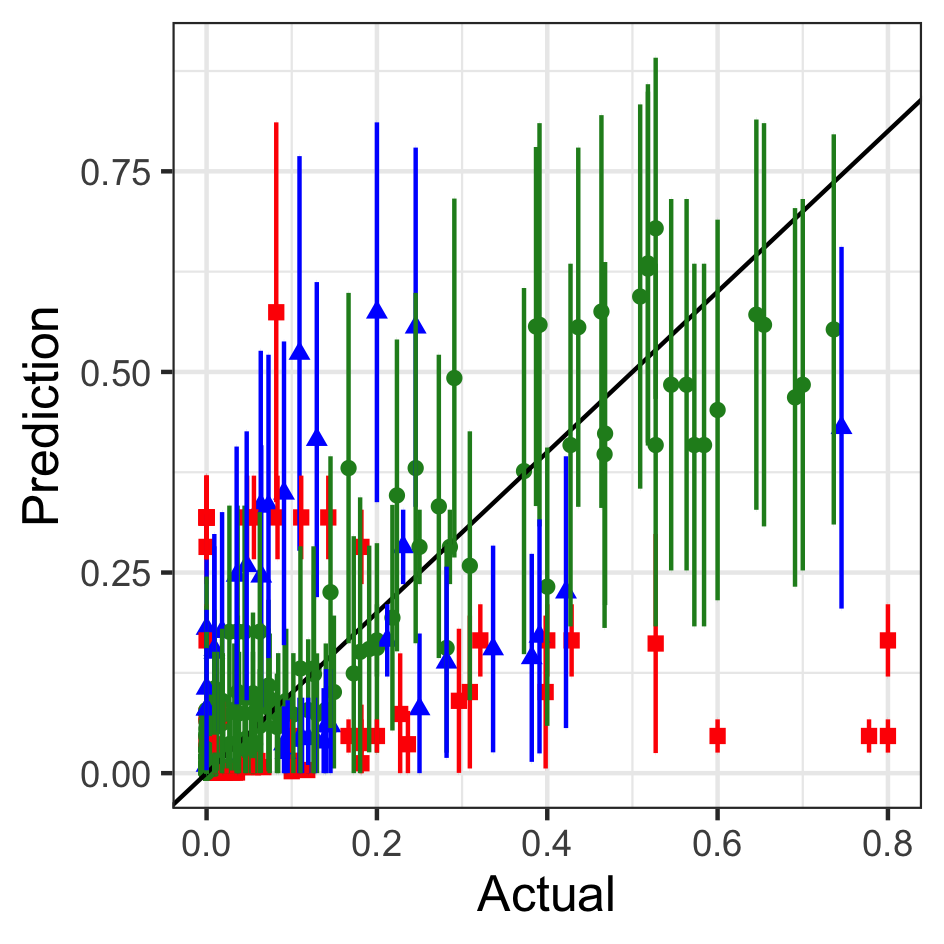}
        \label{fig:FRK50}
    } 
    \captionsetup{width = 0.8\textwidth}
    \caption{Interval predictions for 50-fold cross validation. }
    \label{fig:interval50fold}
\end{figure}

Figure~\ref{fig:actpreds50fold} shows 50-fold cross validation results for the four models while Figure~\ref{fig:interval50fold} shows their interval predictions. 
From Table~\ref{tab:predactuals} we see that GPBoost achieves better results in terms of RMSE and correlation. It has the same performance as INLA for the highest percentage of observations with absolute error less than 0.1.  INLA has the highest percentage of observations with absolute error less than 0.05 and SpRF has the highest percentage of observations with absolute error less than 0.2. From Figure~\ref{fig:actpreds50fold} we see that certain folds perform poorly. These folds match with the locations of the poorly performing folds in the 10-fold CV scenario.  Another interesting observation is that while GPBoost achieves good results for both sets of folds, it performs poorly on high prevalence observations, whereas FRK and SpRF do not appear to have this limitation. 

\subsection{Point predictions by location density}

\begin{figure}[!ht]
    \centering
    \captionsetup{width=.8\linewidth}
    \subfloat[][Malaria prevalence]{
        \includegraphics[width=0.3\textwidth]{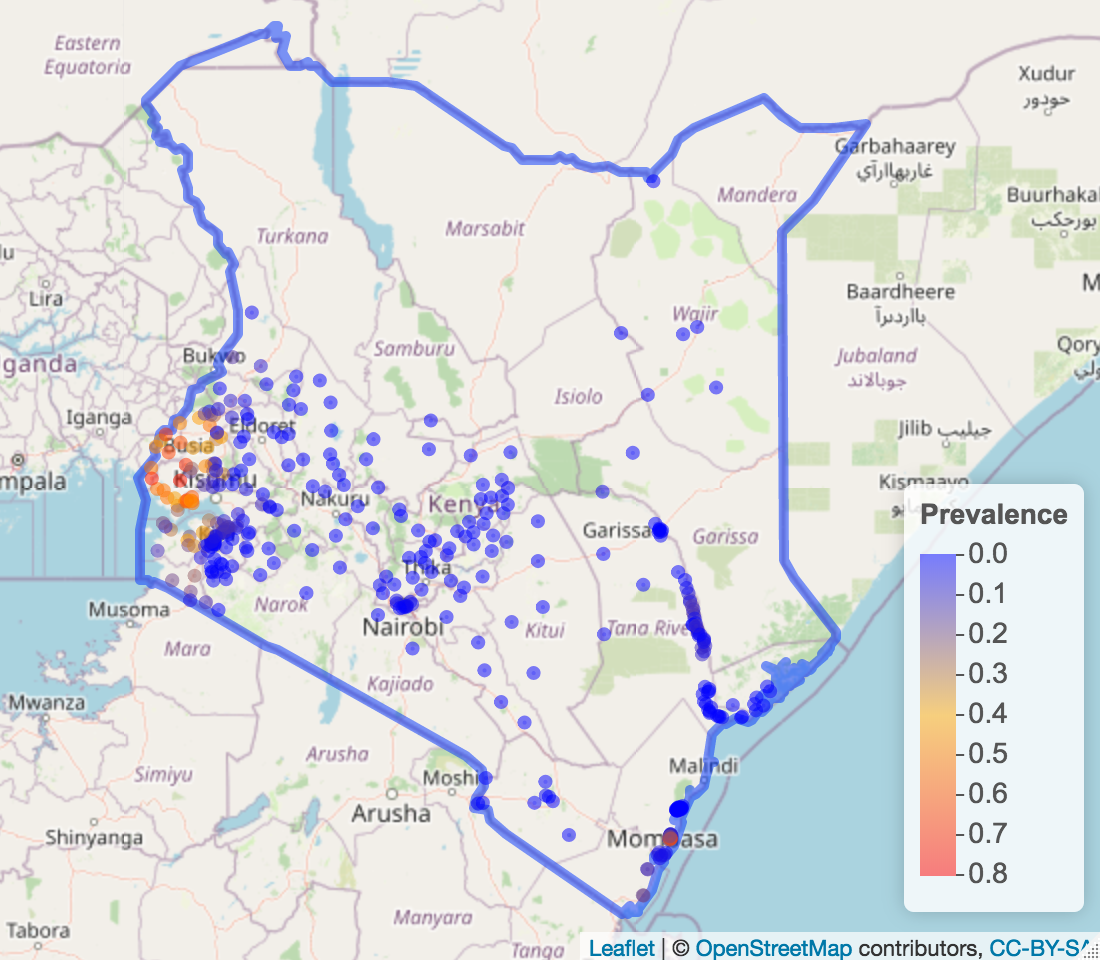}
        \label{fig:prev}
    }
    \hspace{30px}
     \subfloat[][Density of sampled locations]{
        \includegraphics[width=0.3\textwidth]{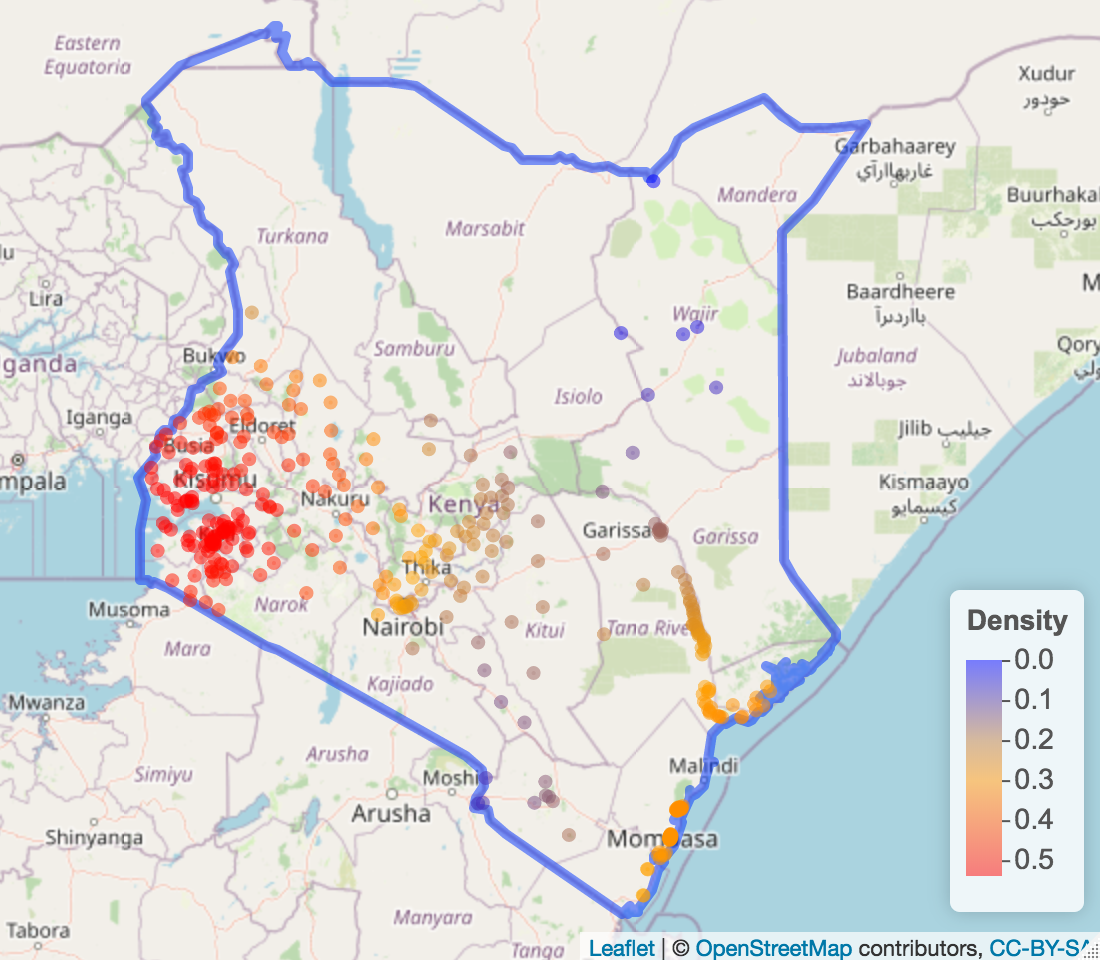}
        \label{fig:density}
    }  
    \caption{\textit{P. falciparum} prevalence in Kenya for 2009 and the kernel density estimates of the sampled locations.  }
    \label{fig:actualanddensityinmap}
\end{figure}

\begin{figure}[!ht]
    \centering
    \subfloat[][10 folds]{
        \includegraphics[height=5.7cm]{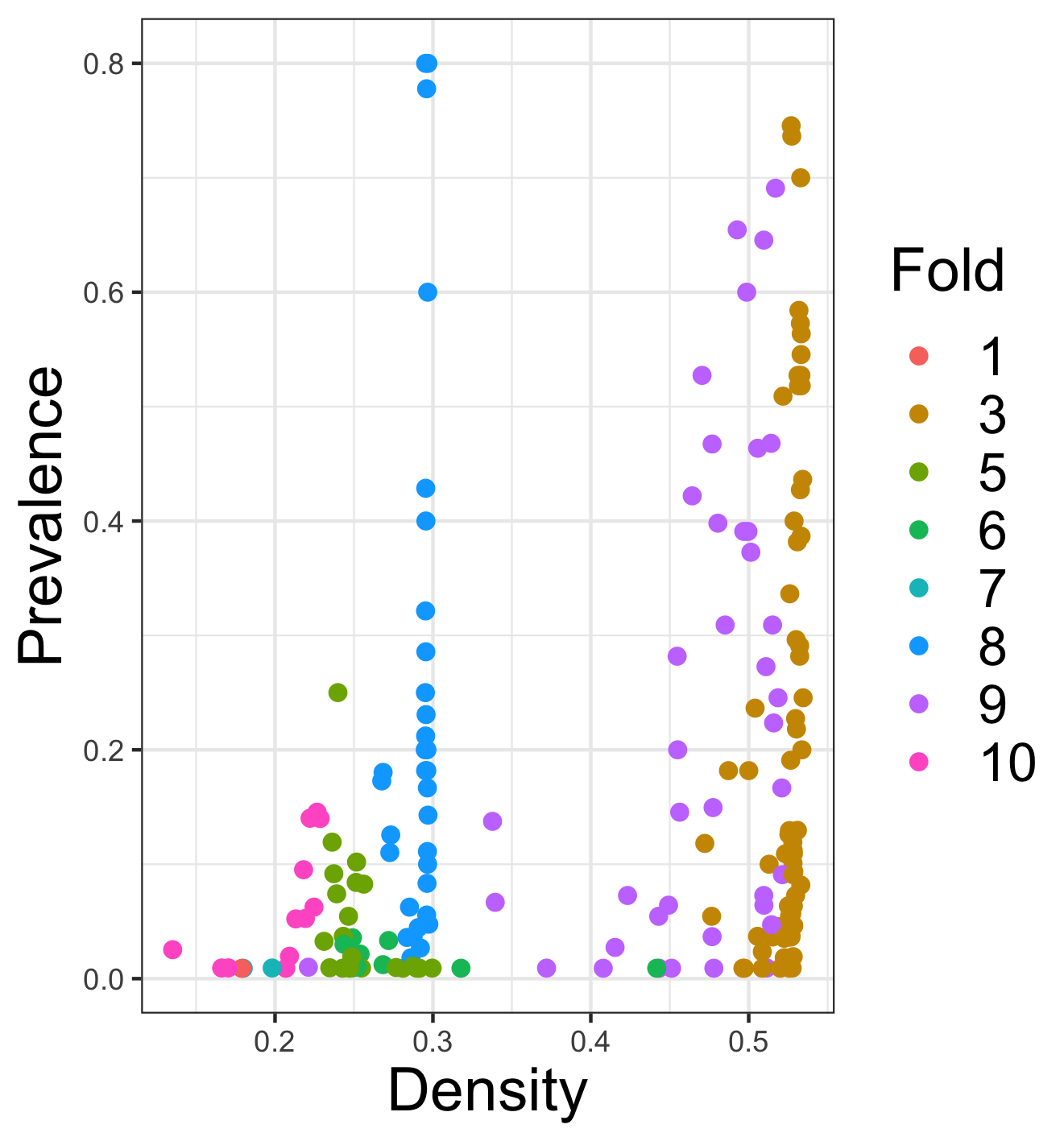}
        \label{fig:actualanddensity10folds}
    }
     \subfloat[][50 folds]{
        \includegraphics[height = 5.7cm]{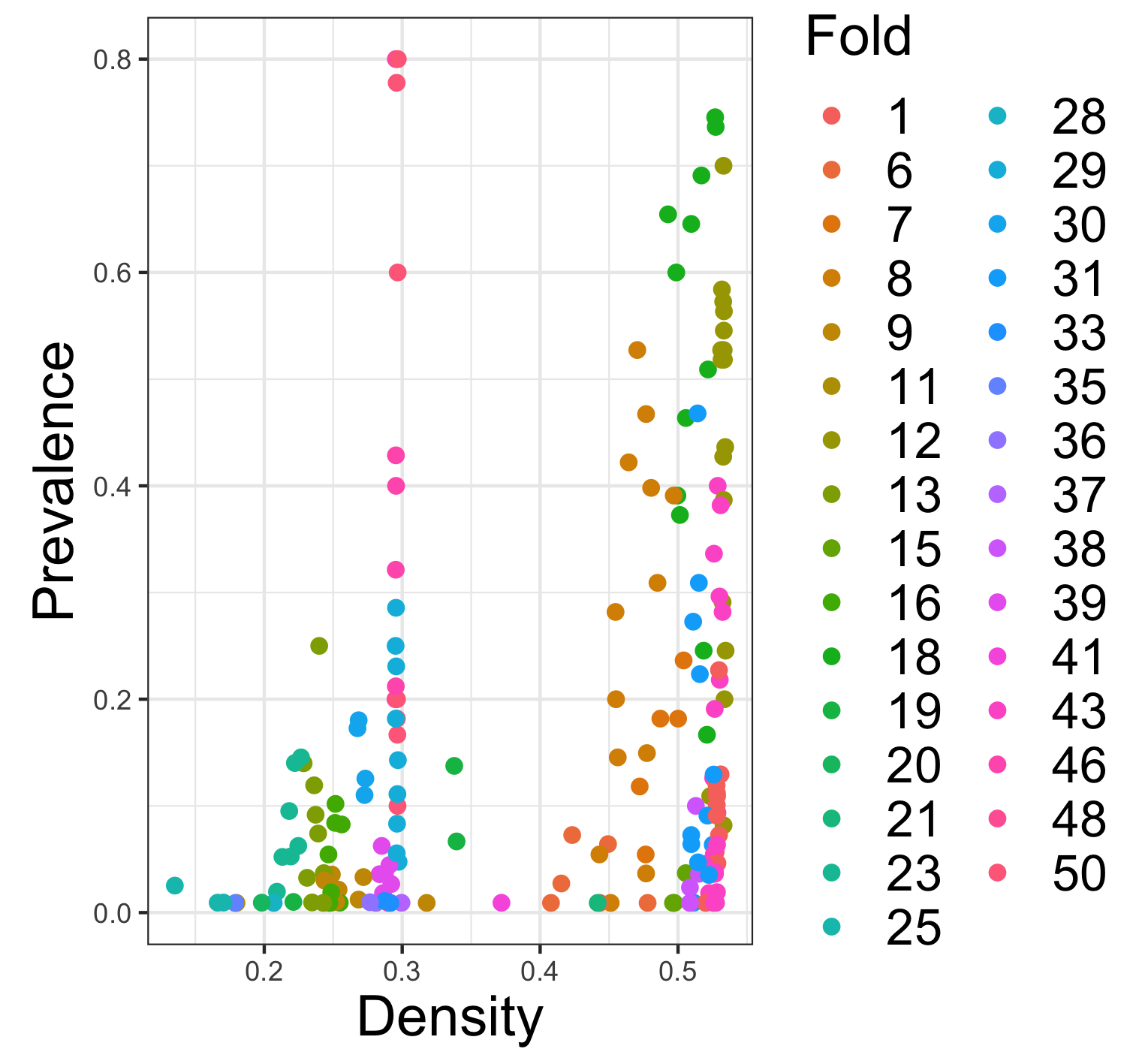}
    }  
    \caption{\textit{P. falciparum} prevalence and kernel density estimates of different clusters (folds) with zero prevalence observations taken out.  }
    \label{fig:actualanddensity}
\end{figure}

We further analyse these results using the density of sampled locations, i.e. do some models find it difficult to predict observations in low density regions? Figure~\ref{fig:actualanddensityinmap} shows the malaria prevalence and kernel density estimates of the sampled locations on two separate maps. Figure~\ref{fig:actualanddensity} shows scatter plots of prevalence and density with points coloured by the fold. For 10-fold CV, we see that fold 8, which is around the city of Mombasa has a broad range of prevalence values while having relatively low density. This  explains the reason behind the high errors for Fold 8 (Figure~\ref{fig:actpreds10fold}). When the  sampled points are away from each other (low density) and the prevalence values have high variation, it is challenging for the models to predict accurately.

\begin{figure}[!ht]
    \centering
     \subfloat[][Density and absolute errors of the four models using 10-fold CV]{
        \includegraphics[height = 8cm]{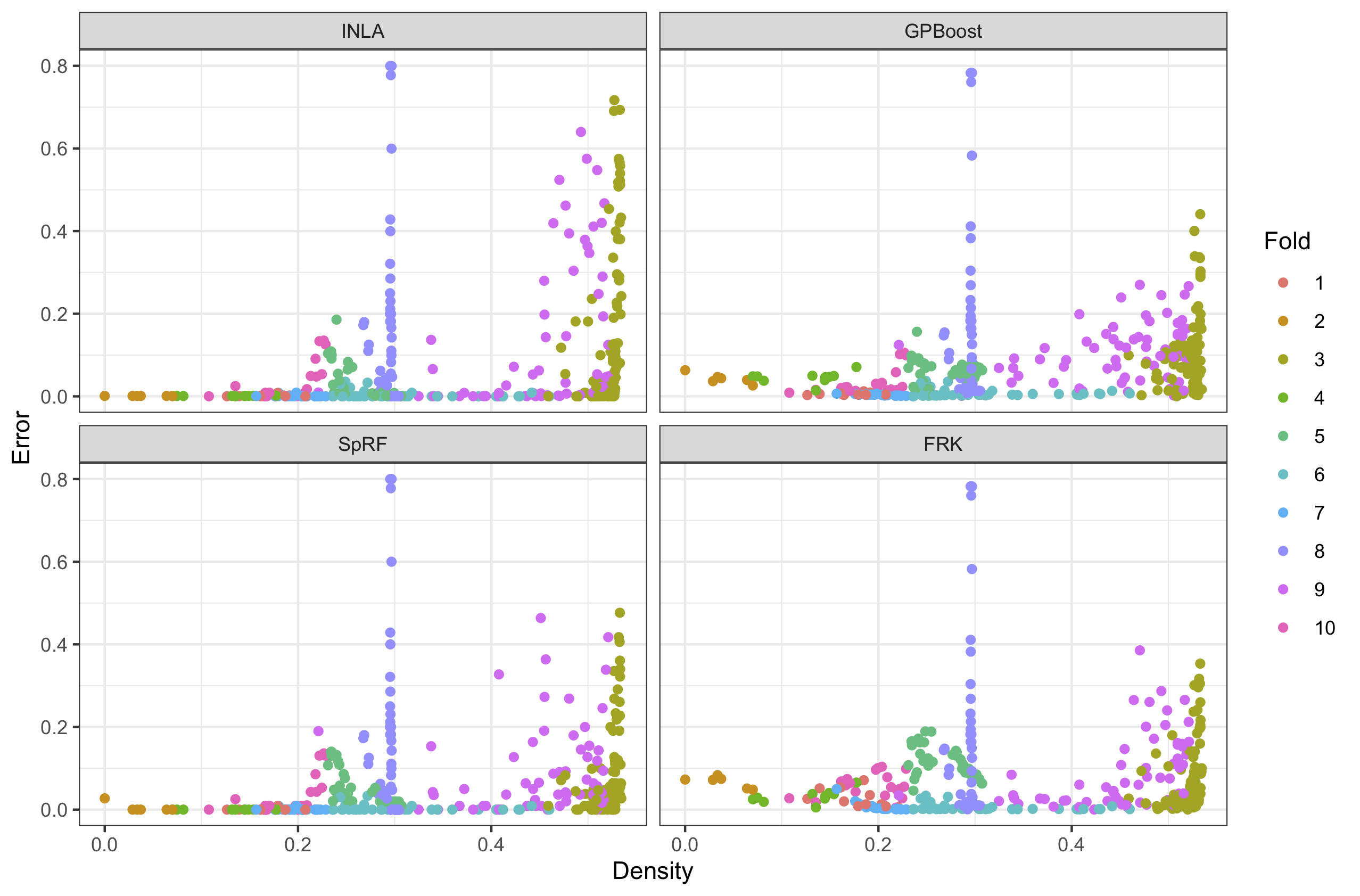}
        \label{fig:errordensity10folds}
    } \\
    \subfloat[][Density and absolute errors of the four models using 50-fold CV]{
        \includegraphics[height = 8cm]{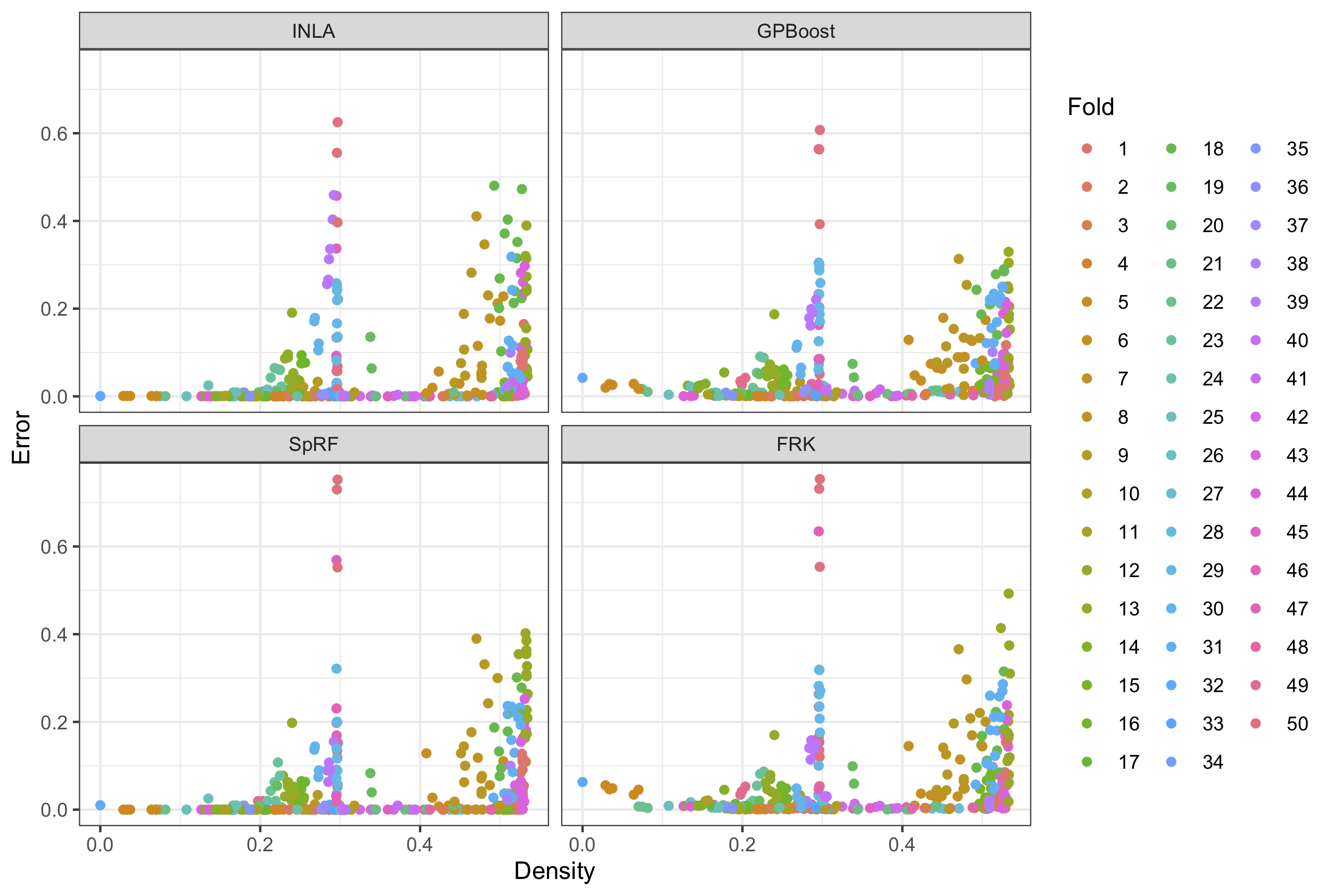}
        \label{fig:errordensity50folds}
    } 
    \caption{Kernel density estimates of the locations and the absolute errors of the four models for 10 and 50-fold cross validation.   }
    \label{fig:errordensity}
\end{figure}

Figure~\ref{fig:errordensity} shows the absolute errors of the four models with respect to density for both 10 and 50-fold cross validation. For 10-fold CV we see that fold 8 exhibits high error rates for all four models. If we consider the same set of points for 50-fold CV, we see that while FRK and SpRF have similar error rates (maximum $\approx 0.75$), INLA and GPBoost have comparatively lower error rates (maximum $\approx 0.6$). Thus, a higher number of folds benefits GPBoost and INLA in this instance more than it benefits FRK or SpRF.  

Moving on to the points with very low density ($< 0.2$) we see from Figure~\ref{fig:actualanddensity} that the prevalence values of these points are relatively low. In Figure~\subref*{fig:errordensity10folds} we see that INLA and SpRF have lower errors for these low density points compared to FRK and GPBoost. A similar outcome can be observed for the 50-fold CV case in Figure~\subref*{fig:errordensity50folds} when the density of points are less than 0.2.   

As seen in Figure~\ref{fig:actualanddensity}, points in high density regions ($> 0.4$) have a higher variation in prevalence ranging from 0 to 0.75. The 10-fold CV results in Figure~\subref*{fig:errordensity10folds} show that FRK performs best for these high density points with an error $ < 0.4$, followed by GPBoost and SpRF. INLA performs poorly on these points for 10-fold CV.  For 50-fold CV (Figure~\subref*{fig:errordensity50folds}) GPBoost performs best on the high density points followed by SpRF, while both FRK and INLA perform similarly. 

Table~\ref{tab:predactualsdiffdensities} gives metrics for different density groups for both 10 and 50-fold CV. We define low density as $\text{density} \leq 0.2$, medium density as  $0.2 < \text{density} \leq 0.4$ and high density as $\text{density} > 0.4$. We see that the absolute errors of all four models are small for low density points.  The percentage of observations with absolute error less than 0.05, 0.1 and 0.2 are quite high for both 10 and 50-fold cross validation sets. In both CV sets, INLA has the lowest RMSE with SpRF following closely. For both 10 and 50-fold CV the correlation coefficients between the actual and the predicted values are negative. This indicates that most prevalence values are close to zero in low density locations. This would also explain why less points were sampled from those regions, as more points are generally sampled from high prevalence regions. 

\begin{table}[!ht]
	\centering
	\caption{Cross validation results grouped by density of sampled locations with best results in boldface.}
	{
	\begin{tabular}{cccp{1.3cm}p{1.5cm}p{1.5cm}p{1.5cm}p{1.5cm}}
		\toprule
   \multirow{2}{*}{Fold} & \multirow{2}{*}{Density} & \multirow{2}{*}{Model} & \multirow{2}{*}{RMSE}  & \multirow{2}{*}{Corr.} & \multicolumn{3}{c}{ \% points with absolute error less than } \\  
   \cmidrule{6-8} & & & &  & $ 0.05$ & $ 0.1$  & $  0.2$  \\
        \midrule
\multirow{12}{*}{10-fold} & \multirow{4}{*}{Low}
    &   INLA    & \textbf{0.005}    & -0.234     & 100       & 100       & 100          \\
   &  & GPBoost & 0.028    & -0.189      & 96        & 100       & 100          \\
   &  & SpRF    & 0.006    & -0.059     & 100       & 100       & 100          \\
   &  & FRK     & 0.048   & -0.021     & 62        & 98      & 100          \\
   \cmidrule{2-8}
   & \multirow{4}{*}{Medium}
    &   INLA    & 0.141     & -0.003        & 75.723        & 82.659    & 93.642        \\
   &  & GPBoost & \textbf{0.138}     & -0.021       & 60.694    & 85.549   & 94.22         \\
   &  & SpRF    & 0.143     & -0.025       & 72.832        & 80.347   & 93.642        \\
   &  & FRK     & 0.145    & -0.035        & 58.382       & 76.301    & 94.22          \\
    \cmidrule{2-8}
   & \multirow{4}{*}{High}
    &   INLA    & 0.24     & 0.305        & 52.229   & 61.783    & 73.248       \\
   &  & GPBoost & 0.134     & 0.788         & 29.936    & 52.866    & 89.809         \\
   &  & SpRF    & 0.14     & 0.737         & 56.051    & 73.885    & 85.35           \\
   &  & FRK     & \textbf{0.119}     & 0.828         & 54.14    & 70.064   & 87.261         \\
        \midrule 
        \midrule
        
\multirow{12}{*}{50-fold}  & \multirow{4}{*}{Low}
    &   INLA    & \textbf{0.005}     & -0.128      & 100       & 100       & 100     \\
  &  & GPBoost  & 0.017     & -0.202        & 98       & 100       & 100 \\
  &  & SpRF     & 0.006      & -0.077        & 100       & 100       & 100  \\
  &  & FRK      & 0.021     & -0.17        & 96        & 100       & 100 \\     
  \cmidrule{2-8}
  & \multirow{4}{*}{Medium}
     &   INLA   &  0.123    & 0.501         & 73.41   & 83.815    & 90.173       \\
  &  & GPBoost  &  \textbf{0.117}    & 0.511         & 73.41    & 84.393    & 91.908 \\
  &  & SpRF     &  0.121    &  0.435        & 71.676        & 85.549   & 96.532 \\
  &  & FRK      &  0.133    & 0.31         & 75.723    & 83.237    & 91.329\\
  \cmidrule{2-8}
   & \multirow{4}{*}{High}
     &   INLA   &   0.143    & 0.776         & 55.414    & 71.338   & 81.529\\
  &  & GPBoost  &   \textbf{0.119}   & 0.809         & 47.134    & 70.701    & 84.713 \\
  &  & SpRF     &   0.14   & 0.742         & 52.866   & 64.968    & 80.892 \\
  &  & FRK      &   0.129   & 0.784         & 45.86     & 68.153    & 85.987 \\  
		 \bottomrule
	\end{tabular} }
	\label{tab:predactualsdiffdensities}
\end{table}

GPBoost, SpRF and FRK have higher RMSE for the medium density point set,  compared to the high density point set for 10-fold CV. A similar behaviour is observed for FRK for 50-fold CV. This is due to the high absolute errors in fold 8 as discussed previously. For the medium density points, GPBoost is preferred in terms of RMSE for both 10-fold CV and 50-fold CV. For high density points FRK is preferred for 10-fold CV while GPBoost is preferred for 50-fold. In terms of the percentage of points with absolute error less than 0.05 and 0.1, SpRF leads the other models for 10-fold CV, while INLA leads for 50-fold. For both 10 and 50-fold CV, INLA and SpRF perform better on low density points compared to the other two methods, while GPBoost and FRK perform better on high density points. 

\subsection{Interval predictions}
As described in Section~\ref{sec:results}, each method has a different uncertainty quantification mechanism, however we have estimated the standard deviation of predictions from each model to allow comparison. 



Table~\ref{tab:intervalpredis} gives the interval prediction results for all four models. Figures~\ref{fig:interval10fold1} and~\ref{fig:interval50fold} show the interval predictions for 10 and 50-fold CV. 
For both 10 and 50-fold CV, SpRF has on average the smallest uncertainty intervals and the smallest number of points within one or two standard deviations of the mean. Surprisingly SpRF’s interval widths are zero for 148 and 164 of the predictions for 10 and 50-fold CV respectively. Each of these points correspond to a prediction (median) of zero prevalence, and the majority correspond to an observed prevalence of zero. However, the mean value of the response at nearly all of these locations is small but non-zero, and so the prevalence at these points does not lie within any number of standard deviations of the mean, contributing to the low percentages for SpRF in Table~\ref{tab:intervalpredis}.

\begin{table}[!ht]
	\centering
	\caption{Interval prediction results of the four models. Mean Width refers to the average of the predicted standard deviations, while Std. Dev. Width refers to their standard deviation.}
	{
	\begin{tabular}{ccp{2.5cm}p{2.5cm}p{1.5cm}p{1.5cm}}
		\toprule
   \multirow{2}{*}{Fold} & \multirow{2}{*}{Model} &  \multirow{2}*{\begin{tabular}{l} Mean \\ Width \end{tabular}} &  \multirow{2}*{\begin{tabular}{l} Std. Dev. \\  Width \end{tabular}}  & \multicolumn{2}{c}{ Points within (\%)} \\  
   \cmidrule{5-6} &  & & &  1SD &  2SD   \\
        \midrule
\multirow{4}{*}{10-fold}  
    &   INLA        & 0.102     & 0.048  & 75    & 87.105  \\
    & GPBoost       & 0.136     & 0.014  & 84.211   & 95.526   \\
     & SpRF          & 0.071     & 0.106  & 37.105   & 55  \\
    & FRK           & 0.1     & 0.072  & 83.421   & 92.368  \\
        \midrule
\multirow{4}{*}{50-fold}  
    &   INLA        & 0.096     & 0.091 & 81.053    & 95.789 \\
    & GPBoost           & 0.112     & 0.014 & 81.053    & 91.053 \\
     & SpRF          & 0.056     & 0.086 & 28.158    & 42.895 \\
    & FRK                 & 0.062     & 0.068 & 74.474    & 85 \\        
		 \bottomrule
	\end{tabular} }
	\label{tab:intervalpredis}
\end{table}
For both 10 and 50-fold CV, GPBoost has the largest mean interval widths but smallest standard deviation, suggesting that it predicts consistently high width intervals for most observations. For 10-fold CV, this results in the highest percentage of points lying within one or two standard deviations. For 50-fold CV INLA has a higher percentage of points lying within each type of interval, which may be accounted for by the higher variation in INLA’s interval widths.

\begin{figure}[!ht]
    \centering
     \subfloat[][Density and interval widths of the four models using 10-fold CV]{
        \includegraphics[height = 8cm]{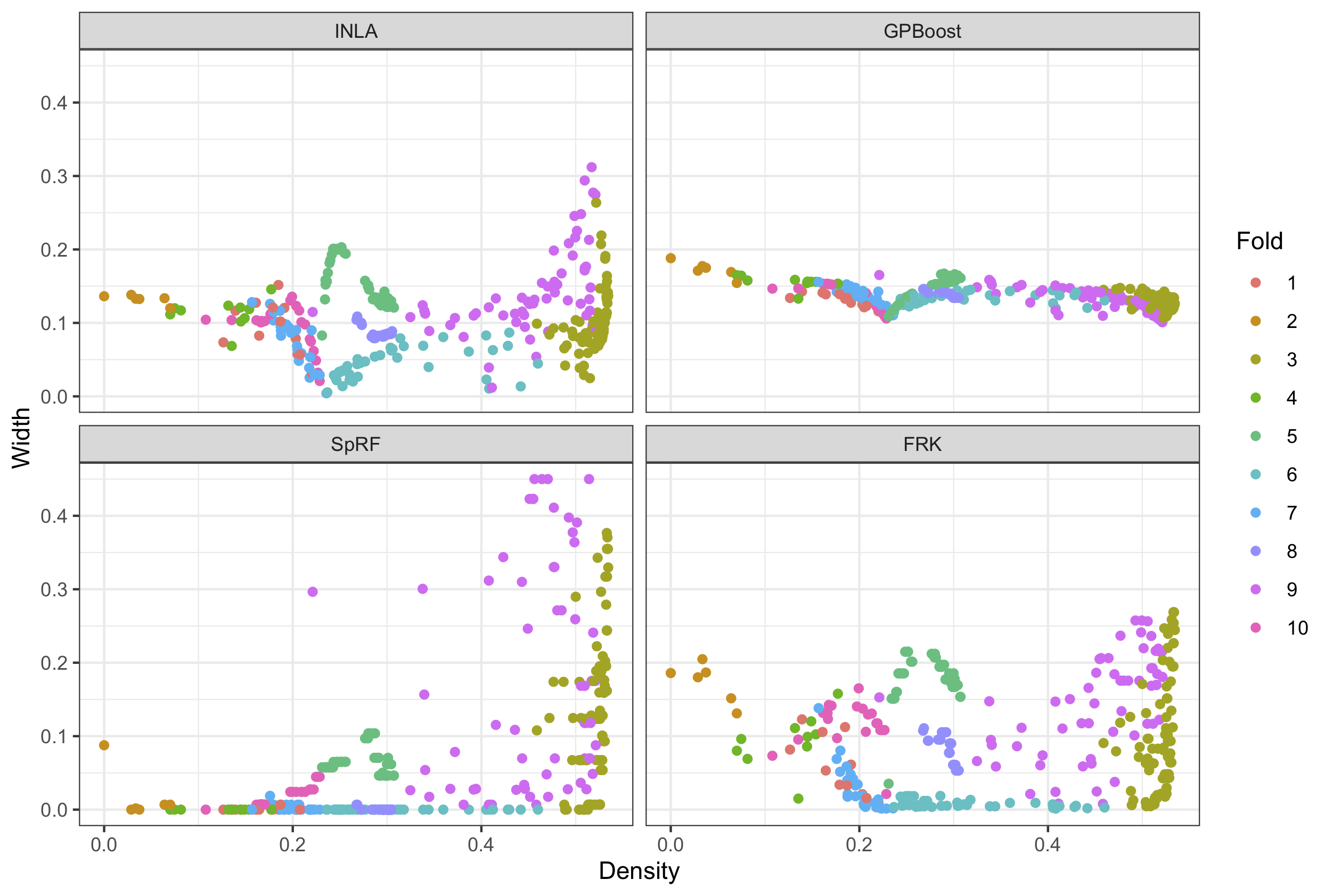}
        \label{fig:widthdensity10folds}
    } \\
    \subfloat[][Density and interval widths of the four models using 50-fold CV]{
        \includegraphics[height = 8cm]{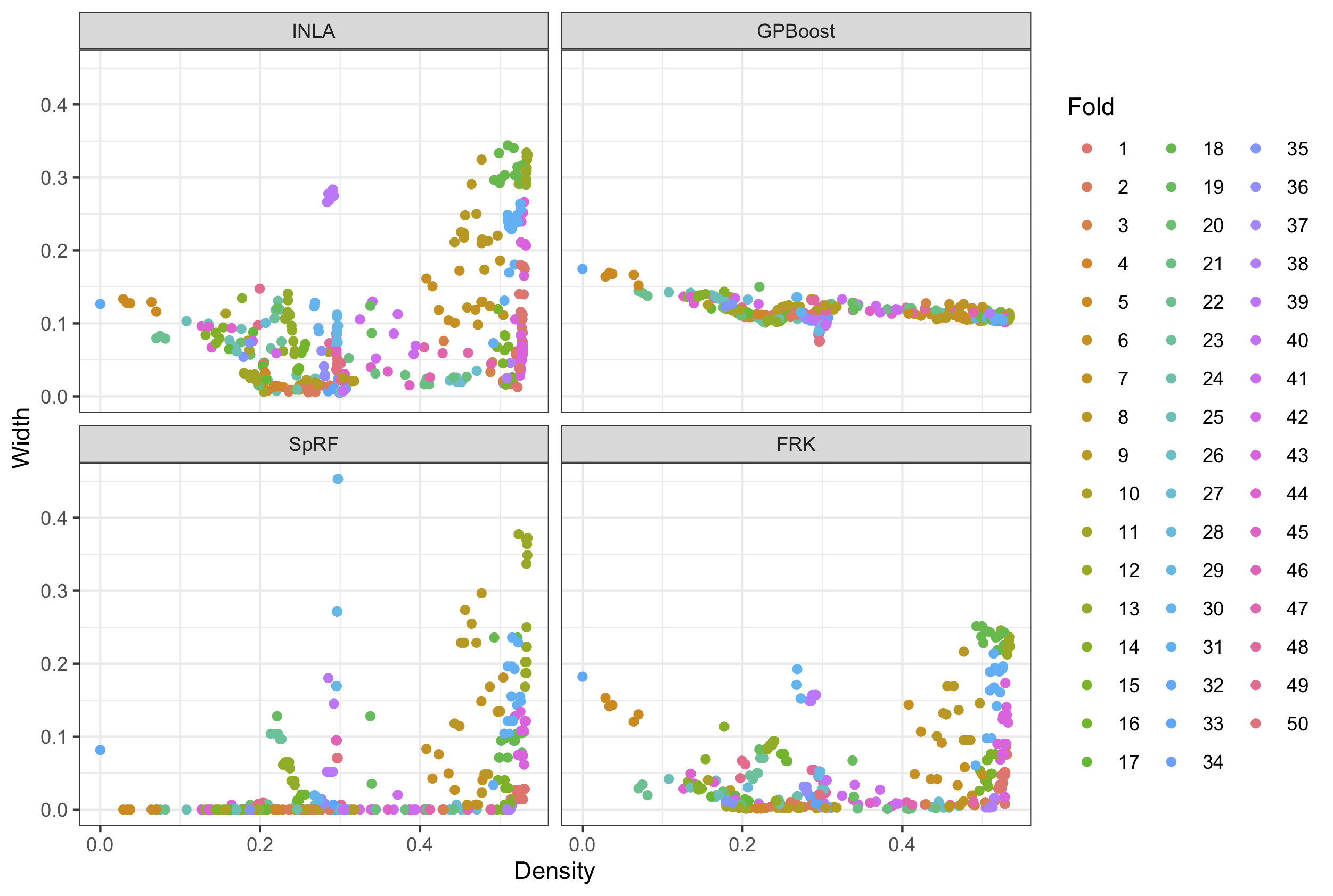}
        \label{fig:widthdensity50folds}
    } 
    \caption{Kernel density estimates of the locations and the interval widths of the four models for 10 and 50-fold cross validation.}
    \label{fig:widthdensity}
\end{figure}

Figure~\ref{fig:widthdensity} shows the kernel density estimates of the locations and the respective interval widths of the four models for both 10 and 50-fold CV. For both 10 and 50-fold CV we see that GPBoost has similar widths for all observations. We observe a slight increase in width for low density points. However, there is not much variation in width with respect to the density.   In contrast, FRK, INLA and SpRF have varying interval widths for different folds. We see that there is high variation for locations with high density, mostly likely because of the variation in prevalence. For both 10 and 50-fold CV, FRK has relatively high width values for low density points. Conversely, INLA and SpRF have low width values for low density points. Similar to the point predictions we see a high variation of width for medium density points ($\text{density} \approx 0.3 $) for FRK, INLA and SpRF.

\section{Effects of input noise on INLA}\label{sec:input_noise}

Figure~\ref{fig:Africa_prevalence_maps}(ai) shows INLA predicting a flat near-zero prevalence over most of Africa when trained on the observation data, a behaviour that is not replicated by fitting the model to either set of simulated data. This behaviour may be due to the noise in the observation data, which is visible in Figure~\subref*{fig:Africa_observation_points} particularly in Uganda. In contrast, the simulated data at the same locations, shown in Figure~\subref*{fig:Africa_observation_binomial_points}, appears much smoother.

We examine this working hypothesis by adding Gaussian noise to the simulated data. Prevalence values sampled from the MAP raster in Figure~\subref*{fig:MAP_raster} with locations based on the observation points were transformed using the logit function. Gaussian white noise with chosen standard deviations was added to the transformed values, before being brought back to values between 0 and 1 via the inverse logit. Binomial samples for the number of positive tests were then drawn using these prevalences, and INLA was fit to this data.

\begin{figure}
\captionsetup{width=.25\linewidth}
    \centering
    \subfloat[][No added noise]{
        \includegraphics[width=0.3\textwidth]{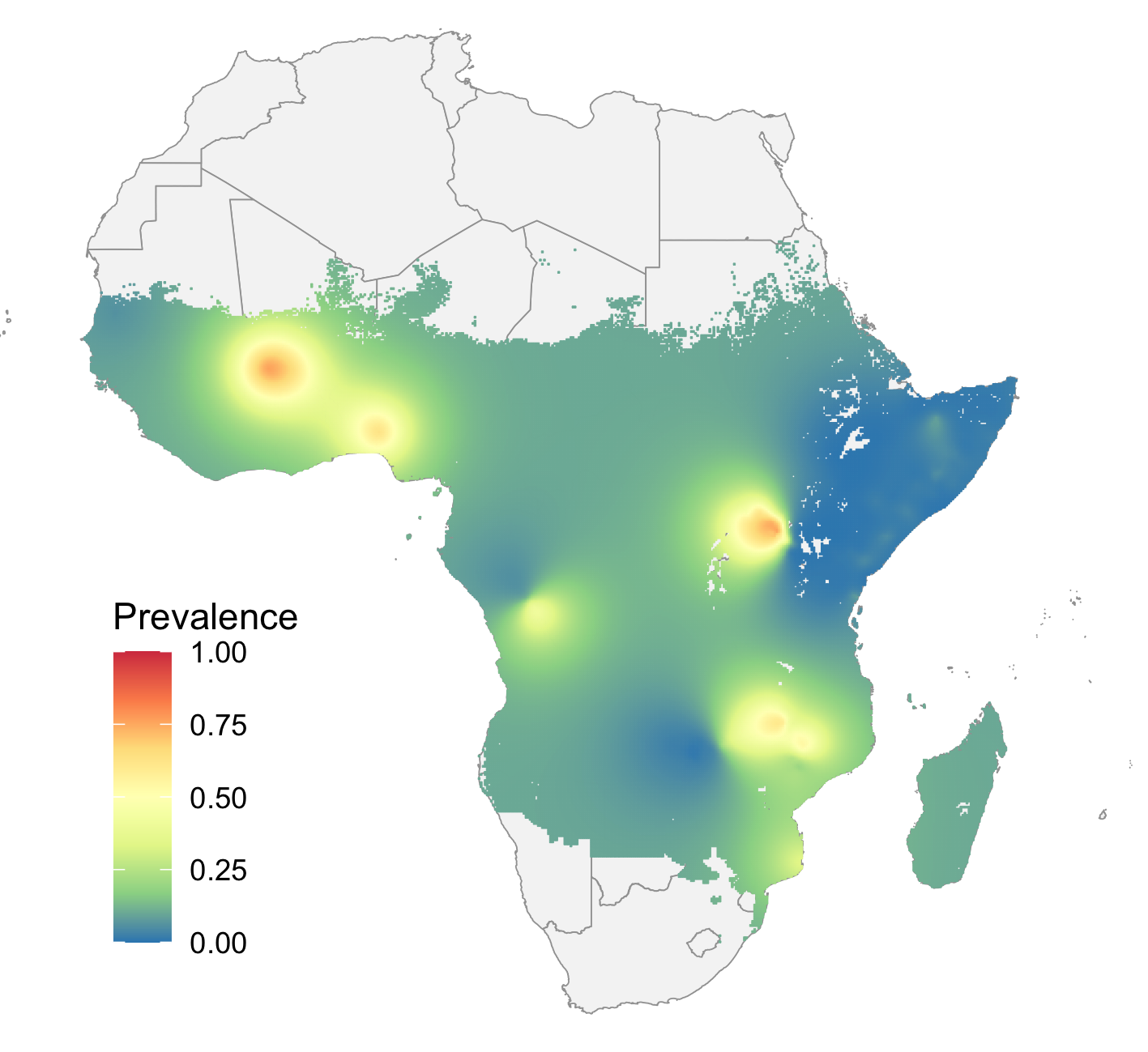}
        \label{fig:no_noise}
    }
    \subfloat[][Gaussian noise with standard deviation 0.4]{
        \includegraphics[width=0.3\textwidth]{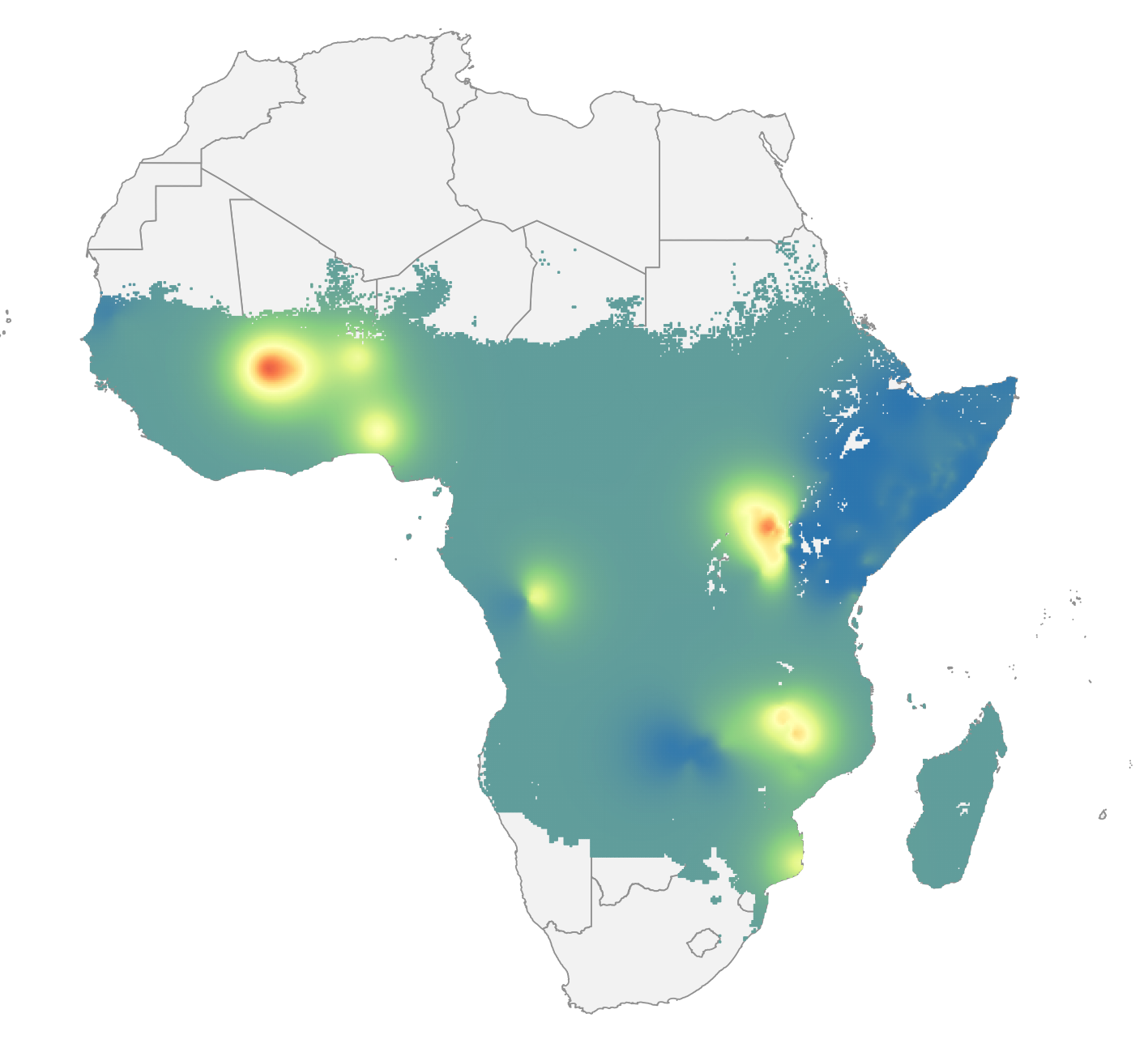}
        
        \label{fig:0.4_noise}
    }
     \subfloat[][Gaussian noise with standard deviation 1.2]{
        \includegraphics[width=0.3\textwidth]{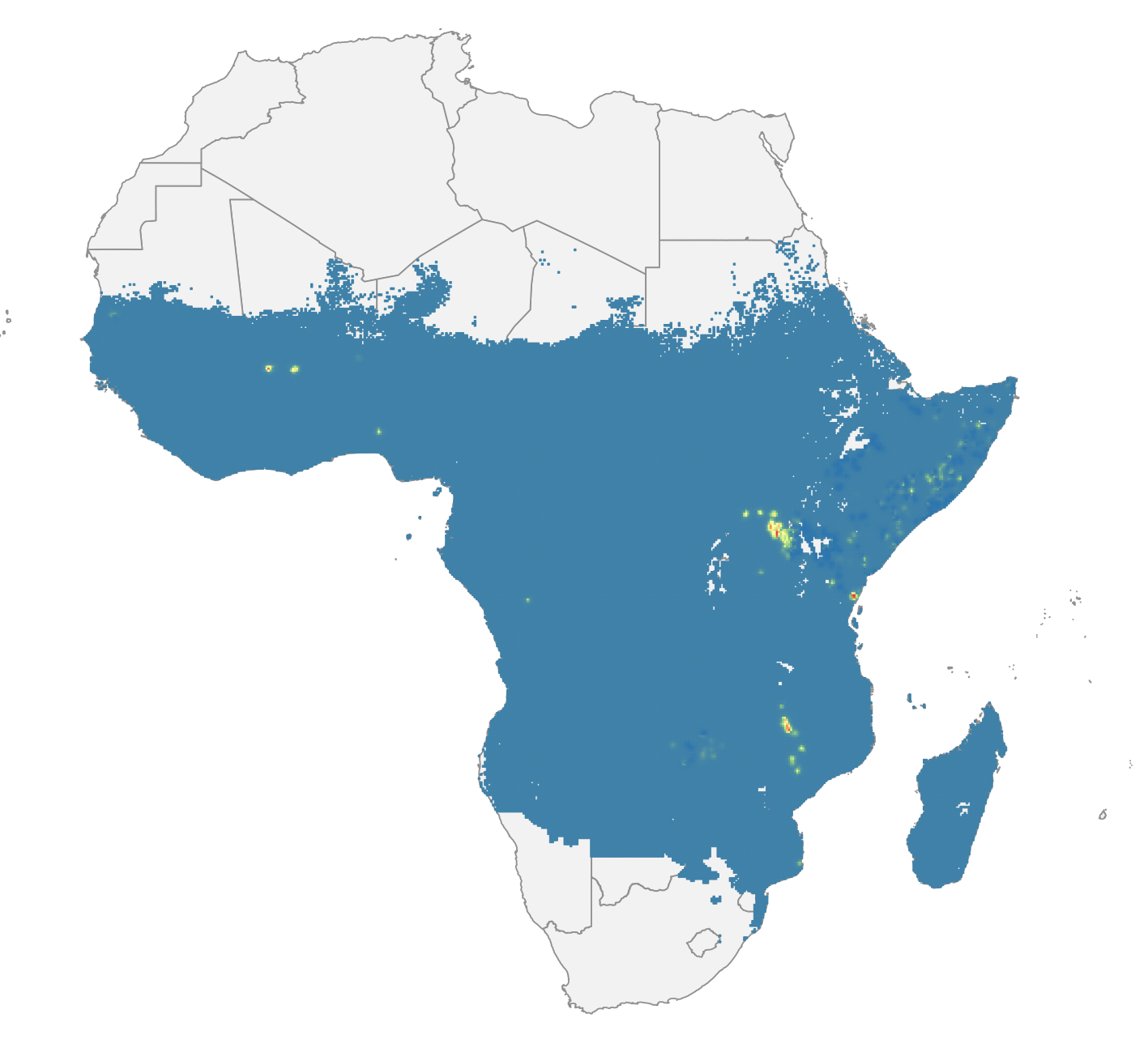}
        \label{fig:1.2_noise}
    }
    \captionsetup{width=.8\linewidth}
    \caption{Predicted prevalence from INLA when fit to simulated data at observation locations with varying amounts of added Gaussian noise. }
    \label{fig:INLA_noise_fits}
\end{figure}

\begin{figure}
    \centering
    \includegraphics[width=0.8\textwidth]{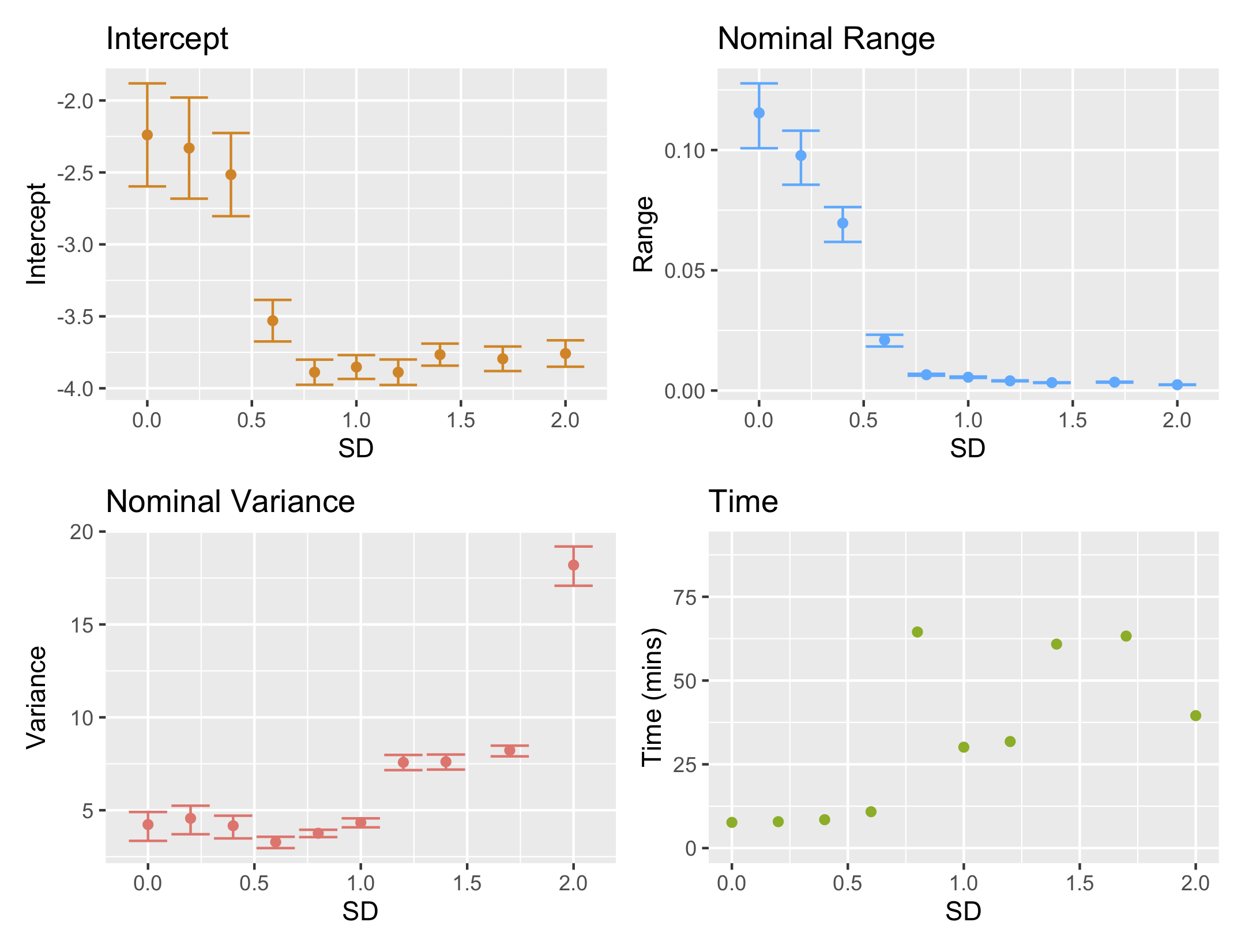}
    \caption{Posterior means of the intercept, range and variance for the INLA model fit using simulated data at the observation locations with added Gaussian noise of varying standard deviation. The bottom right plot shows the time taken to fit the INLA model to each of the datasets. Error bars show posterior interquartile ranges.}
    \label{fig:noise_posterior}
\end{figure}

Predictions from INLA fitted to data with three different levels of noise are shown in Figure~\ref{fig:INLA_noise_fits}. Figure~\ref{fig:noise_posterior} shows posterior means and interquartile ranges for the intercept, range, and variance of the fitted models as the standard deviation of the added noise increases, as well as the time taken to fit each model and generate predictions.

As the standard deviation of added noise increases, both the intercept and spatial range fall, and predictions become less correlated between locations. The time taken jumps for standard deviations above 0.6 and presumably the model has difficulty converging. With greater noise, the model predicts a flat prevalence away from the simulated data dependent on the value of the intercept, and its output in Figure~\subref*{fig:1.2_noise} resembles the predictions in Figure~\ref{fig:Africa_prevalence_maps}(ai) of the model trained on the observation data. These results suggest the presence of overdispersion, and that the INLA model used may be misspecified. Indeed the model in equation \eqref{eq:INLA_model} does not contain an independent error term. Methods to address this include adding an observational random effect to the model, or using a Beta-binomial response.

\subsection{INLA with Gaussian response} \label{sec:INLA_gaussian}
\begin{figure}
    \centering
    \includegraphics[width = 0.4\textwidth]{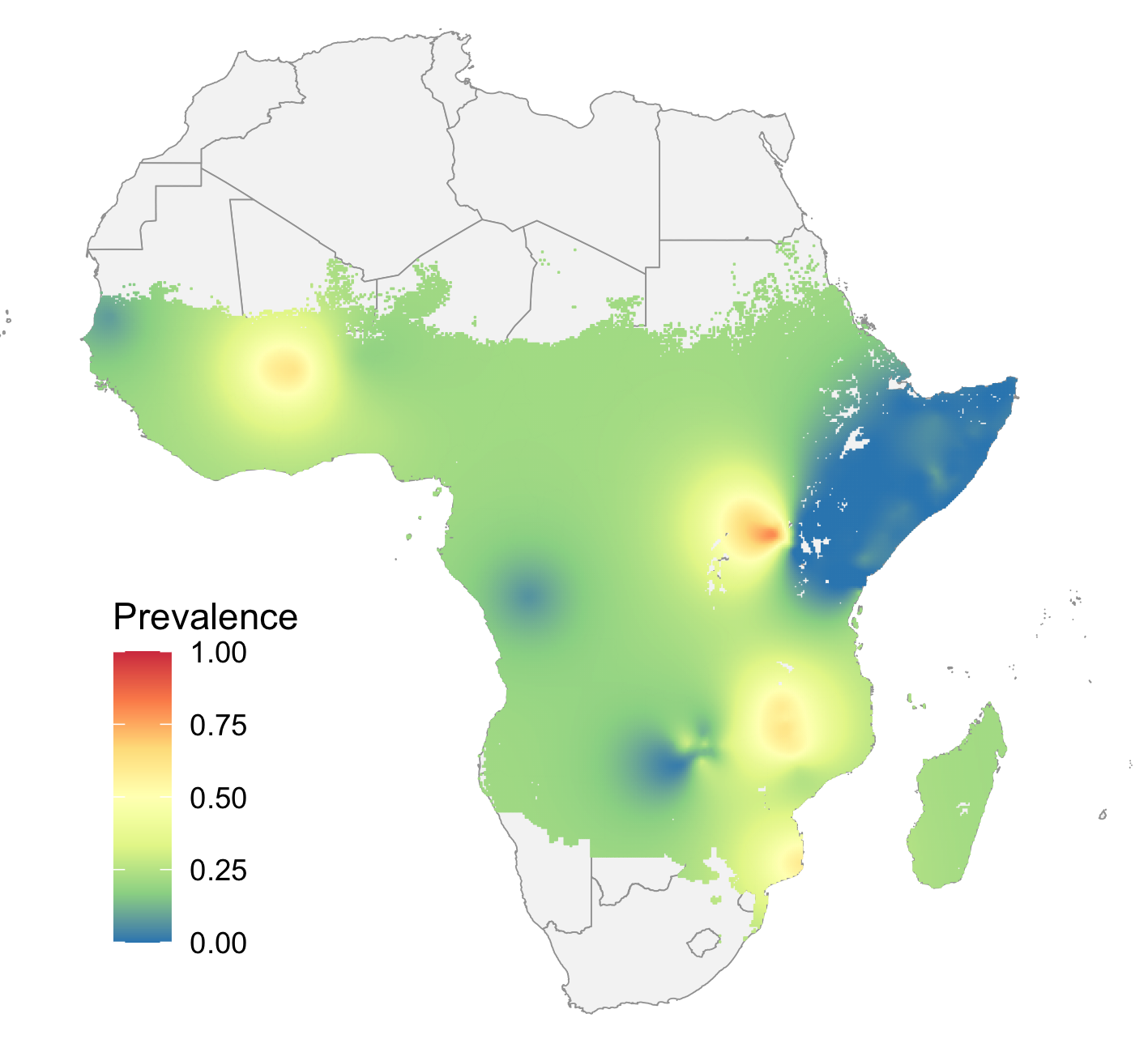}
    \caption{Predictions from an INLA model with a Gaussian response fit to the observation data. Values have been clipped to lie within $[0,1]$.}
    \label{fig:INLA_gaussian}
\end{figure}

Due to the overdispersion when fitting the INLA model to the observation data, we tested an additional INLA model which uses a Gaussian response. Predictions from this model are shown in Figure~\ref{fig:INLA_gaussian}, and the absence of flat predictions suggests that this response is able to resolve the overdispersion.

\section{Prediction Uncertainty} \label{sec:prediction_uncertainty}
\begin{figure}[!p]
\centering
\begin{tabular}{p{2mm}M{0.31\textwidth}M{0.31\textwidth}M{0.31\textwidth}}
        &  (i) Observation data& (ii) Binomial sampling at observation locations & (iii) Binomial sampling with uniform coverage \\
    \rotatebox[origin = c]{90}{(a) INLA} 
    & \includegraphics[width=0.31\textwidth]{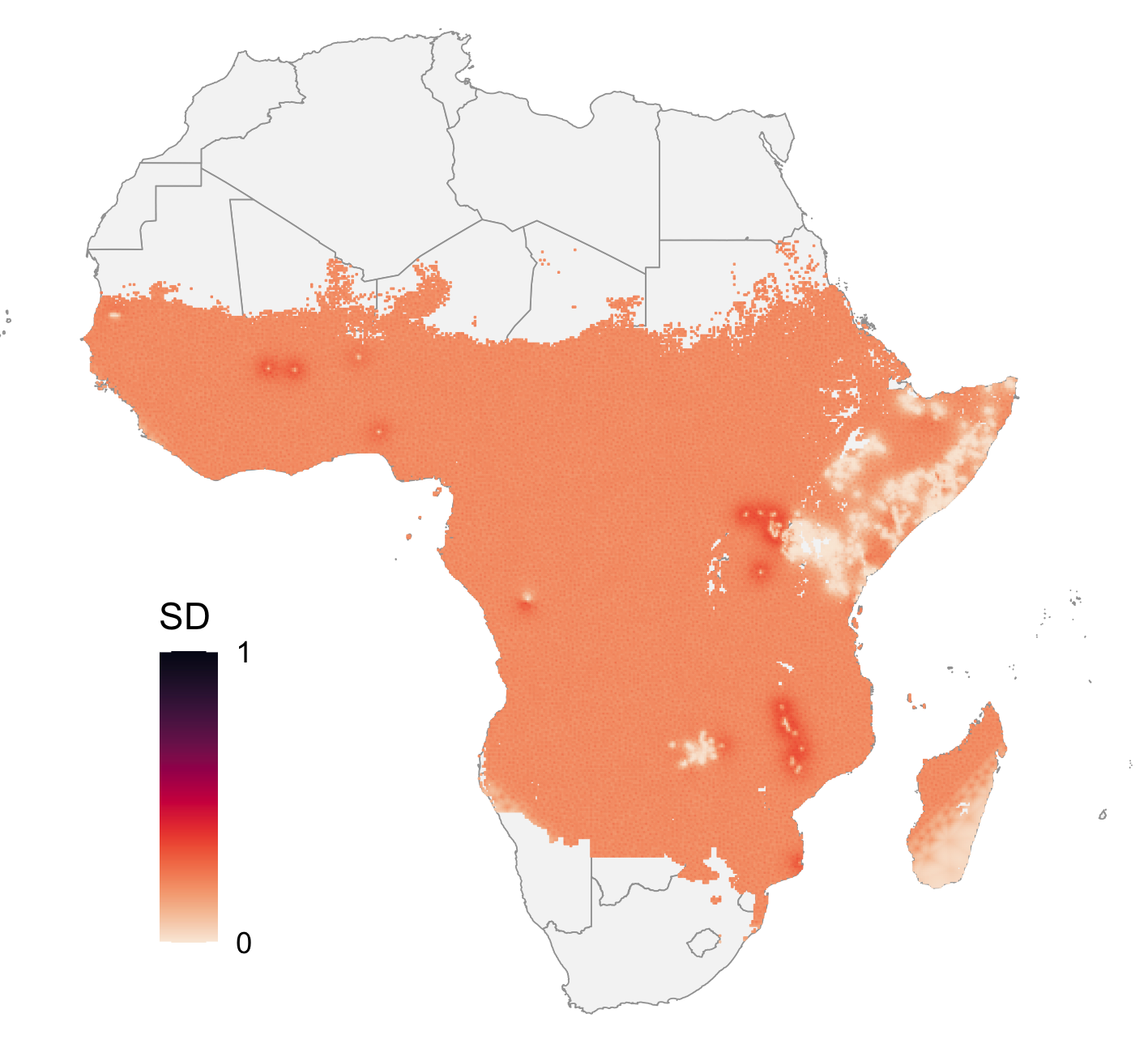}
    & \includegraphics[width=0.31\textwidth]{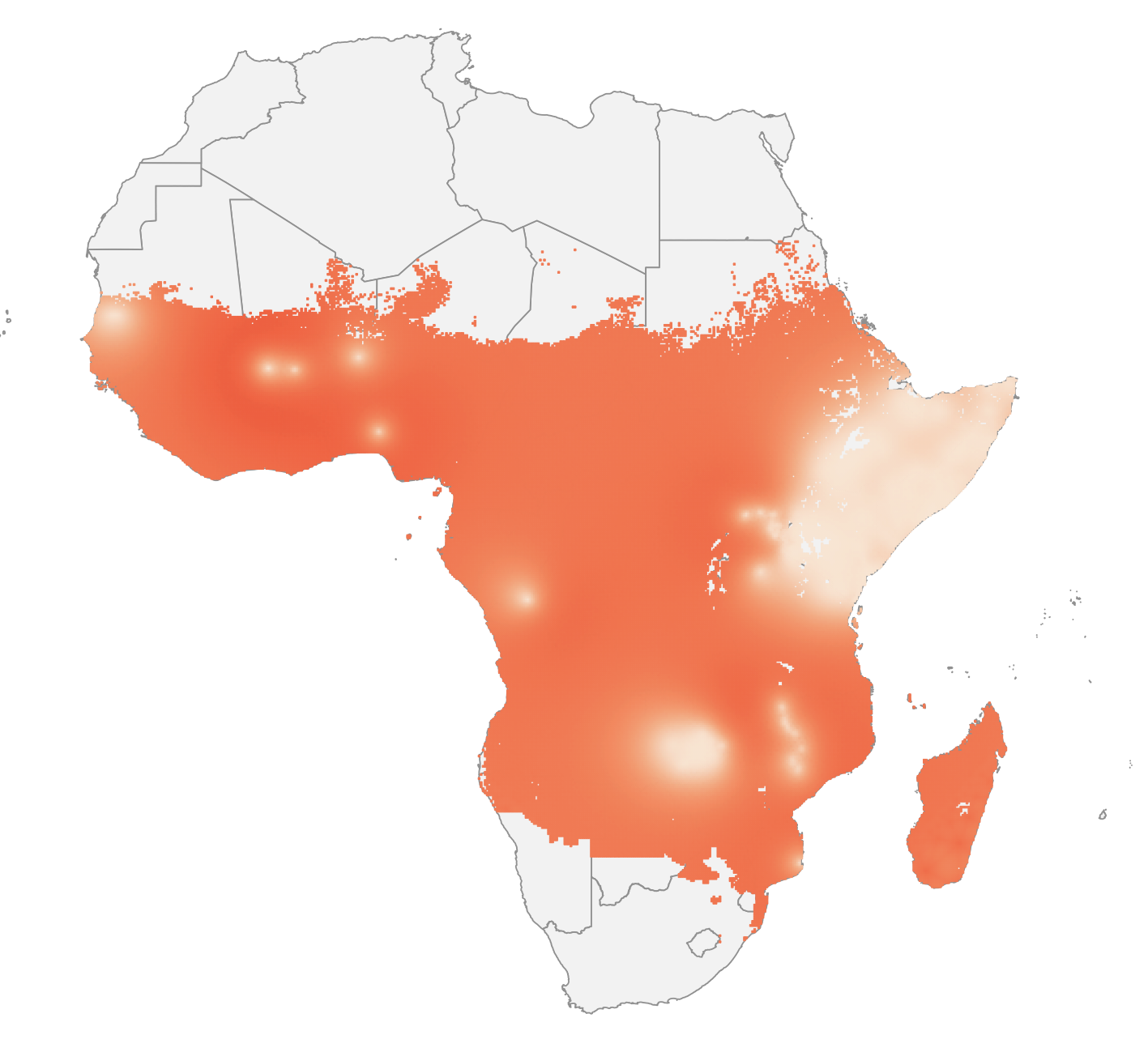}
    & \includegraphics[width=0.31\textwidth]{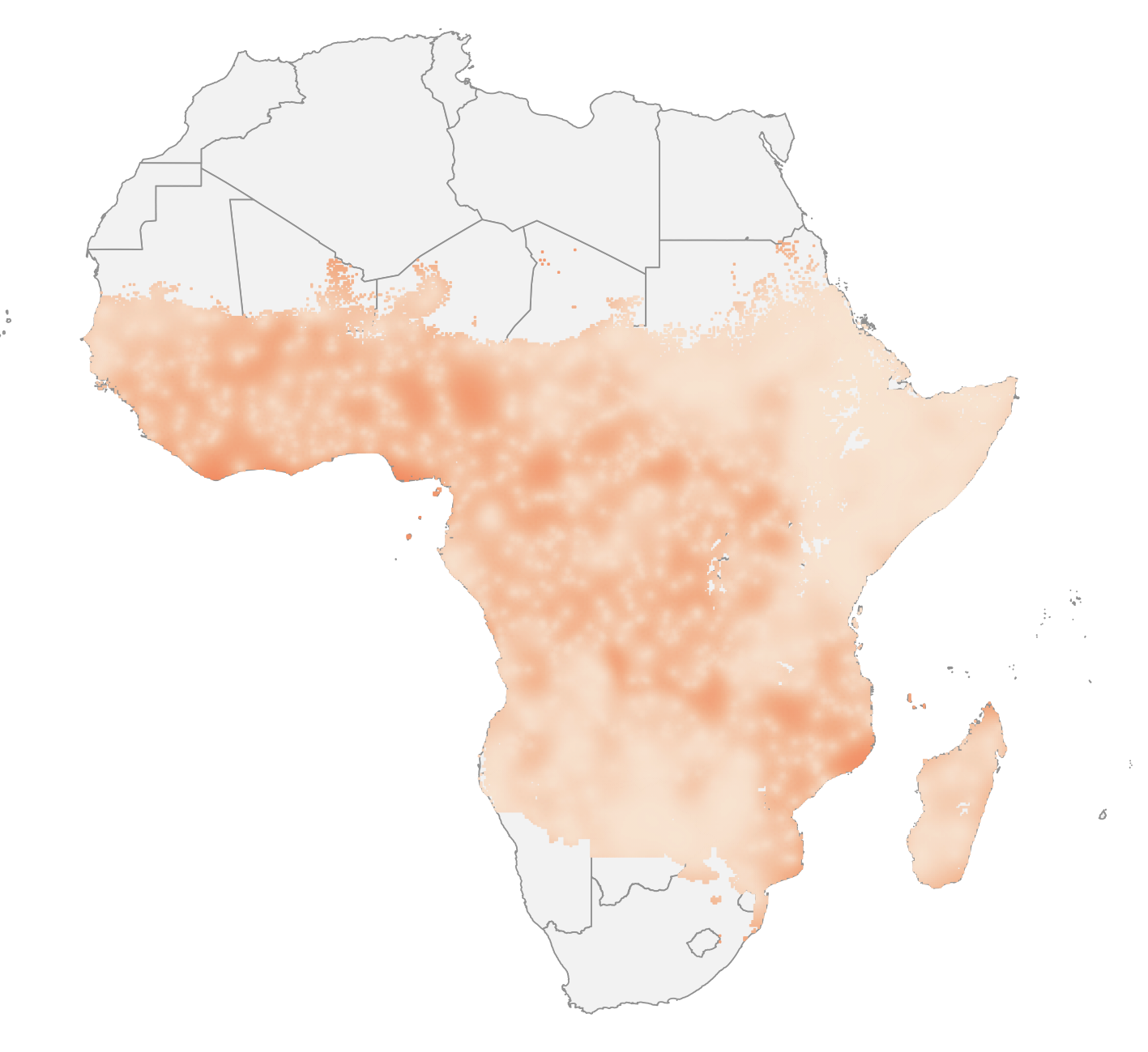}  \\
    \rotatebox[origin = c]{90}{(c) GPBoost} 
    & \includegraphics[width=0.31\textwidth]{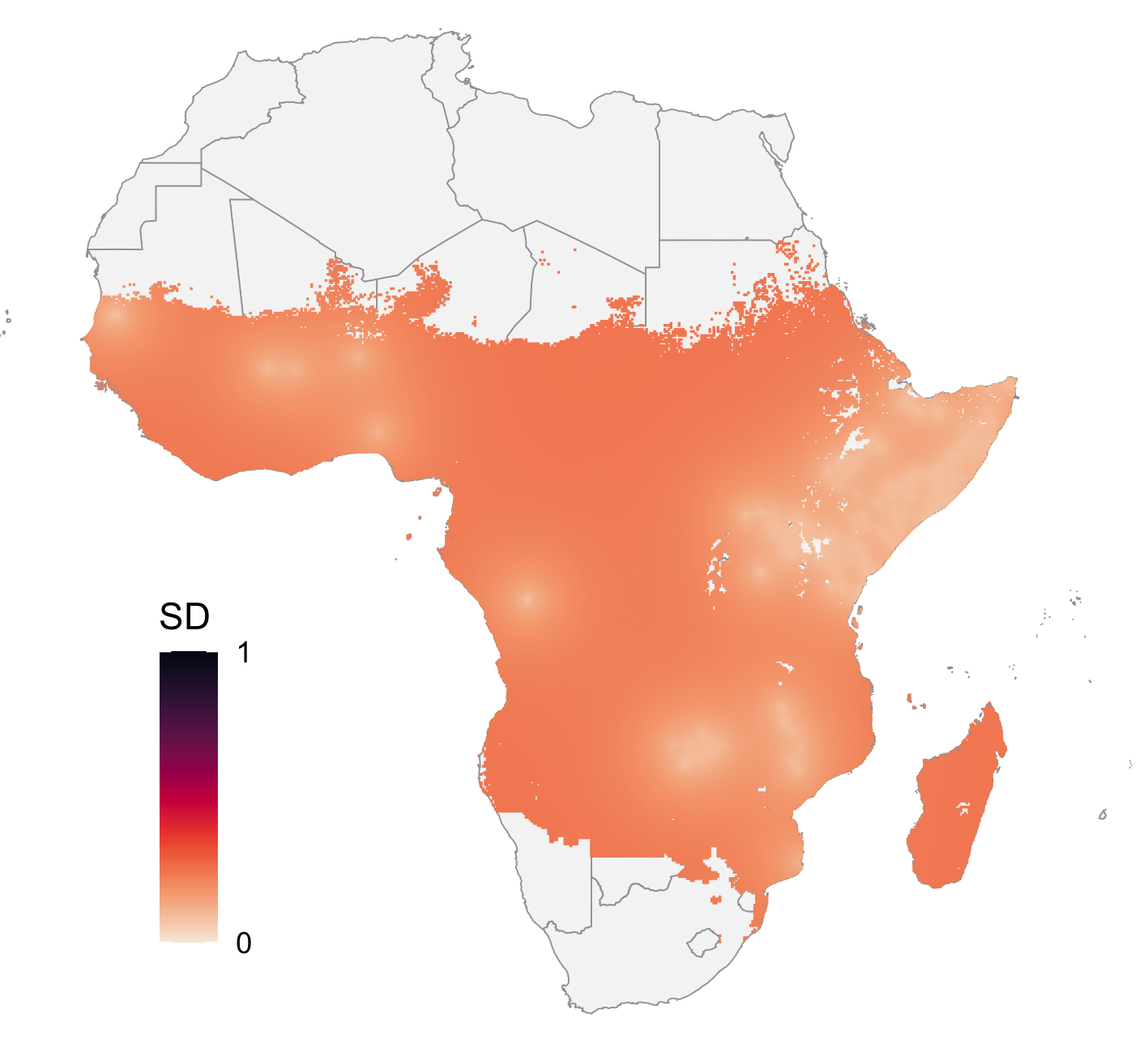} 
    & \includegraphics[width=0.31\textwidth]{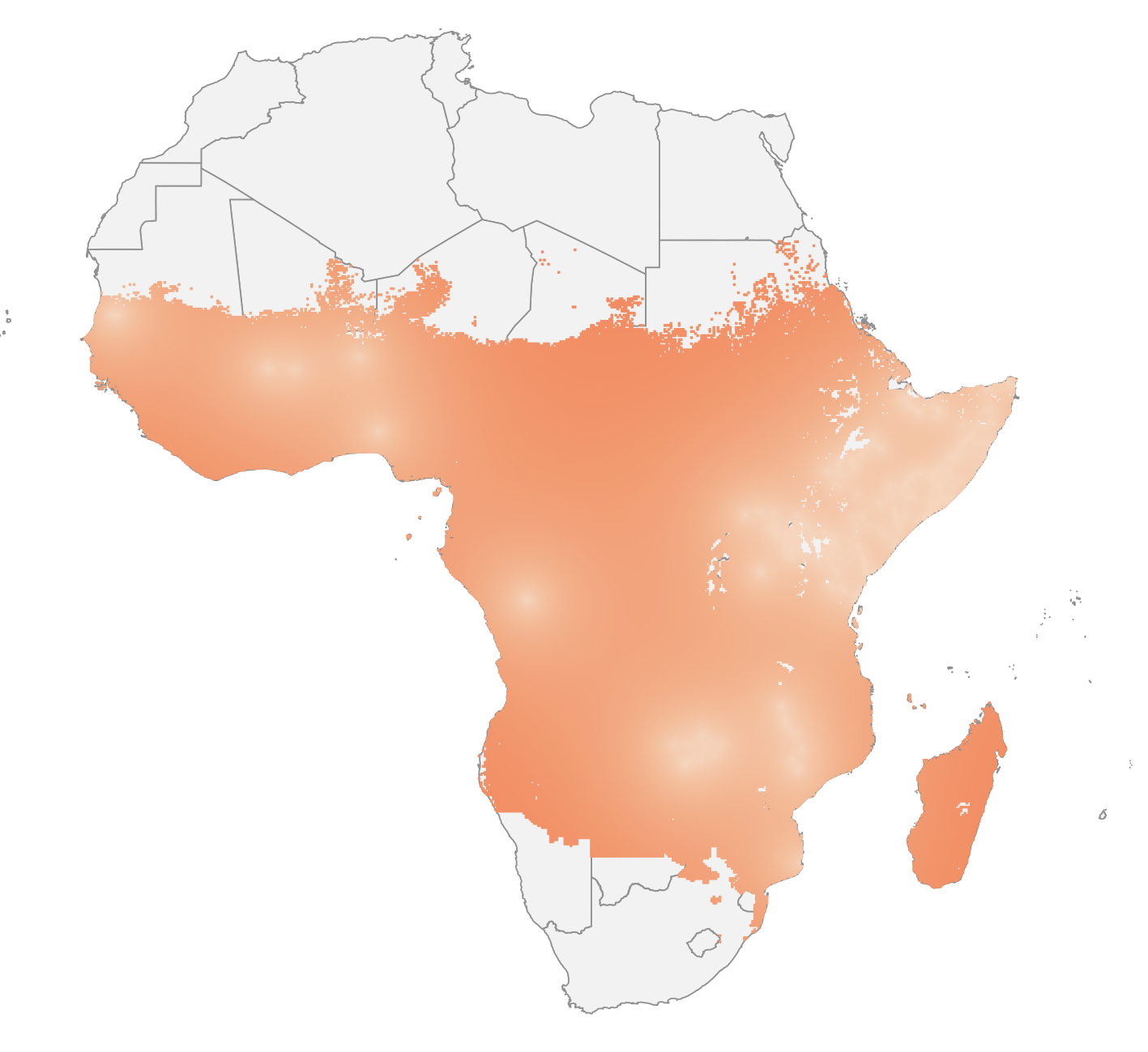} 
    & \includegraphics[width=0.31\textwidth]{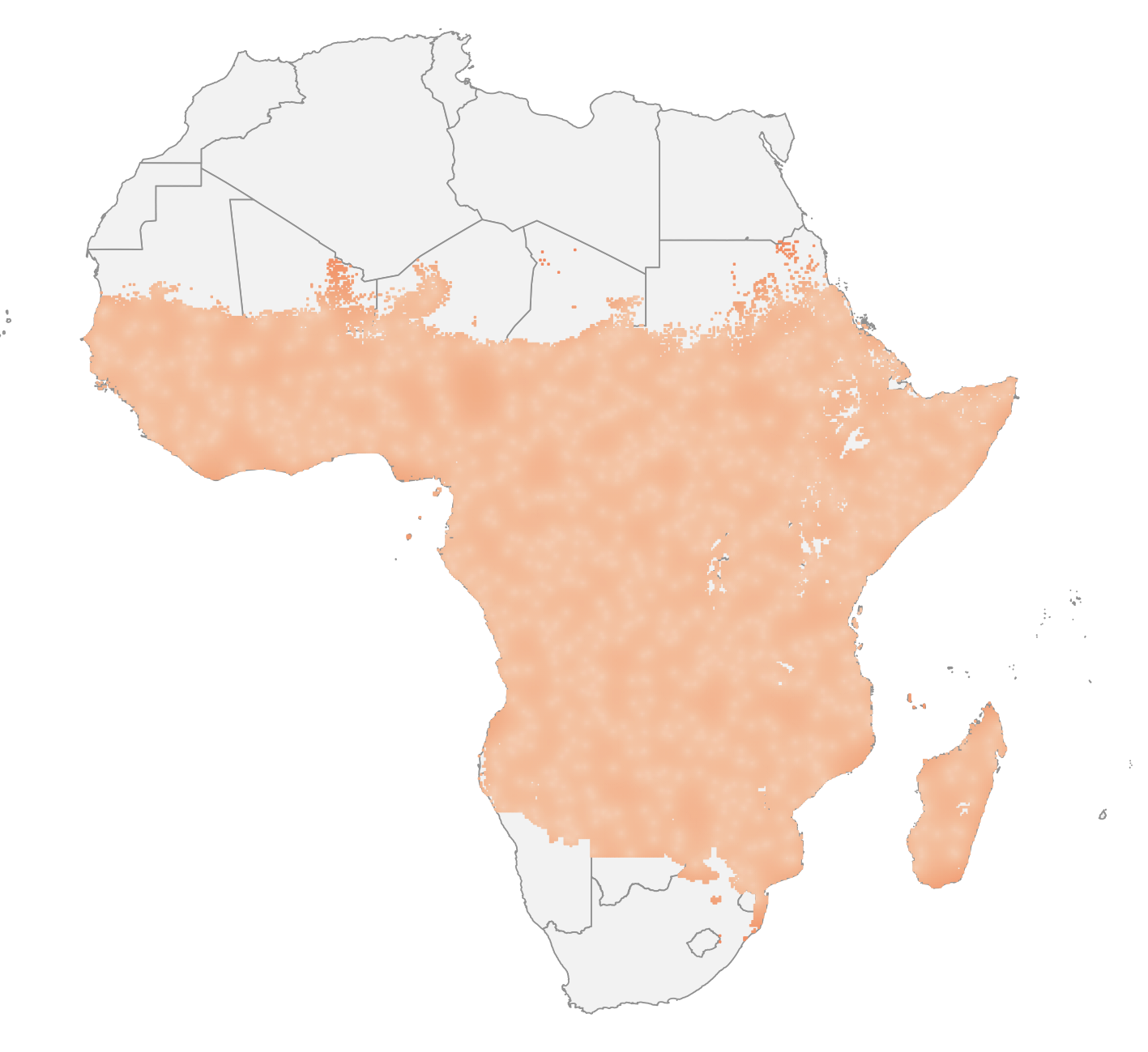} 
    \\
    \rotatebox[origin = c]{90}{(b) SpRF} 
    & \includegraphics[width=0.31\textwidth]{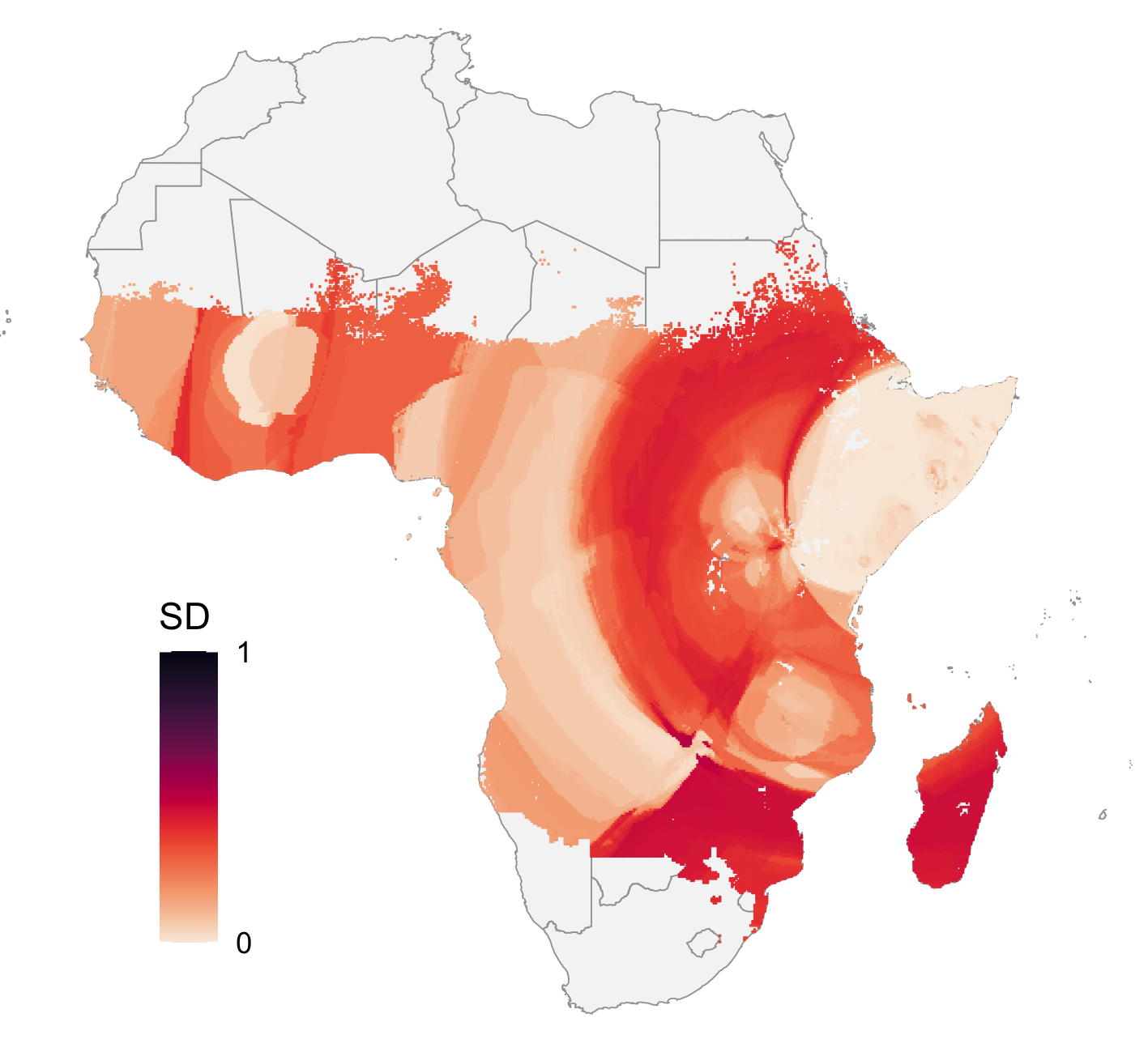}
    & \includegraphics[width=0.31\textwidth]{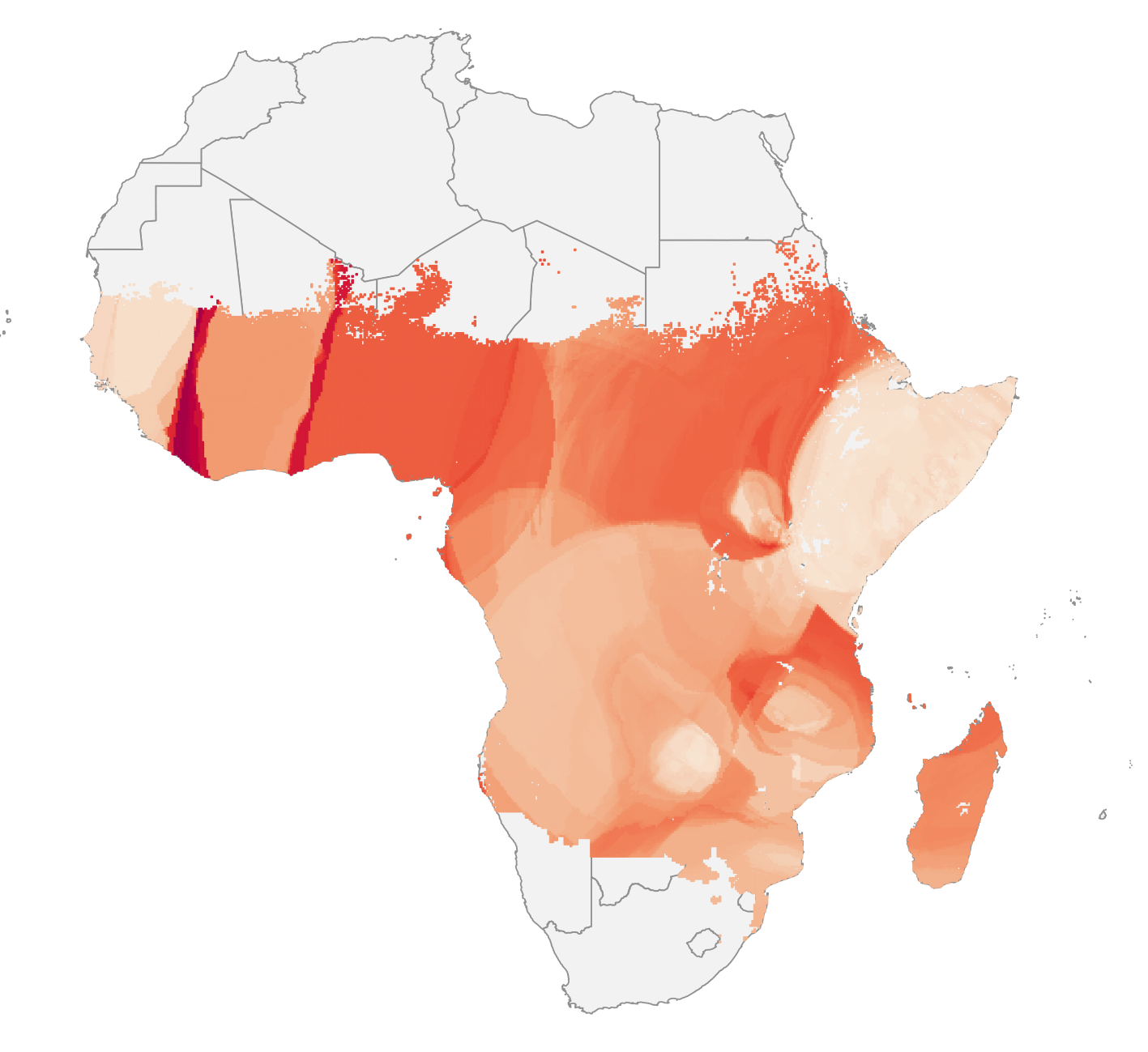} 
    & \includegraphics[width=0.31\textwidth]{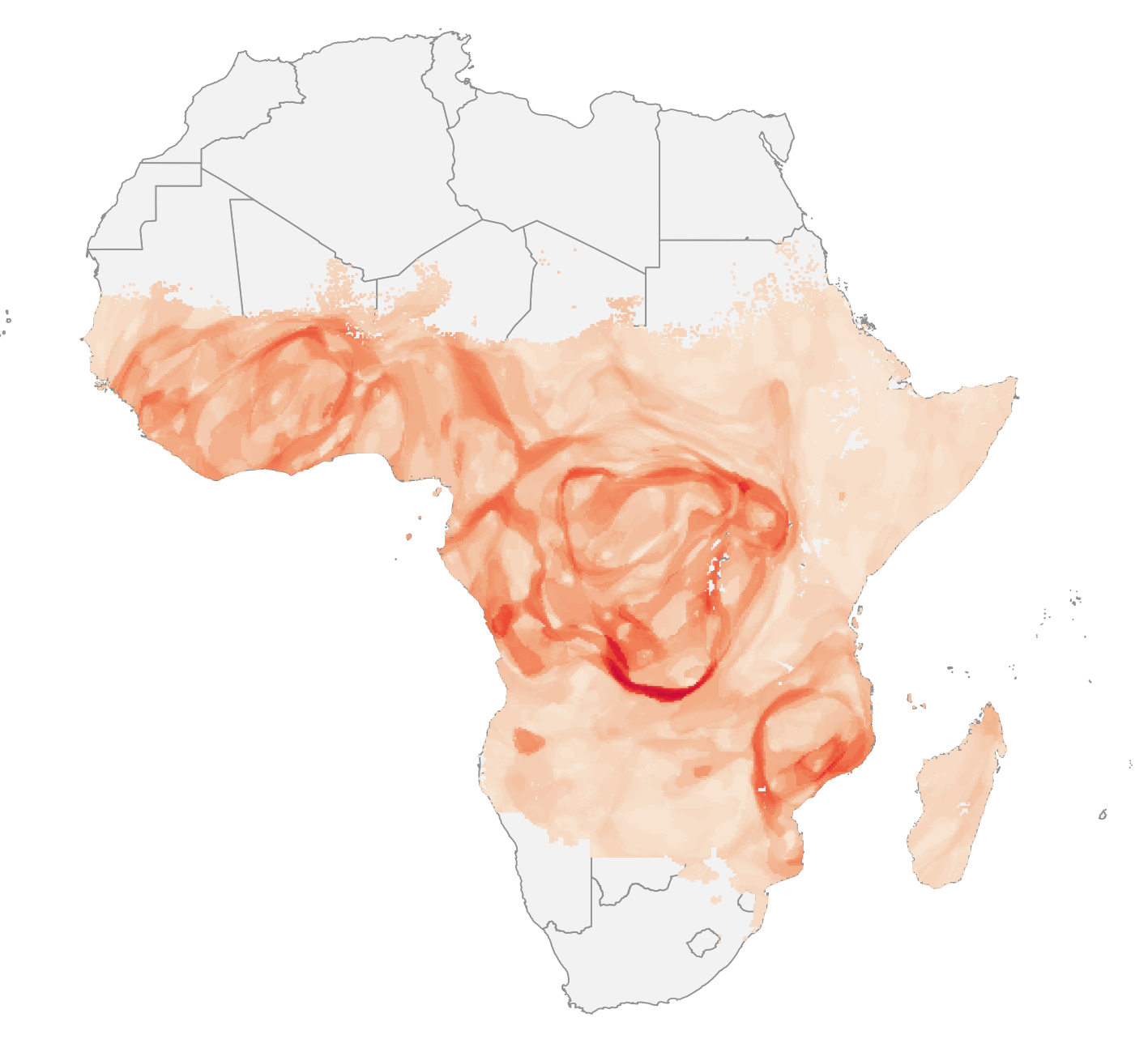} \\
    \rotatebox[origin = c]{90}{(d) FRK} 
    & \includegraphics[width=0.31\textwidth]{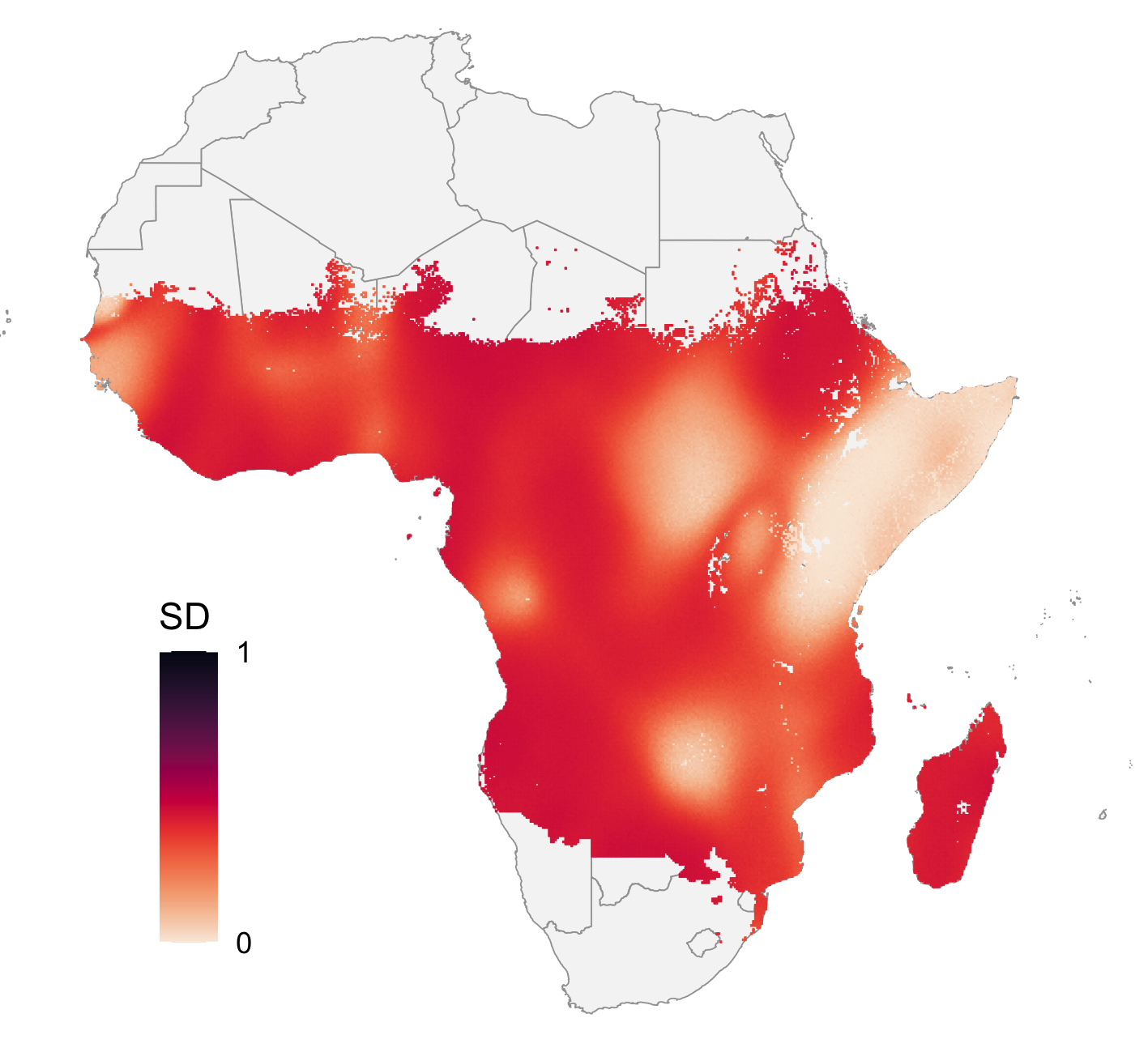}
    & \includegraphics[width=0.31\textwidth]{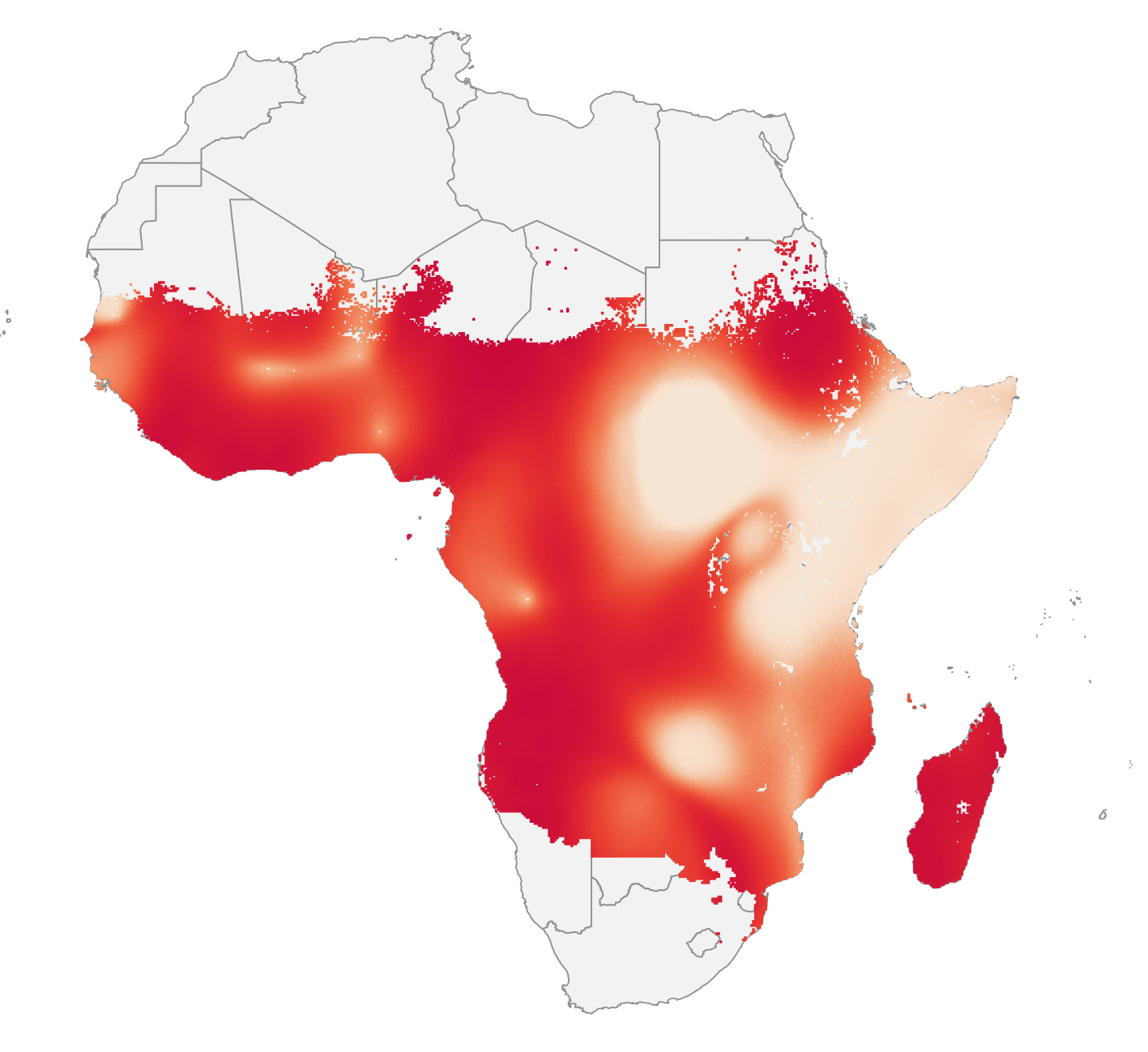} &
    \includegraphics[width=0.31\textwidth]{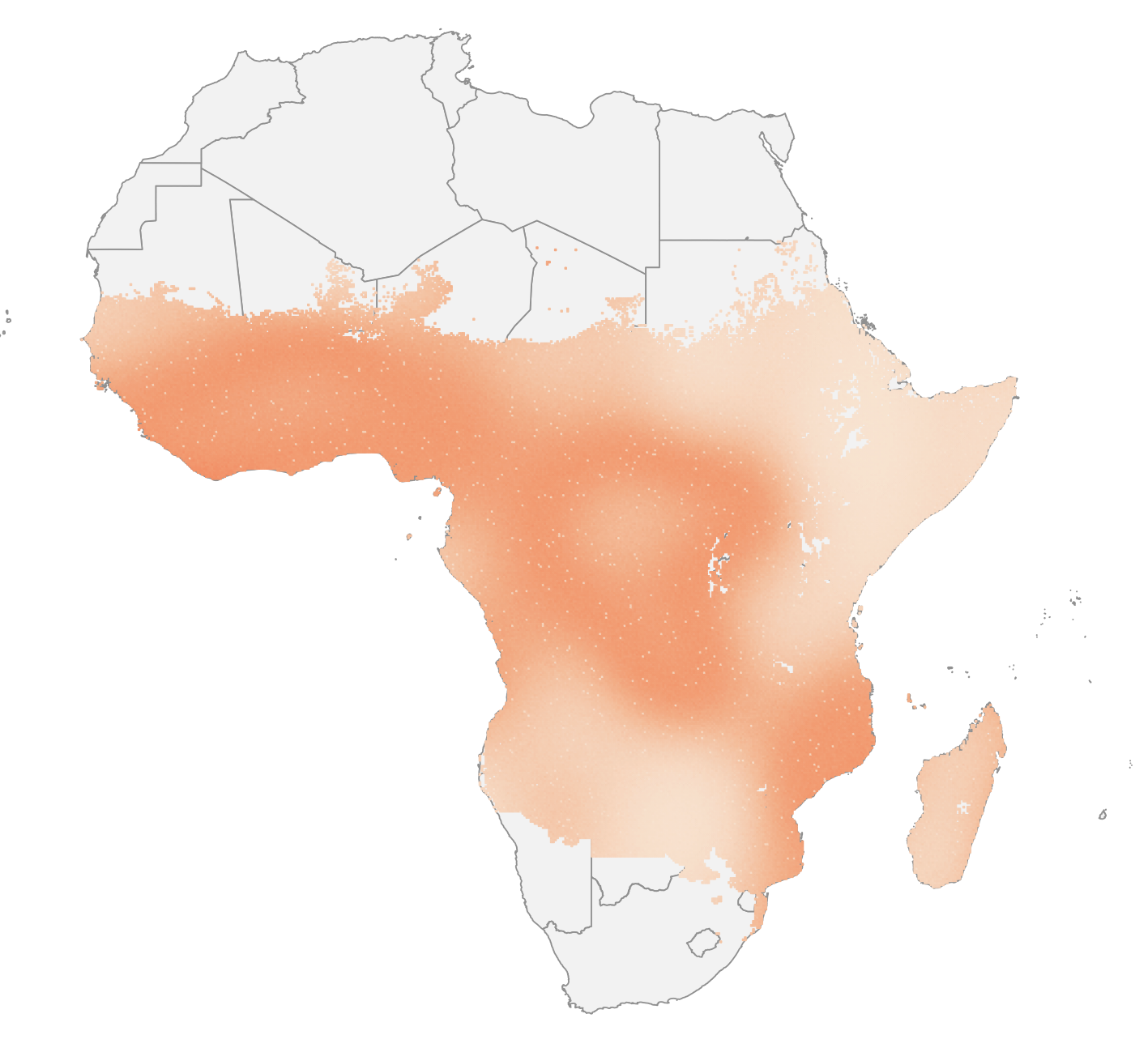}
  
\end{tabular}
  
    \caption{Predicted standard deviations for each of the maps shown in Figure~\ref{fig:Africa_prevalence_maps}.}
    \label{fig:Africa_uncertainty_maps}
\end{figure}

Figure~\ref{fig:Africa_uncertainty_maps} shows the prediction uncertainties corresponding to each of the prevalence maps over Africa in Figure~\ref{fig:Africa_prevalence_maps}.

\section{GPBoost with the Vecchia approximation} \label{sec:Vecchia}
As it uses a full Gaussian process, it is unsurprising that the GPBoost model shows the least favourable computational time for larger datasets. To improve efficiency, a Vecchia approximation is available in the software, which approximates the distribution of the response as 
\begin{align*}
    p(\mathbf{y}|F(X),\boldsymbol{\theta})&= \prod_{i=1}^n p(y_i|y_{i-1},...,y_1, F(X), \boldsymbol{\theta})\\
    &\approx \prod_{i = 1}^N p(y_i | y_{N(i)}, F(X), \boldsymbol{\theta})\,,
\end{align*}

as per \cite{sigrist2020gaussian}. Here $y_{N(i)}$ is the subset of $\{y_1,..., y_{i-1}\}$ containing the $m_v$ nearest neighbours to $y_i$, where ``nearest neighbours'' are determined by the distances between the responses' corresponding locations. The parameter $m_v$ determines the number of neighbours to use during fitting, while a separate parameter, $m_{v,p}$, controls the number of neighbours used for prediction. The approximation additionally requires a choice of ordering of the observed responses $\{y_1,...,y_n\}$, which by default is taken to be the original ordering of the input data. 

\begin{figure}
    \centering
    \subfloat[][]{
        \includegraphics[width=0.28\textwidth]{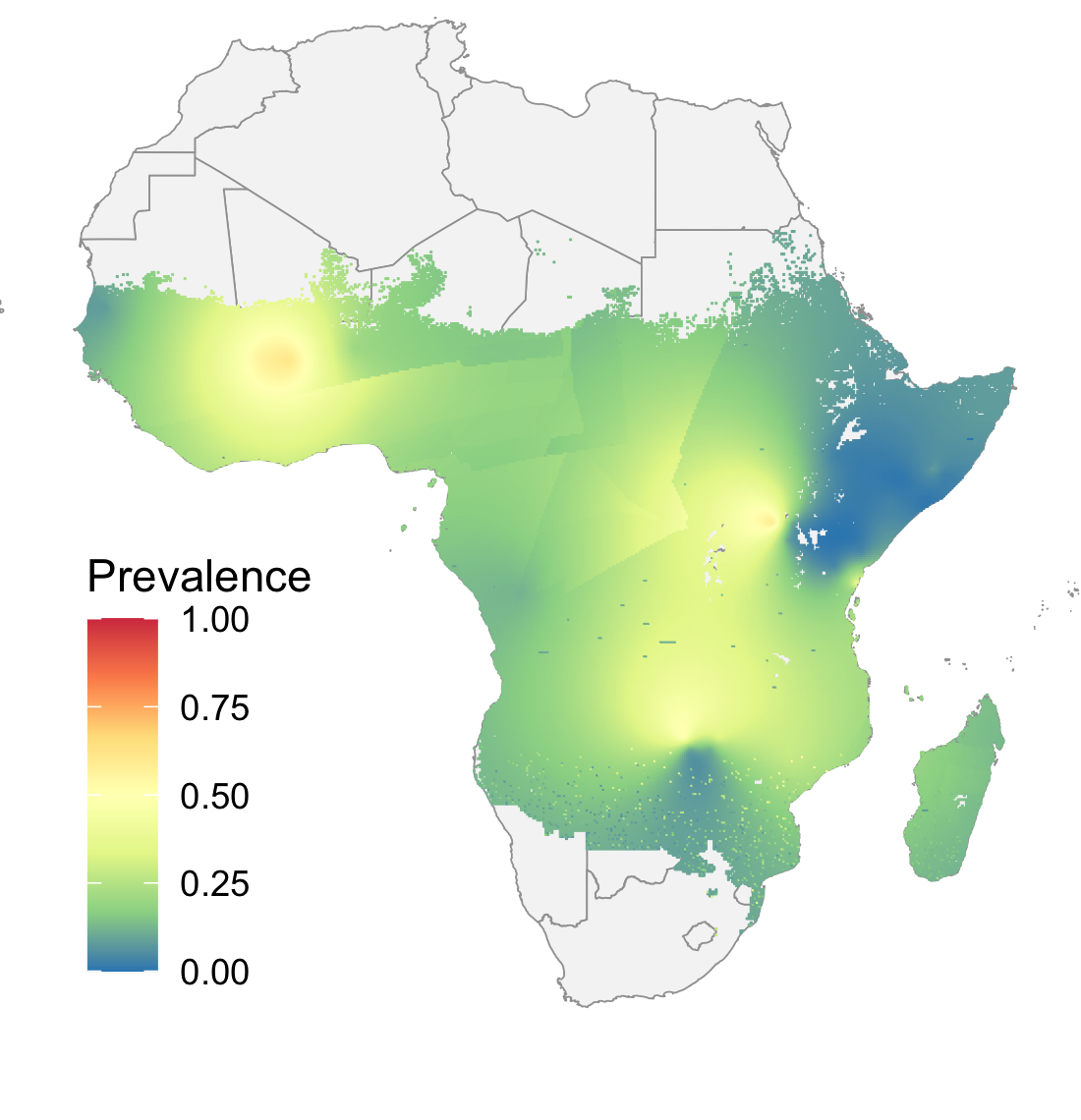}
        \label{fig:vecchia_defaults}
    }
     \subfloat[][]{
        \includegraphics[width=0.28\textwidth]{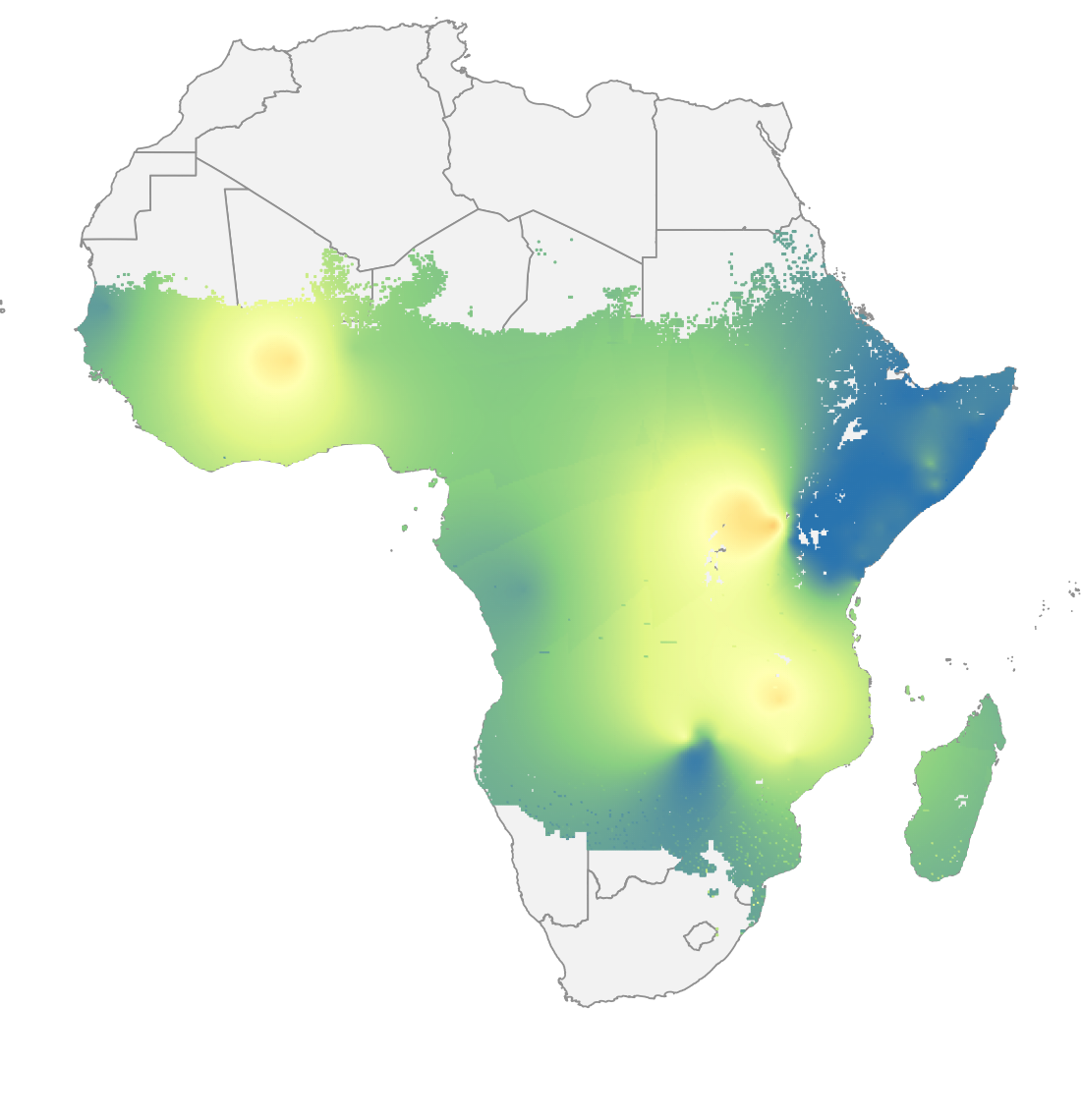}
        \label{fig:vecchia_mvp150}
    } 
     \subfloat[][]{
        \includegraphics[width=0.28\textwidth]{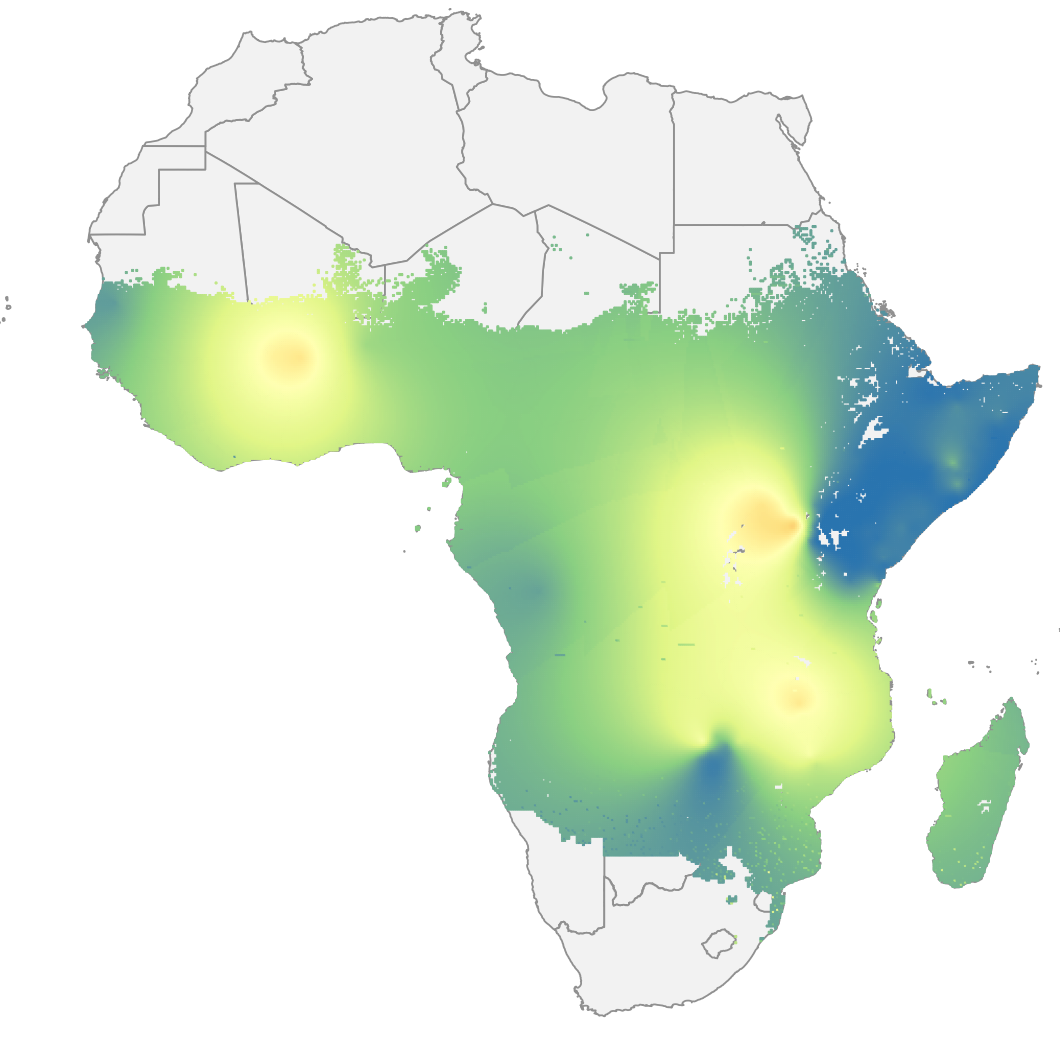}
        \label{fig:vecchia_mv150_mvp150}
    } 
    \\
    \subfloat[][]{
        \includegraphics[width=0.22\textwidth]{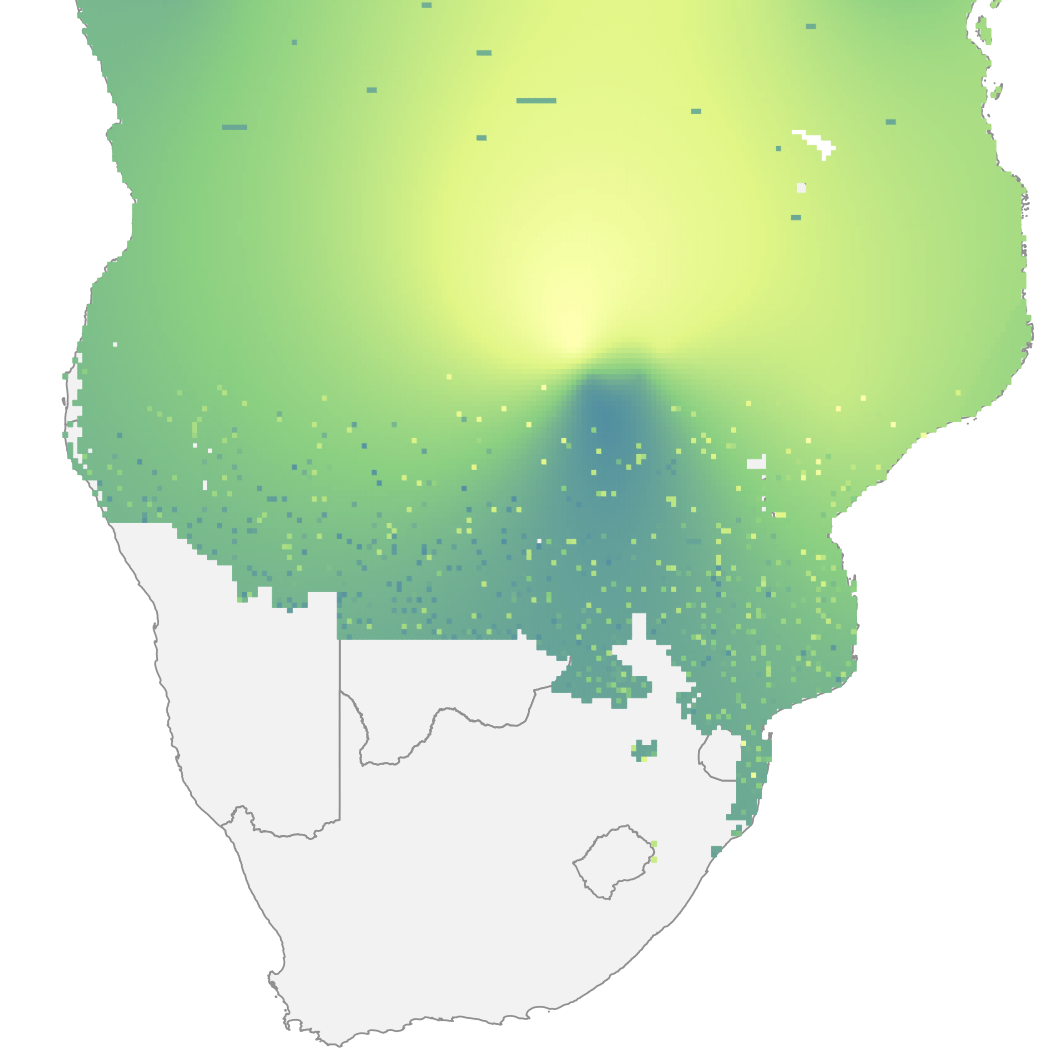}
        \label{fig:vecchia_defaults_detail}
    }\hspace{10px}
    \subfloat[][]{
        \includegraphics[width=0.22\textwidth]{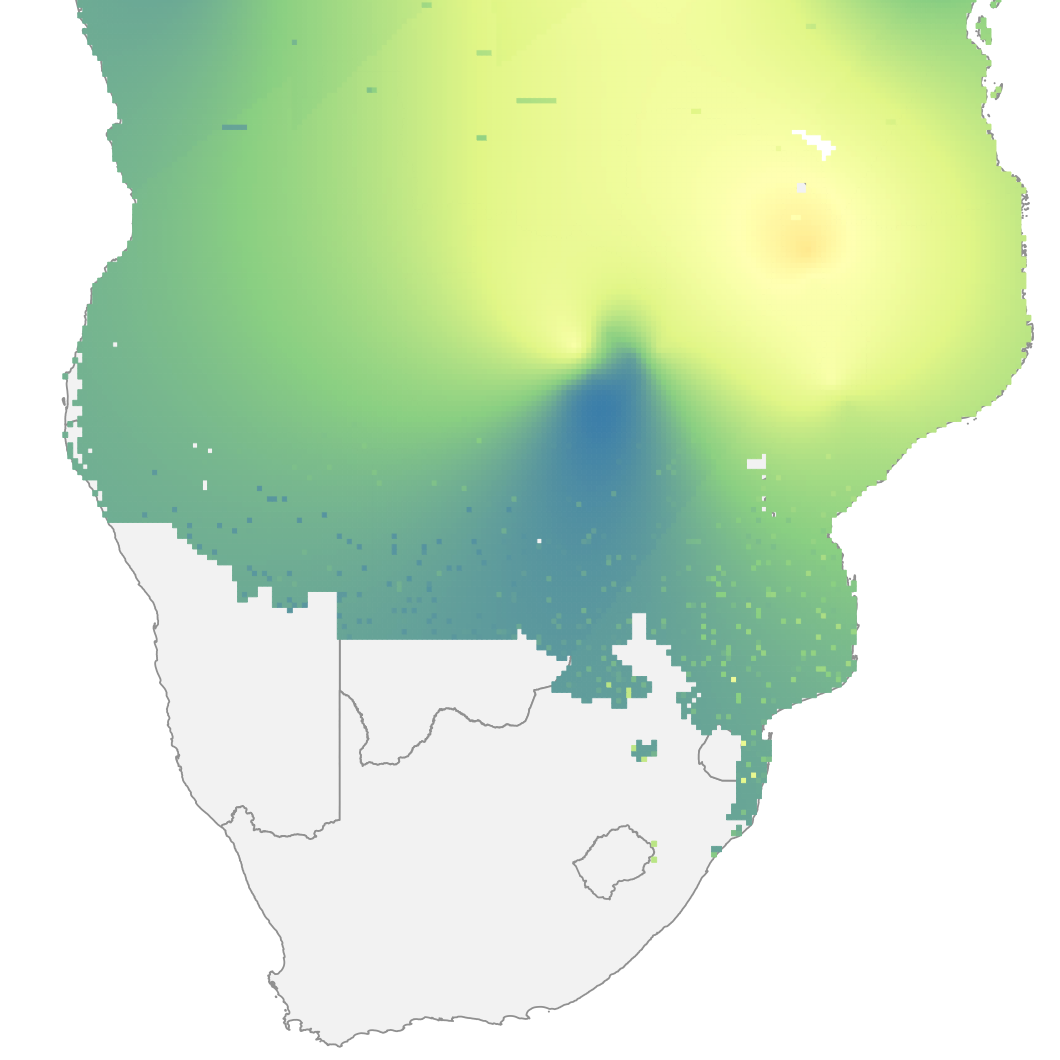}
        \label{fig:vecchia_mvp150_detail}
    } 
    \hspace{10px}
     \subfloat[][]{
        \includegraphics[width=0.22\textwidth]{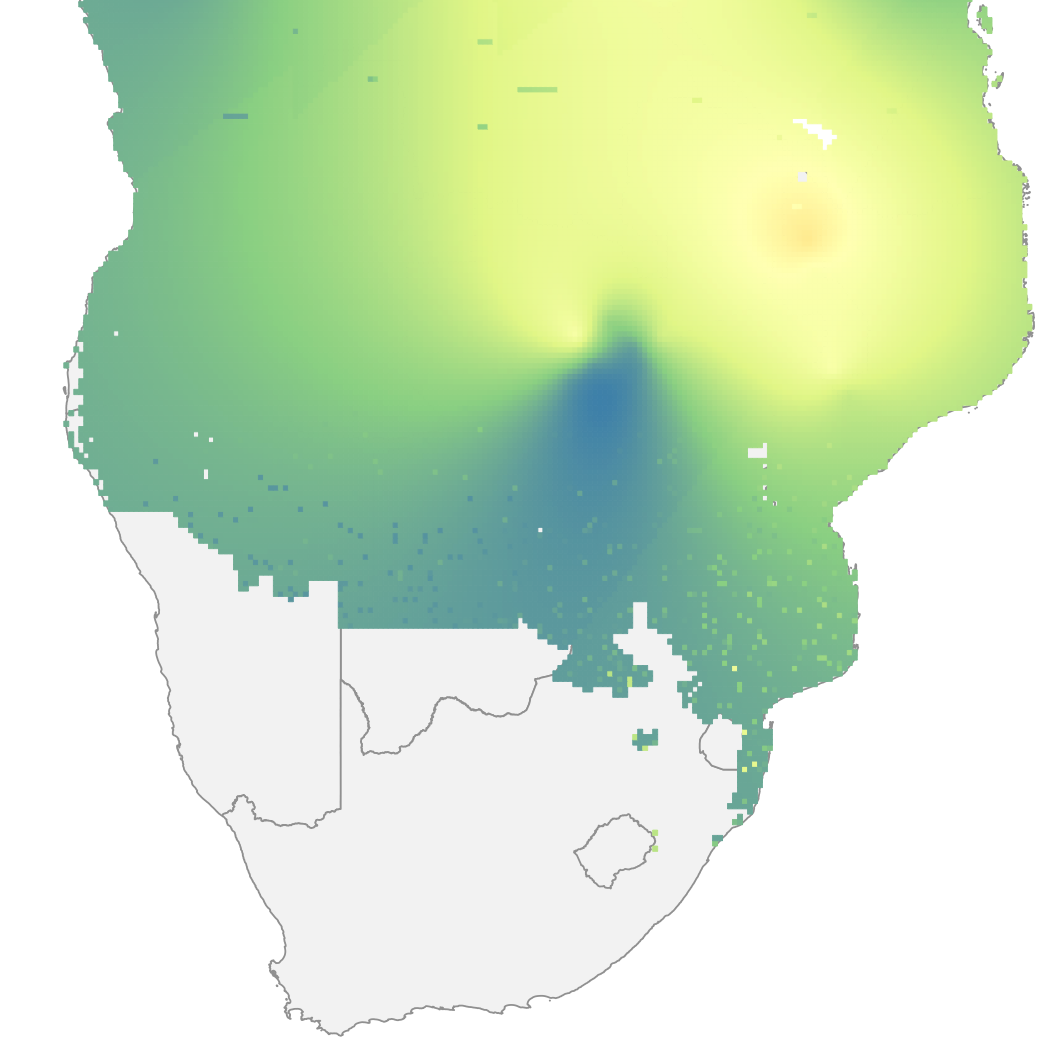}
        \label{fig:vecchia_mv_150_mvp150_detail}
    } 
    \caption{\textit{P. falciparum} prevalence predictions for GPBoost when using the Vecchia approximation for various values of the nearest neighbour parameters, $m_v$ and $m_{v,p}$. (a) uses $m_v = m_{v,p} = 30$,  (b) uses $m_v = 30$ and $m_{v,p} = 150$, while (c) uses $m_v = m_{v,p}$ = 150. (d)-(f) show the southern regions of the above plots, where noise in the predictions is more prominent.}
    \label{fig:GPBoost_vecchia}
    
\end{figure}

Figure~\ref{fig:GPBoost_vecchia} shows GPBoost's predictions when using a Vecchia approximation with several values of $m_v$ and $m_{v,p}$. Uncertainty predictions were not produced as they are not currently well supported in the software when using the Vecchia approximation. 

Applying the Vecchia approximation introduces several artifacts to GPBoost's predictions. Figure~\subref*{fig:vecchia_defaults} and \subref*{fig:vecchia_defaults_detail} show sharp discontinuities and noisy predictions, both of which were prominent whenever low values of $m_v$ and $m_{v,p}$ were used. Experiments using the Kenya data suggested that the discontinuities and noise could be prevented by increasing $m_v$ and $m_{v,p}$ (results not shown), however for the dataset on the continent scale there was a significant computational cost for doing so. Increasing $m_{v,p}$ from $30$ to $150$ while keeping $m_v$ fixed at $30$ had a relatively small impact on the computation time, which increased from $7.4$ to $9$ minutes, but the required memory jumped from 1289MB to 20546MB. Meanwhile, increasing both $m_v$ and $m_{v,p}$ to 150, greatly increased the computation time, requiring over 3.4 hours to run, much longer than when the Vecchia approximation was not applied. Additionally, this model configuration required 20679MB of RAM. These examples suggest that increasing $m_v$ primarily increases the computation time required without affecting the RAM usage, while increasing $m_{v,p}$ increases the required RAM, with a smaller impact on computation time. Despite the increased computational requirements,  neither adjustment to the parameters completely removed the noise and discontinuities. 

\begin{figure}
    \centering
    \includegraphics[width = 0.75\textwidth]{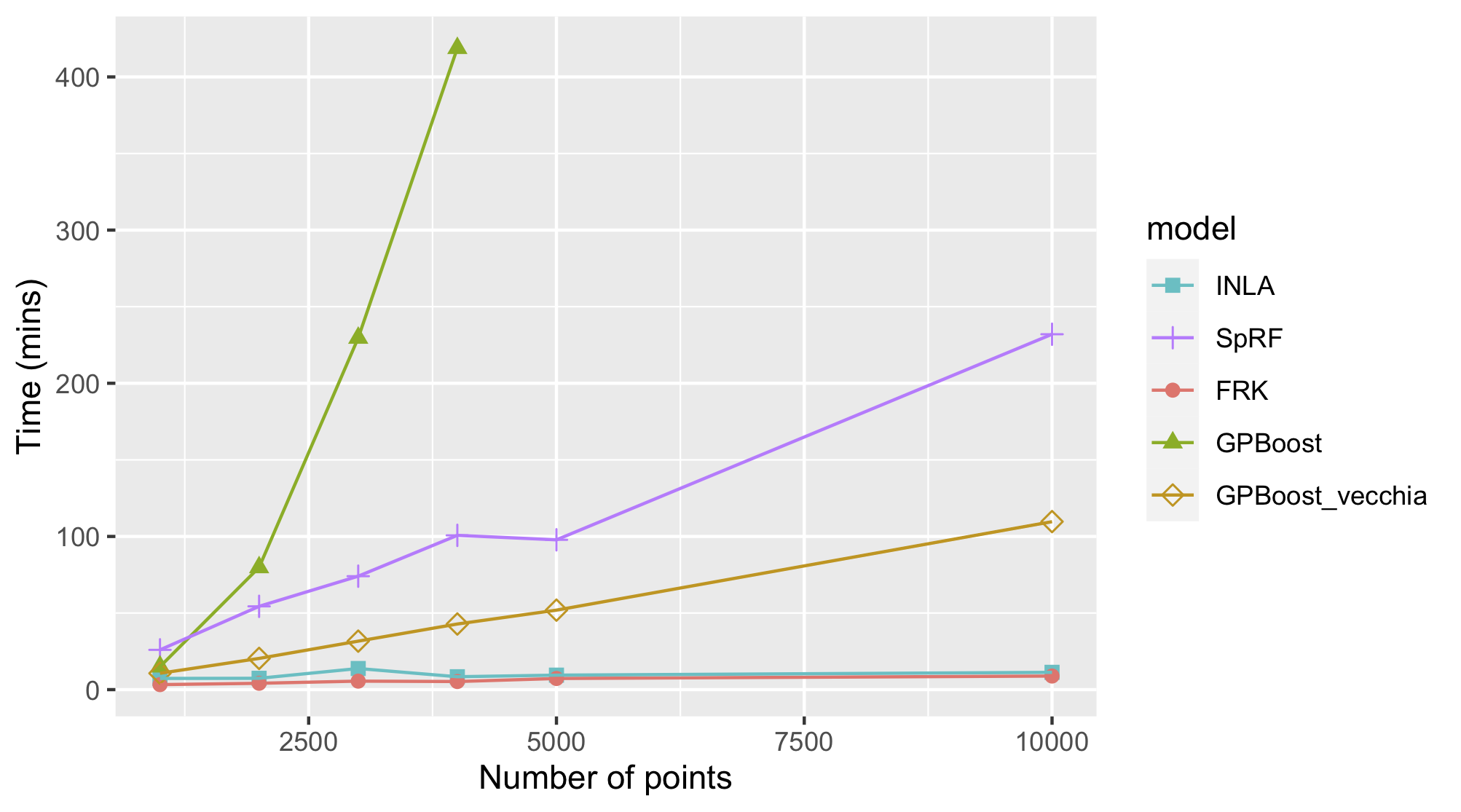}
    \caption{Time taken for GPBoost with the Vecchia approximation applied for simulated datasets of various sizes, compared to the times in Figure~\ref{fig:time_vs_n}.}
    \label{fig:time_vs_n_vecchia}
\end{figure}

One benefit of using the Vecchia approximation is an improvement in scaling behaviour, even if the computational requirements on an individual dataset depend strongly on the choice of $m_v$ and $m_{v,p}$. Figure~\ref{fig:time_vs_n_vecchia} shows the computational results from Figure~\ref{fig:time_vs_n}, with an additional plot for GPBoost using the Vecchia approximation. Parameters $m_v$ and $m_{v,p}$ were held fixed at $30$ and $150$ respectively, and the model shows a linear increase in computation time as the number of observations increases. However, the scaling is still more severe than for INLA and FRK. These results highlight a significant obstacle to applying GPBoost to large scale data. Using the full Gaussian process can result in large computation times, while applying the Vecchia approximation introduces additional artifacts which require sacrifices in computational efficiency to remove.

GPBoost's computation times are amplified by the high value of the \texttt{nrounds} parameter, which has been set to $247$ following available tutorials. As described in Section~\ref{sec:GPBoost_model}, this parameter controls the number of optimisation steps during fitting. When fit to the malaria datasets used throughout this paper, the log-likelihood generally stopped increasing after 5 to 10 steps, suggesting that 247 training steps is unnecessarily high for our data. Reducing \texttt{nrounds} to a value around $10$ would greatly improve the scaling gradient for the GPBoost model with a Vecchia approximation in Figure~\ref{fig:time_vs_n_vecchia}. Additional experimentation however found that reducing the number of rounds had little affect on the high RAM requirements for large values of $m_{v,p}$; something which may be necessary to minimise the discontinuities and noise in the predictions. Reducing \texttt{nrounds} would also improve the efficiency of the GPBoost model when no vecchia approximation is used, however would not change the overall scaling behaviour.

\end{document}